\documentclass[fp,twocolumn]{jpsj3}
\usepackage{txfonts}
\usepackage{color}
\usepackage[dvipdfmx]{graphicx}

\title{
Critical Slowing Down of Quadrupole and Hexadecapole Orderings in Iron Pnictide Superconductor
}

\author{
Ryosuke Kurihara$^1$ \thanks{E-mail : r.kurihara@phys.sc.niigata-u.ac.jp}, 
Keisuke Mitsumoto$^1$, 
Mitsuhiro Akatsu$^2$, 
Yuichi Nemoto$^{1, 5}$,
\\
Terutaka Goto$^{1, 5}$ \thanks{Corresponding author, Emeritus Prof. of Niigata Univ., E-mail : goto@phys.sc.niigata-u.ac.jp}, 
Yoshiaki Kobayashi$^{3, 5}$,
and
Masatoshi Sato$^{3, 4, 5}$ \thanks{Emeritus Prof. of Nagoya Univ., E-mail : msatojun7@gmail.com }
}

\inst{
$^1$ Graduate School of Science and Technology, Niigata University,
Niigata 950-2181, Japan \\
$^2$ Faculty of Science, Niigata University,
Niigata 950-2181, Japan \\
$^3$ Department of Physics, Nagoya University,
Nagoya 464-8602, Japan \\
$^4$ Research Center for Neutron Science and Technology, Comprehensive Research Organization for Science and Society (CROSS), 
Ibaraki 300-0811, Japan \\
$^5$ JST, Transformative Research-Project on Iron Pnictides (TRIP),
Chiyoda, Tokyo 102-0075, Japan
} 

\abst{
Ultrasonic measurements have been carried out to investigate the critical dynamics of structural and superconducting transitions due to degenerate orbital bands in iron pnictide compounds with the formula Ba(Fe$_{1-x}$Co$_x$)$_2$As$_2$. 
The attenuation coefficient $\alpha_{\mathrm{L}[110]}$ of the longitudinal ultrasonic wave for $(C_{11}+C_{12}+2C_{66})/2$ for $x = 0.036$ reveals the critical slowing down of the relaxation time around the structural transition at $T_\mathrm{s} = 65$ K, which is caused by ferro-type ordering of the quadrupole $O_{x'^2-y'^2}$ coupled to the strain  $\varepsilon_{xy}$.
The attenuation coefficient $\alpha_{66}$ of the transverse ultrasonic wave for $C_{66}$ for $x = 0.071$ also exhibits the critical slowing down around the superconducting transition at $T_\mathrm{SC} = 23$ K, which is caused by ferro-type ordering of the hexadecapole
$H_z^\alpha \bigl( \boldsymbol{r}_i, \boldsymbol{r}_j \bigr)
= O_{x'y'}\bigl( \boldsymbol{r}_i \bigr) O_{x'^2 - y'^2}\bigl( \boldsymbol{r}_j \bigr)
 + O_{x'^2 - y'^2}\bigl( \boldsymbol{r}_i \bigr) O_{x'y'}\bigl( \boldsymbol{r}_j \bigr)$
of the bound two-electron state coupled to the rotation $\omega_{xy}$.
It is proposed that the hexadecapole ordering associated with the superconductivity brings about spontaneous rotation of the macroscopic superconducting state with respect to the host tetragonal lattice.
}

\begin{document}
\maketitle

\section{Introduction}
Since the discovery of superconductivity in iron pnictide compounds with the formula LaFeAs(O$_{1-x}$F$_x$) with a high transition temperature of $T_\mathrm{SC} = 26$ K by Hosono and his coworkers in 2008 \cite{Kamihara}, many researchers have been involved in the search for various iron-based composites showing high superconducting transition temperatures and the inherent mechanism of the superconductivity.
Among the many iron-based compounds, the family of LaFeAsO with the 1111-type ZrCuSiAs structure shows transition temperatures as high as $T_\mathrm{SC} \sim 56$ K \cite{Ren}.
The family of BaFe$_2$As$_2$ with the 122-type ThCr$_2$Si$_2$ structure exhibits superconductivity with $T_\mathrm{SC} \sim 38$ K upon the substitution of barium for potassium \cite{Rotter1}.
The other family of LiFeAs with the 111-type PbFCl structure shows superconductivity with $T_\mathrm{SC} \sim 18$ K \cite{Tapp}, and FeSe with the 11-type PbO structure shows superconductivity with $T_\mathrm{SC} \sim 10$ K \cite{Hsu}.
These compounds have a common lattice structure with a two-dimensional square of iron layers, which gives the electronic band structure due to 3$d$ orbitals of Fe$^{2+}$ ions favorable for achieving superconductivity with high transition temperatures. 

The 122-type compounds of BaFe$_2$As$_2$ showing the superconductivity under either chemical doping or applying pressure have received particular attention \cite{Ni1, Drotziger, Fukazawa}, because high-quality single crystals with a fair size are available for experiments.
The end material of BaFe$_2$As$_2$ exhibits a structural phase transition from the tetragonal phase with space group $D_{4h}^{17}$ $(I4/mmm)$ to the orthorhombic phase with space group $D_{2h}^{23}$ $(Fmmm)$ simultaneously accompanied by antiferro-magnetic ordering with a stripe-type spin structure at transition temperatures of $T_\mathrm{s} = T_\mathrm{N} = 140$ K \cite{Rotter1, Rotter2, Huang}.
The chemical doping by substituting Co ions with 3$d^7$ electrons for Fe ions with $3d^6$ electrons in Ba(Fe$_{1-x}$Co$_x$)$_2$As$_2$ reduces both the structural and antiferro-magnetic transition temperatures while slightly splitting the two transition temperatures so that $T_\mathrm{s} > T_\mathrm{N}$ \cite{Ni2, Chu, Lester, Nandi}.
The Co-ion doped compounds with $x > 0.03$ reveal the superconductivity below the successive structural and magnetic transitions.
With increasing the Co dopant concentration $x$ to the quantum critical point (QCP) of $x_\mathrm{QCP} = 0.061$, the structural and antiferro-magnetic orderings disappear and the superconducting phase manifests itself.
Upon further doping over $x_\mathrm{QCP}$, the optimized superconducting transition temperature of as high as $T_\mathrm{SC} = 23$ K emerges in the compound with $x = 0.071$.

In order to clarify the interplay of the structural transition to the superconductivity in the iron pnictide Ba(Fe$_{1-x}$Co$_x$)$_2$As$_2$, the softening of the elastic constant $C_{66}$ as a precursor of the structural transition has been intensively investigated.
By using resonance ultrasonic spectroscopy, Fernandes \textit{et al}. first showed the softening of $C_{66}$ in Ba(Fe$_{1-x}$Co$_x$)$_2$As$_2$ for the end material and the over-doped compound with $x = 0.08$, the latter exhibiting a superconducting transition at $T_\mathrm{SC} = 16$ K, and proposed that the softening of $C_{66}$ is caused by spin nematic fluctuations of the Fe $3d^6$ electrons \cite{Fernandes}.
By using the ultrasonic pulse-echo method, Goto \textit{et al}. showed the softening of $C_{66}$ by $21\%$ for $x = 0.071$ from 300 K down to the optimized superconducting transition temperature $T_\mathrm{SC} = 23$ K, and deduced that the electric quadrupole $O_{x'^2-y'^2}$ associated with the degenerate $y'z$ and $zx'$ orbitals of the Fe$^{2+}$ ion coupled to the elastic strain $\varepsilon_{xy}$ of the transverse ultrasonic wave plays a role in the appearance of the superconductivity \cite{Goto}.
Here, the $x$ and $y$ ($x'$ and $y'$) coordinates for neighboring Ba-Ba (Fe-Fe) directions are adopted.
Yoshizawa \textit{et al}. systematically investigated the elastic constant $C_{66}$ for compounds with various Co concentrations and clarified the quantum criticality of the softening of $C_{66}$ around the QCP of  $x_\mathrm{QCP} = 0.061$ \cite{Yoshizawa}. 

Angle-resolved photoemission spectroscopy experiments on Ba$_{0.6}$K$_{0.4}$Fe$_2$As$_2$ \cite{ARPES_Ding}, muon resonance experiments on Ba$_{1-x}$K$_x$Fe$_2$As$_2$ \cite{Muon}, microwave penetration depth measurements of PrFeAsO$_{1-y}$ \cite{Microwave}, and NMR relaxation rate measurements of LnFe$_{1-y}$M$_y$AsO$_{1-x}$F$_x$  \cite{M. Sato} commonly show mostly isotropic and nodeless superconducting energy gaps with an s-like shape.  
The robustness of the superconducting transition temperature $T_\mathrm{SC}$ for nonmagnetic impurity doping to the system is favorable for the sign-conserving s-wave state of s$_{++}$ \cite{M. Sato, S. C. Lee, Kawamata}.
Taking the quadrupole fluctuations associated with the elastic softening of $C_{66}$ into account, the s$_{++}$ state for the superconducting energy gap has been theoretically deduced \cite{Yanagi_1, Kontani, Onari}.
On the other hand, neutron scattering experiments on Ba(Fe$_{0.925}$Co$_{0.075}$)$_2$As$_2$ have shown a magnetic excitation peak around $\boldsymbol{q} = (1/2, 1/2, 1)$, indicating a role of the antiferro-magnetic fluctuations in the superconductivity \cite{Inosov}.
Accordingly, the sign-reversing s$_\pm$-wave state has been presented while emphasizing the spin fluctuation effects due to the antiferro-magnetic interaction \cite{STM, Mazin1, Kuroki}.
The clarification of the superconducting mechanism of the iron pnictides is still an important issue in solid-state physics.  

Since ultrasonic waves with frequencies as high as 100 MHz easily penetrate into metals, attenuation measurements are useful for examining order parameter dynamics around phase transitions in metals.
There are a few reports on ultrasonic attenuation measurements around the structural and superconducting transitions of the iron pnictides \cite{Simayi, Saint-Paul}.
However, the attenuation of ultrasonic waves relating the elastic constant $C_{66}$ has not been reported so far to the best of our knowledge.
In the present paper, we show the critical dynamics around the structural and superconducting transitions in the iron pnictide Ba(Fe$_{1-x}$Co$_x$)$_2$As$_2$ by means of ultrasonic measurements. 

This paper is organized as follows.
In Sect. 2, the experimental procedures of sample preparation and ultrasonic measurements are shown.
In Sect. 3, we present experimental results of the critical slowing down of order parameters around the superconducting transition for $x = 0.071$ and the structural transition for $x = 0.036$.
The critical temperatures characterizing the phase diagram of the system are shown.
In Sect. 4, we demonstrate the theory treating the quantum degrees of freedom of the electric quadrupoles $O_{x'^2-y'^2}$ and $O_{x'y'}$ and the angular momentum $l_z$ associated with the degenerate $y'z$ and $zx'$ orbitals and also their interactions with the transverse acoustic phonons accompanying rotation and strain.
In addition, we show plausible models of the quadrupole ordering for the structural transition and the hexadecapole ordering for the superconductivity in the iron pnictide  Ba(Fe$_{1-x}$Co$_x$)$_2$As$_2$. 
We show that the superconducting Hamiltonian due to the quadrupole interaction gives the superconducting ground state bearing the hexadecapole.
In Sect. 5, we present conclusions.

\section{Experiments}

Single crystals of Ba(Fe$_{1-x}$Co$_x$)$_2$As$_2$ with Co concentration $x$ were grown by the Bridgman method.
Energy-dispersive X-ray spectrometry showed that nominal concentrations of $x_\mathrm{nominal} = 0.03, 0.07$, and $0.10$ corresponded to actual concentrations of $x = 0.017, 0.036$, and $0.071$, respectively.
A Laue X-ray camera was used to determine the crystallographic orientation of the crystals.
Samples with sizes of $2.30 \times 1.05 \times 0.44$ mm$^3$ for the end material $x = 0$, $3.11 \times 5.35 \times 0.77$ mm$^3$ for $x = 0.017$, $2.13 \times 2.06 \times 0.26$ mm$^3$ for $x = 0.036$, and $1.52 \times 2.61 \times 0.83$ mm$^3$ for $x = 0.071$ were used.
The elastic constant $C = \rho_\mathrm{M}(x)v^2$ was calculated from the ultrasonic velocity $v$.
Mass densities of $\rho_\mathrm{M}(x) = 6.50$ g/cm$^3$ for $x =$ 0, 0.017, and 0.036 and $\rho_\mathrm{M}(x = 0.071) = 6.51$ g/cm$^3$ were estimated using the Vegard's law for the lattice constants in Ba(Fe$_{1-x}$Co$_x$)$_2$As$_2$ \cite{Sefat}.
Piezoelectric ultrasonic transducers with a 36$^\circ$ Y-cut LiNbO$_3$ plate for the generation and detection of longitudinal waves and an X-cut plate for transverse waves were glued on plane-parallel surfaces of the samples.
LiNbO$_3$ plates with thicknesses of 200 and 100 $\mu$m were used to generate ultrasonic waves with frequencies as high as 260 MHz. 
Because the present crystal shows a large amount of softening with a considerably small elastic constant $C_{66}$, even a small external stress $\sigma_{xy}$ might easily deform the sample with the tetragonal lattice to an orthorhombic deformed lattice with finite $\varepsilon_{xy}$.
This unwanted lattice deformation might easily lift the degenerate $y'z$ and $zx'$ orbital states.
This lifting due to careless sample treatment would prevent accurate measurements of the softening of $C_{66}$ and the critical divergence of the ultrasonic attenuation across the structural and superconducting transitions.
In order to avoid applying undesired stress $\sigma_{xy}$ to the samples, we carefully kept each sample in a brass disk during the ultrasonic experiments.
A vector-type detector based on the ultrasonic pulse-echo method was used for simultaneous measurements of the attenuation coefficient $\alpha$ and velocity $v$.
A $^3$He cryostat (Oxford Heliox TL) was employed for the ultrasonic measurements down to 250 mK.

\section{Results}

\subsection{Attenuation around superconducting transition for $x = 0.071$}

\begin{figure}[htbp]
\begin{center}
\includegraphics[angle=0,width=0.50\textwidth, bb=10 14 439 525]{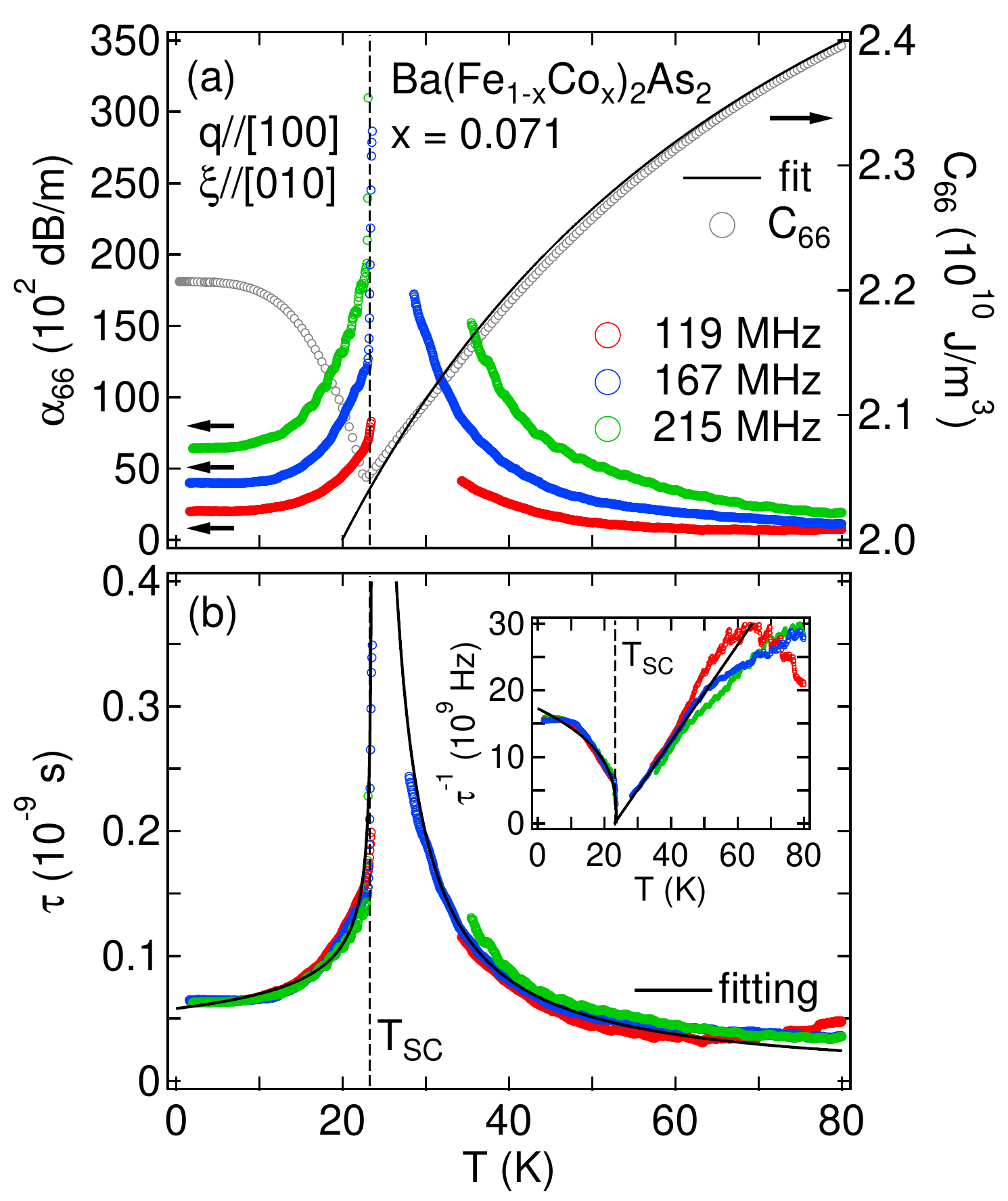}
\end{center}
\caption{
(Color online)
(a) Temperature dependence of the ultrasonic attenuation coefficient $\alpha_{66}$ measured using transverse ultrasonic waves with frequencies of 119, 167, and 215 MHz, and the elastic constant $C_{66}$ in Ba(Fe$_{1-x}$Co$_x$)$_2$As$_2$ with $x = 0.071$ having the superconducting transition temperature $T_\mathrm{SC} = 23$ K.
The solid line is the softening of $C_{66}$ in the normal phase fit by
$C_{66} = C_{66}^0 \bigl( T - T_\mathrm{s}^0 \bigr) / \bigl( T - \mathit{\Theta}_\mathrm{Q} \bigr)$
with $T_\mathrm{s}^0 = -26.5$ K and $\mathit{\Theta}_\mathrm{Q} = -47$ K.
(b) Temperature dependence of the relaxation time $\tau$ obtained by the attenuation coefficient $\alpha_{66}$ for frequencies 119, 167, and 215 MHz. Solid lines show fits by $\tau = \tau_0 \varepsilon^{- z \nu}$ for the reduced temperature
$\varepsilon = \left| T - T_\mathrm{c}^0 \right| / T_\mathrm{c}^0$ with $T_\mathrm{c}^0 = 23$ K and $z \nu = 1$.
The inset in (b) shows the temperature dependence of the relaxation rate $\tau^{-1}$.
Solid lines in both the normal and superconducting phases show fits by $\tau^{-1} = \tau_0^{-1}  \varepsilon^{z \nu}$.
}
\label{Fig1}
\end{figure}

In order to investigate the dynamical features of the superconducting transition while clearly distinguishing them from the structural and antiferro-magnetic orderings, we focused on the over-doped compound $x = 0.071$, which exhibits the optimized superconducting transition temperature $T_\mathrm{SC} = 23$ K but neither a structural transition nor antiferro-magnetic ordering.
We measured the attenuation coefficient $\alpha_{66}$ using the transverse ultrasonic waves with the propagation vector $\boldsymbol{q} // \left[ 100 \right]$ and polarization vector $\boldsymbol{\xi} // \left[ 010 \right]$ with frequencies of 119, 167, and 215 MHz. 
The attenuation coefficient $\alpha_{66}$ for $x = 0.071$ in Fig. \ref{Fig1}(a) increases with decreasing temperature below 80 K in the normal phase and reveals critical divergence with approaching the superconducting transition point of $T_\mathrm{SC} = 23$ K.
With further lowering the temperature below $T_\mathrm{SC} = 23$ K, the attenuation coefficient $\alpha_{66}$ rapidly decreases.
In both the normal and superconducting phases, the frequency dependence of the attenuation coefficient $\alpha_{66}$ obeys the $\omega^2$ law, consistent with the low-frequency regime of the Debye formula.

The elastic constant $C_{66}$ in the normal phase for $x = 0.071$ exhibits the softening, which is again shown in Fig. \ref{Fig1}(a) \cite{Goto}. 
This softening is caused by the interaction of the quadrupole $O_{x'^2-y'^2}$ to the strain $\varepsilon_{xy}$ expressed as
\begin{align}
\label{HQS exp.}
H_\mathrm{QS} = -g_{x'^2-y'^2}O_{x'^2-y'^2} \varepsilon_{xy}.
\end{align}
Here, $g_{x'^2-y'^2}$ is the quadrupole-strain coupling constant.
We expect the Curie-type behavior of the quadrupole susceptibility proportional to the reciprocal temperature as $\chi_\mathrm{Q} = | \langle \psi_l | O_{x'^2-y'^2} | \psi_l \rangle |^2 / T$ for suffixes $l = y'z$ and $zx'$ of the degenerate orbitals.
The softening of $C_{66}$ in the normal phase is written by \cite{Kataoka and Kanamori}
\begin{align}
\label{Temp. dep. C66}
C_{66}
 = C_{66}^0 - \frac{ n_\mathrm{Q} g_{x'^2-y'^2}^2\chi_\mathrm{Q} } {1 - g'_{x'^2-y'^2}\chi_\mathrm{Q}} 
= C_{66}^0\left(\frac{T - T_\mathrm{s}^0} {T - \mathit{\Theta}_\mathrm{Q} }\right).
\end{align}
Here, $C_{66}^0$ is the background elastic constant when softening is absent and $g'_{x'^2-y'^2}$ is the mutual interaction coefficient between the quadrupoles $O_{x'^2-y'^2}$ at different sites.
The solid line in Fig. \ref{Fig1}(a) is the softening of $C_{66}$ fit by Eq. (\ref{Temp. dep. C66}) in the normal phase.
We obtain negative values of the quadrupole interaction energy $\mathit{\Theta}_\mathrm{Q} = - 47.0$ K and the critical temperature $T_\mathrm{s}^0 = \mathit{\Theta}_\mathrm{Q} + \mathit{\Delta}_\mathrm{Q}  = -26.5$ K.
This fitting gives the quadrupole-strain interaction energy $\mathit{\Delta}_\mathrm{Q} = 20.5$ K, which represents the coupling energy between the quadrupoles at different sites mediated by the strain $\varepsilon_{xy}$.

The negative quadrupole interaction energy $\mathit{\Theta}_\mathrm{Q} = - 47.0$ K indicates the antiferro-type quadrupole interaction for $x = 0.071$.
The negative critical temperature of $T_\mathrm{s}^0 = -26.5$ K corresponds to the fictitious critical temperature of the structural instability for $C_{66} \rightarrow 0$.
Consequently, the superconducting transition temperature $T_\mathrm{SC} = 23$ K is definitely distinguished from the fictitious critical temperature $T_\mathrm{s}^0 = -26.5$ K.
We conclude that the marked increase in the attenuation coefficient $\alpha_{66}$ for $x = 0.071$ in Fig. \ref{Fig1}(a) is caused by the critical slowing down of the order parameter around the superconducting transition, which is strictly distinguished from the quadrupole $O_{x'^2-y'^2}$ interacting to the strain $\varepsilon_{xy}$ in Eq. (\ref{HQS exp.}).

Because two electrons are accommodated in the degenerate $y'z$ and $zx'$ orbitals of an Fe$^{2+}$ ion, we tentatively adopt $n_\mathrm{Q} = 2N_\mathrm{Fe} = 3.92 \times 10^{22}$ cm$^{-3}$ in Eq. (\ref{Temp. dep. C66}) as the maximum number of electrons, that play a role in the softening of $C_{66}$.
Here, $N_\mathrm{Fe}$ is the number of Fe$^{2+}$ ions per unit volume.
The solid line in Fig. \ref{Fig1}(a) for the softening of $C_{66}$ fit by Eq. (\ref{Temp. dep. C66}) gives the quadrupole-strain coupling constant $g_{x'^2-y'^2} = 1045$ K per electron.
This coupling constant is comparable with the results for  manganese compounds of $g = 1167$ K for La$_{0.88}$Sr$_{0.12}$MnO$_3$ and $g = 1020$ K for Pr$_{0.65}$Ca$_{0.35}$MnO$_3$ with considerably extended 3$d$ orbitals \cite{Hazama La, Hazama Pr}, but is considerably larger than $g \sim100$ K for rare-earth compounds of the form RB$_6$ with well-screened 4$f$ orbitals in the inner shell \cite{Nakamura}.

In our attempt to clarify the order parameter dynamics around the superconducting transition, we analyzed the frequency dependence of the attenuation coefficient $\alpha_{66}$ in Fig. \ref{Fig1}(a) in terms of the Debye formula expressed as
\begin{align}
\label{Debye-type}
\alpha_{66}
 = \frac{C_{66} \left( \infty \right) - C_{66} \left( 0 \right)} {2\rho_\mathrm{M} {v_{66} \left( \infty \right) }^3} \frac{\omega^2\tau} {1 + \omega^2\tau^2}.
\end{align}
Here, $C_{66} \left( \infty \right)$ and $C_{66} \left( 0 \right)$ stand for the high- and low-frequency limits of the elastic constant $C_{66}$, respectively.
We regard the elastic constant experimentally observed in Fig. \ref{Fig1}(a) as the low-frequency limit $C_{66} \left( 0 \right)$ in Eq. (3).
The background $C_{66}^0$ used in the analysis based on the quadrupole susceptibility of Eq. (\ref{Temp. dep. C66}) corresponds to the high-frequency limit  of the elastic constant $C_{66} \left( \infty \right)$ for the sound velocity $v_{66} \left( \infty \right) = \sqrt{C_{66}\left( \infty \right) / \rho_\mathrm{M} }$ in Eq. (\ref{Debye-type}) \cite{Goto}.
$\omega$ is the angular frequency of the employed ultrasonic waves, and $\tau$ is the relaxation time of the order parameter fluctuation of the system.
We show the temperature dependence of the relaxation time $\tau$ obtained by Eq. (\ref{Debye-type}) in Fig. \ref{Fig1}(b) and that of the relaxation rate $\tau^{-1}$ in the inset of Fig. \ref{Fig1}(b).
The rapid increase in the relaxation time $\tau$ around $T_\mathrm{SC} = 23$ K indeed indicates the critical slowing down of the order parameter accompanying the superconducting transition in the system.  

The correlation length $\zeta$ of the order parameter $Q \left( \boldsymbol{r} \right)$ for an appropriate phase transition is widely used to describe the correlation function as
$G \left( \boldsymbol{r} \right)
= \left\langle Q \left( \boldsymbol{r} \right) Q \left( 0 \right) \right\rangle
\propto r^{- \left( d - 1 \right)} e^{- r / \zeta}$.
Here, $| \boldsymbol{r} | = r$ stands for the spatial distance between the order parameters and $d$ denotes the dimension of the system.
In the vicinity of the critical temperature $T_\mathrm{c}^0$ of the phase transition, the correlation length $\zeta$ becomes infinite due to the increase in the local order as $\zeta = \zeta_0 \varepsilon ^{- \nu}$ for a reduced temperature of $\varepsilon = \bigl| T - T_\mathrm{c}^0 \bigr| / T_\mathrm{c}^0 $.
The critical index $\nu =1/2$ is expected from mean field theory \cite{Nishimori}. 

When an ultrasonic pulse wave enters a crystal, the equilibrium state of the system is instantaneously perturbed to a nonequilibrium state.
After the relaxation time $\tau$, the system returns to the equilibrium state.
In the vicinity of the critical point, however, the relaxation time $\tau$ becomes infinite due to the divergence of the correlation length as $\tau \propto \zeta^z$ for dynamical critical index $z$.
Presumably, the critical slowing down of the relaxation time $\tau$ is explained by the critical index $z\nu$ as \cite{Suzuki, Mori, Ma, Halperin and Hohenberg}
\begin{align}
\label{Relaxation time}
\tau
 = \tau_0 \left| \frac{T - T_\mathrm{c}^0} {T_\mathrm{c}^0} \right| ^{-z\nu} 
 = \tau_0\varepsilon^{-z\nu}.
\end{align}
In mean field theory, the dynamical critical index $z = 2$ is expected. 
Actually, the temperature dependence of the relaxation time $\tau$ in the normal phase of Fig. \ref{Fig1}(b) is well reproduced by the solid line for the critical index $z\nu = 1$, the critical temperature $T_\mathrm{c}^{0+} = 23.0$ K, and the attempt time $\tau_0^+ = 6.0 \times 10^{-11}$ s.
In the superconducting phase, however, we obtain $z\nu = 1/3$, $T_\mathrm{c}^{0-} = 23.5$ K, and $\tau_0^- = 5.8 \times 10^{-11}$ s. 
The distinct deviation of the critical index of $z\nu = 1/3$ from $z\nu = 1$ of mean field theory may be caused by the inherent property that the hexadecapole ordering appears in accordance with the superconductivity in the present iron pnictide.
The analysis of the experimental results in Fig. \ref{Fig1}(b)  gives a ratio of the attempt times of $ \tau_0^+ / \tau_0^- = 1.03$, which is distinguished from the ratio of $ \tau_0^+ / \tau_0^- = 2$ expected from mean field theory.
Because the ultrasonic echo signal almost disappears in the vicinity of the superconducting transition point due to the critical slowing down, the absolute values of the attenuation coefficients inevitably include experimental errors.

\subsection{Attenuation around structural transition for $x = 0.036$}

 Next, we examine the critical slowing down around the structural transition from the tetragonal to orthorhombic phases of the under-doped compound $x = 0.036$.
This is worth comparing with the critical slowing down around the superconducting transition of the over-doped compound $x = 0.071$ presented in Sect. 3.1. 

\begin{figure}[h]
\begin{center}
\includegraphics[angle=0,width=0.50\textwidth, bb=11 19 443 707]{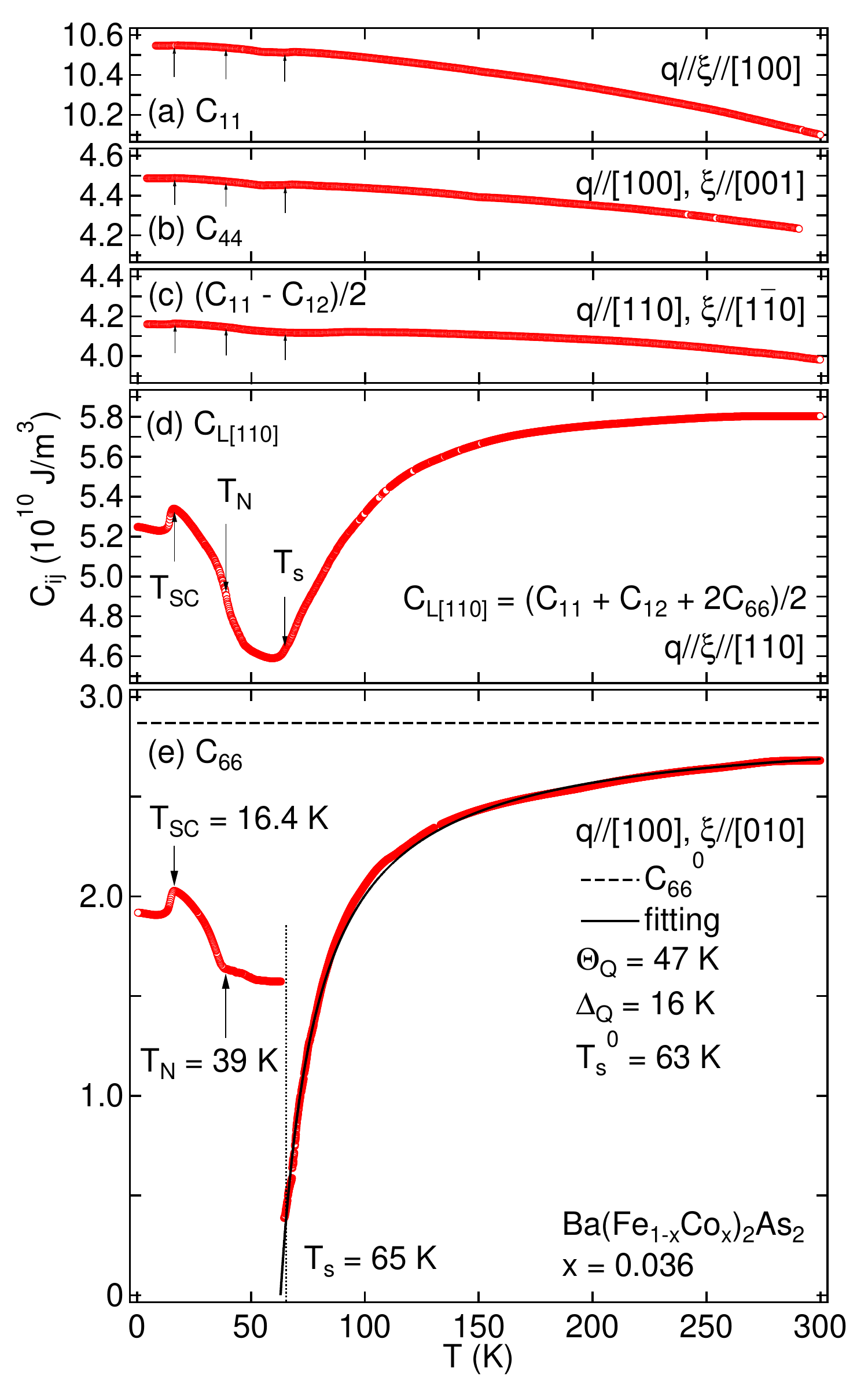}
\end{center}
\caption{
(Color online)
Temperature dependence of elastic constants of Ba(Fe$_{1-x}$Co$_x$)$_2$As$_2$ with $x = 0.036$.
The considerable softening of $C_{66}$ in (e) and $C_{ \mathrm{L}[110]} = \left( C_{11} + C_{12} + 2C_{66} \right) / 2$ in (d) are found to be a precursor of the structural transition at $T_\mathrm{s} = 65$ K.
Anomalies associated with the antiferro-magnetic transition at $T_\mathrm{N} = 39$ K and the superconducting transition at $T_\mathrm{SC} = 16.4$ K are observed.
The softening of $C_{66}$ in (e) is reproduced by a solid line fit by
$C_{66} = C_{66}^0 \bigl( T - T_\mathrm{s}^0 \bigr) / \bigl( T - \mathit{\Theta}_\mathrm{Q} \bigr)$
with $T_\mathrm{s}^0 = 63$ K and $\mathit{\Theta}_\mathrm{Q} = 47$ K.
}
\label{Fig2}
\end{figure}

The temperature dependence of the elastic constants for $x = 0.036$ is shown in Fig. \ref{Fig2}.
We denote the propagation vector as $\boldsymbol{q}$ and the polarization vector as $\boldsymbol{\xi}$ for the ultrasonic waves measured in Fig. \ref{Fig2}.
The elastic constant $C_{66}$ in Fig. \ref{Fig2}(e) measured by the transverse ultrasonic wave reveals considerable softening of 85$\%$ with decreasing temperature from 300 K down to the structural transition temperature of $T_\mathrm{s} = 65$ K.
As indicated by arrows in Fig. \ref{Fig2}(e), the antiferro-magnetic ordering at $T_\mathrm{N} = 39$ K and the superconducting transition at $T_\mathrm{SC}  = 16.4$ K are also observed as anomalies in $C_{66}$.
The softening of $C_{66}$ in the tetragonal phase is reproduced by the solid line in Fig. \ref{Fig2}(e), which is obtained by Eq. (\ref{Temp. dep. C66}) for the quadrupole interaction energy of $\mathit{\Theta}_\mathrm{Q} = 47$ K, the quadrupole-strain interaction energy $\mathit{\Delta}_\mathrm{Q} = 16$ K, and the critical temperature of $T_\mathrm{s}^0 = \mathit{\Theta}_\mathrm{Q} + \mathit{\Delta}_\mathrm{Q} = 63$ K.
The background of $C_{66}^0 = 2.87 \times 10^{10}$ J/m$^3$ is shown by the dashed line in Fig. 2(e).
The critical temperature $T_\mathrm{s}^0 = 63$ K is in agreement with the experimentally observed structural transition temperature $T_\mathrm{s} = 65$ K.
We deduce the quadrupole-strain coupling constant to be $g_{x'^2 - y'^2} = 920$ K per electron by adopting the electron number $n_\mathrm{Q} = 2N_\mathrm{Fe}$ in Eq. (\ref{Temp. dep. C66}).
The positive values of the quadrupole interaction energy $\mathit{\Theta}_\mathrm{Q} = 47$ K and the critical temperature $T_\mathrm{s}^0 = 63$ K are consistent with the ferro-quadrupole ordering $O_{x'^2 - y'^2}$ accompanying the structural transition from the tetragonal to orthorhombic phase. 

 The elastic constant $C_{\mathrm{L} \left[110 \right]} = \left( C_{11} + C_{12} + 2C_{66} \right)/2$ in Fig. \ref{Fig2}(d) measured by a longitudinal ultrasonic wave with the propagation direction $\boldsymbol{q} // \left[110 \right]$ parallel to the polarization direction $\boldsymbol{\xi}$ also shows marked softening with decreasing temperature down to the structural transition at $T_\mathrm{s} = 65$ K.
With further decreasing temperature in the distorted orthorhombic phase, the graph of the elastic constant $C_{\mathrm{L} \left[110 \right]}$ slightly bend at the antiferro-magnetic transition temperature $T_\mathrm{N} = 39$ K and distinctly decreases at the superconducting transition temperature $T_\mathrm{SC} = 16.4$ K.
The longitudinal ultrasonic wave for $C_{\mathrm{L} \left[110 \right]}$ induces elastic strain of $\varepsilon_{ \mathrm{L} \left[110 \right] } = \varepsilon_\mathrm{B}/3 - \varepsilon_u/ \bigl( 2\sqrt{3} \bigr) + \varepsilon_{xy}$.
The coupling of the strain $\varepsilon_{xy}$ to the quadrupole $O_{x'^2 - y'^2}$ induces the softening in $C_{\mathrm{L} \left[110 \right]}$, partly consisting of $C_{66}$.
The volume strain $\varepsilon_\mathrm{B} = \varepsilon_{xx} + \varepsilon_{yy} + \varepsilon_{zz}$ and the tetragonal strain $\varepsilon_u = \bigl( 2\varepsilon_{zz} - \varepsilon_{xx} - \varepsilon_{yy} \bigr) / \sqrt{3}$ scarcely affect the softening. 

\begin{figure}[h]
\begin{center}
\includegraphics[angle=0,width=0.50\textwidth, bb= 12 17 450 529]{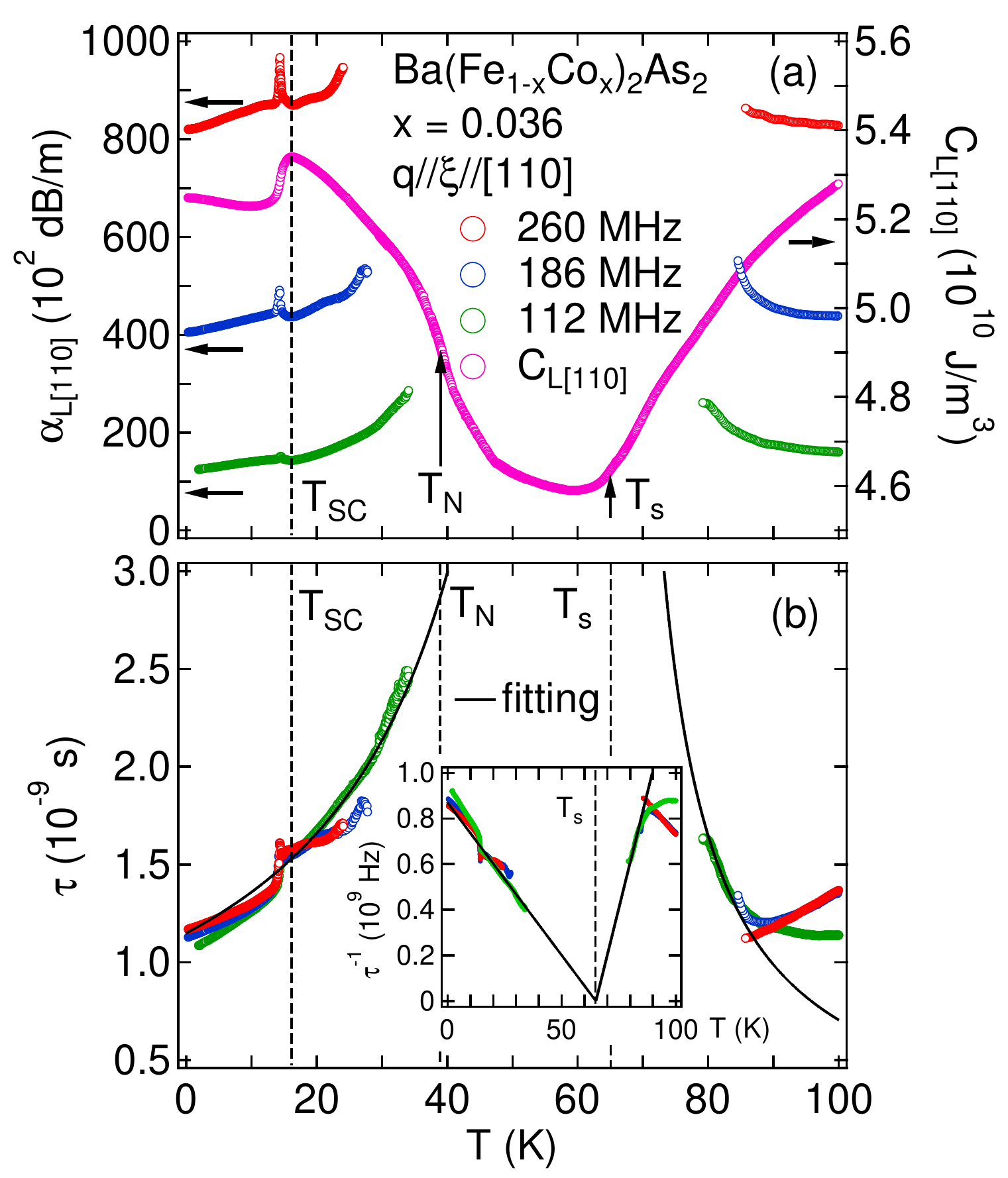}
\end{center}
\caption{
(Color online)
(a) Temperature dependence of the ultrasonic attenuation coefficient $\alpha_{ \mathrm{L}[110]}$ measured using longitudinal ultrasonic waves with frequencies of 112, 186, and 260 MHz, and the elastic constant $C_{ \mathrm{L}[110]} = \left( C_{11} + C_{12} + 2C_{66} \right) / 2$ in Ba(Fe$_{1-x}$Co$_x$)$_2$As$_2$ with $x = 0.036$. 
(b) Temperature dependence of the relaxation time $\tau$ obtained from the results for 112, 186, and 260 MHz ultrasonic waves.
Solid lines indicate fits by $\tau = \tau_0 \varepsilon^{- z \nu}$ for the reduced temperature $\varepsilon = \left| T - T_\mathrm{c}^0 \right| / T_\mathrm{c}^0$ with $T_\mathrm{s}^0 = 65$ K and the critical index $z\nu = 1$ for both tetragonal and orthorhombic phases.
The inset shows the temperature dependence of the relaxation rate $\tau^{-1}$ and the solid lines are fits by $\tau^{-1} = \tau_0^{-1}  \varepsilon^{z \nu}$.
}
\label{Fig3}
\end{figure}

We initially attempted to measure the attenuation coefficient $\alpha_{66}$ by using the pure transverse ultrasonic wave.
However, the attenuation of  $\alpha_{66}$ near the structural transition was too large to measure.
In the present experiments, therefore, we used the longitudinal ultrasonic wave, which shows relatively moderate damping, to measure the attenuation coefficient $\alpha_{\mathrm{L} \left[110 \right]}$.
In Fig. \ref{Fig3}(a), we show the attenuation coefficient $\alpha_{\mathrm{L} \left[110 \right]}$ for $x = 0.036$ acquired using longitudinal ultrasonic waves with frequencies of 112, 186, and 260 MHz together with the elastic constant $C_{\mathrm{L} \left[110 \right]}$.
The attenuation coefficient $\alpha_{\mathrm{L} \left[110 \right]}$ increases considerably with approaching the structural transition point of $T_\mathrm{s} = 65$ K from both sides.
With decreasing temperature in the orthorhombic phase, the attenuation $\alpha_{\mathrm{L} \left[110 \right]}$ exhibits a distinct peak at the superconducting transition point of $T_\mathrm{SC} = 16.4$ K.
The missing ultrasonic echo signal due to considerable damping of the longitudinal ultrasonic wave prevents us from acquiring $\alpha_{\mathrm{L} \left[110 \right]}$ approaching the structural transition.
Nevertheless, the distinct tendency of the divergence of $\alpha_{\mathrm{L} \left[110 \right]}$ around the structural transition at $T_\mathrm{s} = 65$ K indicates the critical slowing down of the relaxation time $\tau$ for the order parameter fluctuation.

The frequency dependence for $\alpha_{\mathrm{L} \left[110 \right]}$ for the under-doped compound $x = 0.036$ in Fig. \ref{Fig3}(a) is analyzed in terms of the Debye formula in Eq. (\ref{Debye-type}), where we read the attenuation as $\alpha_{\mathrm{L} \left[110 \right]}$ and the elastic constant as $C_{\mathrm{L} \left[110 \right]}$.
We depict the temperature dependence of the relaxation time $\tau$ in Fig. \ref{Fig3}(b) and the relaxation rate $\tau^{-1}$ in the inset of Fig. \ref{Fig3}(b).
The increase in the relaxation time $\tau$ around the structural transition at $T_\mathrm{s} = 65$ K is caused by the critical slowing down of the quadrupole $O_{x'^2 - y'^2}$, which is the order parameter of the structural transition.
The ultrasonic frequencies of $\omega / 2\pi$ up to 260 MHz used in the present experiments are much higher than the relaxation rate $\tau^{-1} < 100$ MHz in the vicinity of the structural transition point $T_\mathrm{s}$ for the narrow reduced temperature region of $\varepsilon = | T - T_\mathrm{c}^0|/T_\mathrm{c}^0 < 0.038$.
As a result, the ultrasonic echo signal disappears on both sides around the structural transition point of $T_\mathrm{s} = 65$ K as depicted in Fig. 3(a).
The solid lines in Fig. \ref{Fig3}(b) are fits for the relaxation time $\tau$ in terms of Eq. (\ref{Relaxation time}).
Supposing that the critical temperature $T_\mathrm{c}^0 = 65$ K coincides with the structural transition point and the critical indices for both phases are consistent with $z\nu = 1$ of mean field theory, we obtain the fit shown by the solid line in Fig. \ref{Fig3}(b) using Eq. (\ref{Relaxation time}).
This gives the attempt time $\tau_0^+ = 0.380 \times 10^{-9}$ s for the tetragonal phase and $\tau_0^- = 1.15 \times 10^{-9}$ s for the orthorhombic phase.

The quadrupole-strain interaction of Eq. (\ref{HQS exp.}) induces the ferro-quadrupole ordering accompanying the structural transition.
It is expected that the long-range Coulomb force between electrons bearing the quadrupole exhibits critical phenomena described by mean field theory.
Actually, the adoption of the critical index $z\nu = 1$ for both tetragonal and orthorhombic phases consistent with mean field theory reproduces the experimental results in Fig. \ref{Fig3}.
The small-echo signal due to the considerable damping of the longitudinal ultrasonic waves gives inevitable errors in the absolute-value of the attenuation coefficient $\alpha_{\mathrm{L}[110]}$ in Fig. \ref{Fig3}.
The dispersive scattering of the ultrasonic wave by the domain wall due to the orthorhombic distortion is also included in the attenuation. 
These ambiguity might have resulted in the deviation of the experimentally determined ratio of $\tau_0^+ / \tau_0^- = 0.33$ from $\tau_0^+ / \tau_0^- = 2$ expected from mean field theory.
The attempt time $\tau_0^+ = 0.55 \times 10^{-9}$ s of the structural transition for $x = 0.036$ in Fig. \ref{Fig3}(b) is one order of magnitude larger than the attempt time $\tau_0^+ = 6.0 \times 10^{-11}$ s of the superconducting transition for $x = 0.071$ in Fig. \ref{Fig1}(b).
This notable result suggests that the order parameter showing the critical slowing down around the superconducting transition for $x = 0.071$ is distinguished from the ferro-quadrupole ordering accompanying the structural transition for $x = 0.036$.

\subsection{Critical temperatures of the system}

\begin{figure}[h]
\begin{center}\includegraphics[angle=0,width=0.46\textwidth, bb=5 20 384 472]{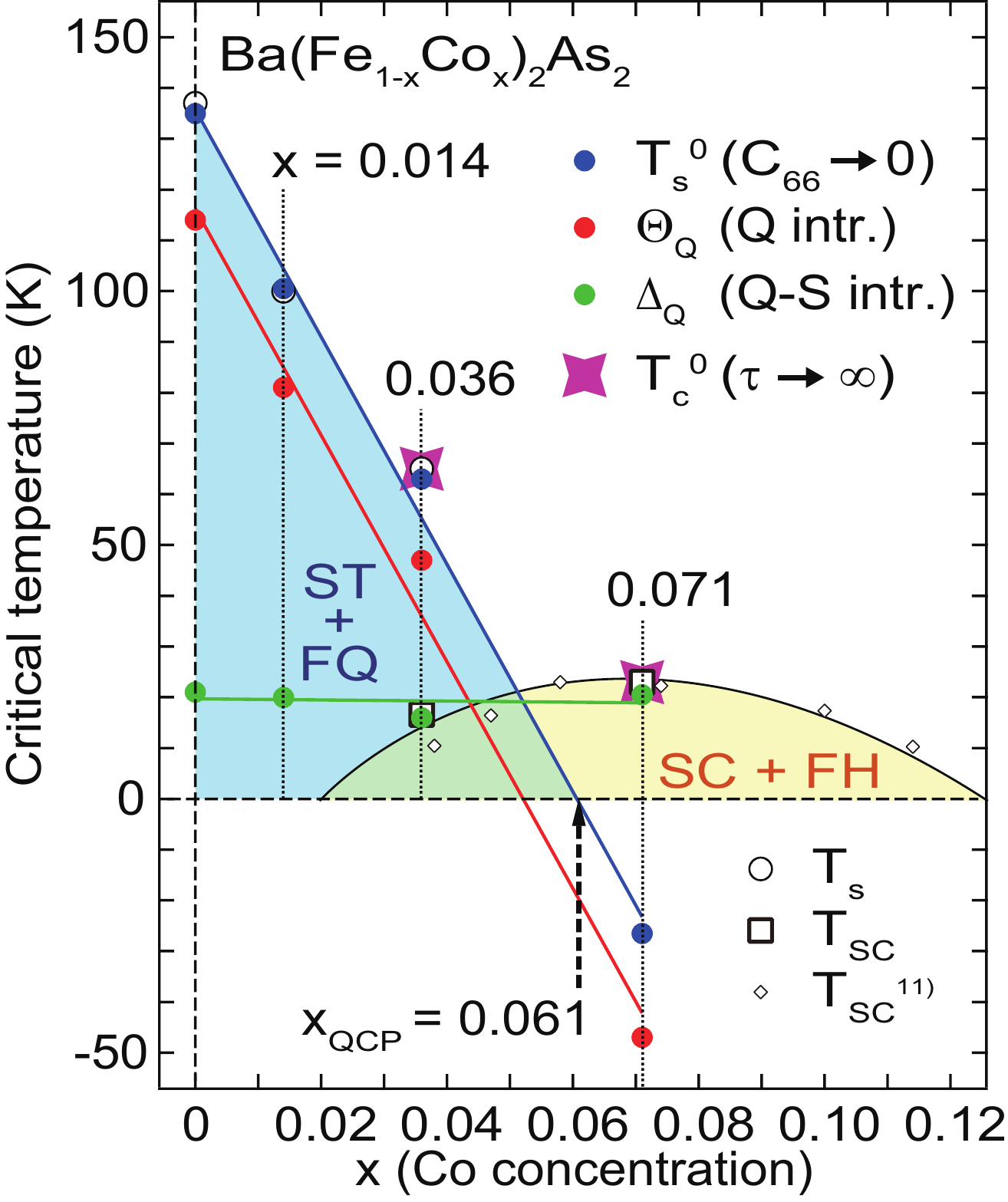}
\end{center}
\caption{
(Color online)
Critical temperatures relating to structural (ST) and superconducting (SC) transitions in Ba(Fe$_{1-x}$Co$_x$)$_2$As$_2$ determined by the present ultrasonic measurements.
The quadrupole interaction (Q intr.) energy $\mathit{\Theta}_\mathrm{Q}$, quadrupole-strain interaction (Q-S intr.) energy $\mathit{\Delta}_\mathrm{Q}$, and structural transition temperature $T_\mathrm{s}^0 = \mathit{\Theta}_\mathrm{Q} + \mathit{\Delta}_\mathrm{Q}$ associated with the ferro-quadrupole ordering are determined by the elastic constant $C_{66} \rightarrow 0$.
The critical temperature $T_\mathrm{c}^0$ is determined by the critical slowing down of the relaxation time $\tau \rightarrow \infty$ measured by the ultrasonic attenuation coefficient $\alpha$.
Structural transition temperatures $T_\mathrm{s}$ associated with ferro-quadrupole (FQ) ordering are indicated by black open circles.
Superconducting transition temperatures $T_\mathrm{SC}$ associated with ferro-hexadecapole (FH) ordering are indicated by black open squares. 
Superconducting transition temperatures indicated by black open rhombuses are results of Ni $et$ $al$. \cite{Ni2}. 
\label{Fig4}
}
\end{figure}

The ultrasonic measurements of the critical divergence in the attenuation coefficients and the softening in the elastic constants provided us with the critical temperatures, which characterize the structural and superconducting transitions of the present iron pnictide Ba(Fe$_{1-x}$Co$_x$)$_2$As$_2$.
In Fig. \ref{Fig4}, we plot the critical temperatures obtained by the experimental results for $x = 0.036$ and 0.071 together with those for $x = 0.017$ and the end material $x = 0$ \cite{Kurihara Dr}. 

The softening of the elastic constant $C_{66}$ analyzed by the quadrupole susceptibility $\chi_\mathrm{Q}$ in Eq. (\ref{Temp. dep. C66}) for $O_{x'^2 - y'^2}$ gives the critical temperature $T_\mathrm{s}^0 = \mathit{\Theta}_\mathrm{Q} + \mathit{\Delta}_\mathrm{Q}$  corresponding to the structural instability point as $C_{66} \rightarrow 0$.
As shown by blue closed circles and the blue solid line in Fig. \ref{Fig4}, the critical temperature of $T_\mathrm{s}^0 = 135$ K of the end material $x = 0$ decreases to $T_\mathrm{s}^0 = 100$ K for $x = 0.017$ and $T_\mathrm{s}^0 = 63$ K for $x = 0.036$.
Beyond the QCP of $x_\mathrm{QCP} = 0.061$, the critical temperature changes its sign to $T_\mathrm{s}^0 = -26.5$ K for $x = 0.071$.
The quadrupole interaction energy $\mathit{\Theta}_\mathrm{Q}$ shown by red closed circles and the red solid line decreases from $\mathit{\Theta}_\mathrm{Q} = 115$ K for the end material $x = 0$ to $\mathit{\Theta}_\mathrm{Q} = 80$ K for $x = 0.017$, $\mathit{\Theta}_\mathrm{Q} = 47$ K for $x = 0.036$, and changes to a negative value of $\mathit{\Theta}_\mathrm{Q} = -47$ K for $x = 0.071$.
On the other hand, the quadrupole-strain interaction energy $\mathit{\Delta}_\mathrm{Q}$ shown by green closed circles and the green solid line remains at the positive value $\mathit{\Delta}_\mathrm{Q} \sim 20$ K.
The positive critical temperatures $T_\mathrm{s}^0$ of the under-doped compounds are comparable with the ferro-quadrupole ordering $O_{x'^2 - y'^2}$ associated with the structural transition from the tetragonal to orthorhombic phase, while the negative critical temperature $T_\mathrm{s}^0 = -26.5$ K for the over-doped compound $x = 0.071$, suggesting absence of the ferro-quadrupole ordering, implies the fictitious lattice instability point.

The infinite divergence of the relaxation time $\tau$ due to the critical slowing down provides the critical temperature $T_\mathrm{c}^0$ at which the long-range ordering appears due to freezing of the order parameter fluctuation \cite{Halperin and Hohenberg, Ma}.
In the case of the structural transition for $x = 0.036$, the divergence of the relaxation time $\tau$ gives the critical temperature of $T_\mathrm{c}^0 = 65$ K, which coincides with the structural transition temperature $T_\mathrm{s} = 65$ K experimentally observed, and is in agreement with the critical temperature $T_\mathrm{s}^0 = \mathit{\Theta}_\mathrm{Q} + \mathit{\Delta}_\mathrm{Q} = 63$ K obtained by fitting the elastic softening of $C_{66}$ by Eq. (\ref{Temp. dep. C66}).
This result confirms that the critical divergence of the ultrasonic attenuation coefficient $\alpha_{\mathrm{L} \left[110 \right]}$ for $x = 0.036$ is caused by the critical slowing down of the ferro-quadrupole order parameter $O_{x'^2 - y'^2}$.
The critical temperature for $T_\mathrm{c}^0 = 65$ K for $x = 0.036$ is shown in Fig. \ref{Fig4} by a purple closed asterisk together with the critical temperature of $T_\mathrm{s}^0 = 63$ K. 
Note that the quadrupole-strain interaction energy $\mathit{\Delta}_\mathrm{Q} \sim 20$ K, which is almost independent of the Co concentration, also promotes the appearance of the structural transition at the critical temperature $T_\mathrm{s}^0 = \mathit{\Theta}_\mathrm{Q} + \mathit{\Delta}_\mathrm{Q}$.
The quadrupole-strain interaction energy $\mathit{\Delta}_\mathrm{Q}$ is shown by green closed circles in Fig. \ref{Fig4}.

In the case of the superconducting transition for $x = 0.071$, the critical temperature $T_\mathrm{c}^0 = 23$ K determined by the critical slowing down of the relaxation time $\tau$ is in good agreement with the superconducting transition temperature $T_\mathrm{SC} = 23$ K.
In Fig. \ref{Fig4}, we show the former critical temperature $T_\mathrm{c}^0 = 23$ K for the relaxation time $\tau$ by a purple closed asterisk together with the latter superconducting transition temperature  $T_\mathrm{SC}$ = 23 K by a black open square.
The critical temperature  $T_\mathrm{s}^0 = -26.5$ K indicated by a blue filled circle corresponds to the fictitious lattice instability point as $C_{66} \rightarrow 0$.
The fictitious instability point of $T_\mathrm{s}^0 = -26.5$ K is definitely distinguished from the critical temperature $T_\mathrm{c}^0 = 23$ K of the relaxation time $\tau$ coinciding with the superconducting transition temperature $T_\mathrm{SC}$.
This result evidences that the order parameter exhibiting the critical slowing down around the superconducting transition does not correspond to the electric quadrupole $O_{x'^2-y'^2}$ but to other quantum degrees of freedom, which interact with the transverse ultrasonic waves used in the experiments.
It is of great importance to identify the order parameter, which brings about the critical slowing down of the relaxation time $\tau$ around the superconducting transition.
In Sect. 4.6, we give the electric hexadecapole carried by a two-electron state as a plausible order parameter associated with the superconducting transition.

\section{Theory}

\subsection{Transverse acoustic wave}

The relaxation time $\tau$ determined by the ultrasonic attenuation reveals the critical slowing down around the superconducting transition for the compound $x = 0.071$ as well as the structural transition for $x = 0.036$.
In our attempt to properly explain the ultrasonic experiments, we treat the couplings between the ultrasonic waves and electrons accommodated in the degenerate $y'z$ and $zx'$ orbitals of Fe ion.
We will show plausible order parameters associated with the superconducting phase and the structurally distorted phase in the system. 

The transverse ultrasonic wave with the propagation vector $\boldsymbol{q} = \bigl( q_x, 0, 0 \bigr)$ and the polarization vector $\boldsymbol{\xi} = \bigl( 0, \xi_y, 0 \bigr)$ employed in the present experiments is expressed in terms of the plane wave as 
\begin{align}
\label{Deformation_y}
\xi_{y} = \xi_{y}^0 \exp \left[i \left( q_{x} x - \omega t \right) \right]
.
\end{align}
Here, $\xi_y^0$ is the amplitude of the ultrasonic wave, $q_x$ is the wavevector component, and $\omega$ is the angular frequency.
The elastic constant is determined by the sound velocity $v_x = \omega / q_x = v_{66}$ as $C_{66} = \rho v_x^2$.
The transverse ultrasonic wave induces the deformation tensor
\begin{align}
\label{Dif. Deformation_y}
\frac{ \partial \xi_y } { \partial x } = iq_x \xi_y^0 \exp \left[ i \left(q_x x - \omega t \right) \right] =  iq_x \xi_y
.
\end{align}
The deformation tensor of Eq. (\ref{Dif. Deformation_y}) associated with the transverse ultrasonic wave consists of the strain $\varepsilon_{xy}$ expressed as a symmetric tensor and the rotation $\omega_{xy}$ as an antisymmetric tensor
\begin{align}
\label{Strain and rotation}
\left\{
\begin{array}{c}
\varepsilon_{xy} \\
\omega_{xy} \\
\end{array}
\right\}
= \frac{1}{2} \left( \frac{ \partial \xi_{y} } { \partial x } \pm \frac{ \partial \xi_{x} } { \partial y } \right)
.
\end{align}
In experiments using the transverse ultrasonic wave with $\boldsymbol{q} = (q_x, 0, 0)$ and $\boldsymbol{\xi} = (0, \xi_y, 0)$, we tune the perturbation parameter of $\delta = q_x \xi_y^0 = (\omega / v_x) \xi_y^0 = (2 \pi / \lambda_x ) \xi_y^0$ to an infinitesimal value by controlling the amplitude of the generated ultrasonic wave $\xi_y^0$ and the frequency $\omega$ for a given phase velocity $v_x = v_{66}$.
The rotation $\omega_{xy} = q_x \xi_y^0 = \sin \theta$ and the strain $\varepsilon_{xy} = q_x \xi_y^0 = \sin \theta$ are caused by slight twisting of the $x$- and $y$-axes by an angle of $\pi/2 \pm 2\theta$.
This will later be illustrated in Figs. \ref{Fig7}(a) and \ref{Fig7}(b).  
The transverse ultrasonic wave simultaneously induces the rotation $\omega_{xy}$ and strain $\varepsilon_{xy}$, which are strictly distinguished from each other from a symmetrical viewpoint.
Concerning the space group $D_{4h}^{17}$ of the tetragonal lattice, the rotation $\omega_{xy}$ is compatible with $A_2$ symmetry, while the strain $\varepsilon_{xy}$ is compatible with $B_2$ \cite{Lax} symmetry.
Thus, it is expected that the rotation $\omega_{xy}$ and strain $\varepsilon_{xy}$ interact with the electronic states of the system in different manners.
When the electron-phonon interaction is absent, both the rotation $\omega_{xy}$ and strain $\varepsilon_{xy}$ propagate in the lattice with the same velocity $v_{66}$.

In our attempt to explain the critical divergence of the ultrasonic attenuation coefficients and the marked elastic softening, we suppose that the interactions of thermally excited transverse acoustic phonons with electronic states play a role in the appearance of the structural and superconducting transitions.
In the interactions of the transverse acoustic phonons with the electronic states of the iron pnictide, we take the Hamiltonian for the harmonic oscillators with polarization vectors in the $xy$ plane as
\begin{align}
\label{Phonon energy}
H_\mathrm{ph}
&= \sum_{\boldsymbol{q} }
\left[
\hbar \omega_{x} \left( \boldsymbol{q} \right) \left( a_{x,\boldsymbol{q}}^\dagger a_{x,\boldsymbol{q}}  + \frac{1}{2} \right)
\right.
\nonumber \\
&\left. \qquad \qquad 
+ \hbar \omega_{y} \left( \boldsymbol{q} \right) \left( a_{y,\boldsymbol{q}}^\dagger a_{y,\boldsymbol{q}} + \frac{1}{2} \right)
\right]
.
\end{align}
Here, $a_{i, \boldsymbol{q} }$ and $a_{i,\boldsymbol{q} }^\dagger$ respectively denote annihilation and creation operators of the transverse acoustic phonon with the polarization direction $\xi_i$ for $i = x$ and $y$ and momentum $\hbar \boldsymbol{q}$.
$\hbar \omega_i \left( \boldsymbol{q} \right)$ stands for the phonon energy.
The strain $\varepsilon_{xy} \left( \boldsymbol{r}, t \right)$ and rotation $\omega_{xy} \left( \boldsymbol{r}, t \right)$ at position $\boldsymbol{r}$ and time $t$ are expressed in terms of the phonon operators $a_{i, \boldsymbol{q} }$ and $a_{i,\boldsymbol{q} }^\dagger$ as \cite{QTS}
\label{Strain and Rotation 2nd quant.}
\begin{align}
\left\{
\begin{array}{c}
\varepsilon_{xy} \left( \boldsymbol{r}, t \right) \\
\omega_{xy} \left( \boldsymbol{r}, t \right) \\
\end{array}
\right\}
&= \frac{1}{2} \left[ \frac{ \partial \xi_{y}\left( \boldsymbol{r}, t \right) } { \partial x }
\pm \frac{ \partial \xi_{x}\left( \boldsymbol{r}, t \right) } { \partial y } \right]
\nonumber \\
&= \frac{i}{2} \sum_{\boldsymbol{q}} \sqrt{\mathstrut \frac{\hbar} {2V\rho_\mathrm{M} \omega_{y} \left( \boldsymbol{q} \right)} } q_{x}
\nonumber \\
&\quad \times 
\left[ a_{y,\boldsymbol{q}} e^{i\boldsymbol{q}\cdot \boldsymbol{r}}e^{-i\omega_{y}\left( \boldsymbol{q} \right) t} 
- a_{y,\boldsymbol{q}}^{\dagger}   e^{-i\boldsymbol{q}\cdot \boldsymbol{r}}e^{i\omega_{y}\left( \boldsymbol{q} \right)t} \right]
\nonumber \\
&\pm
\frac{i}{2} \sum_{\boldsymbol{q} } \sqrt{\mathstrut \frac{\hbar} {2V\rho_\mathrm{M} \omega_{x} \left( \boldsymbol{q} \right)} } q_{y}
\nonumber \\
&\quad \times 
\left[ a_{x,\boldsymbol{q}} e^{i\boldsymbol{q}\cdot \boldsymbol{r}}e^{-i\omega_{x}\left( \boldsymbol{q} \right)t} 
- a_{x,\boldsymbol{q}}^{\dagger} e^{-i\boldsymbol{q}\cdot \boldsymbol{r} }e^{i\omega_{x}\left( \boldsymbol{q} \right) t} \right]
.
\end{align}
Here, $\rho_M$ is the mass density and $V$ is the volume of the system.

\subsection{Degenerate $\psi_{y'z}$ and $\psi_{zx'}$ orbitals}

Many reports of band-structure calculations on the iron pnictide compounds have shown inherent 3$d$ electronic structures favorable for the superconductivity as well as the structural transition \cite{Nekrasov, Mazin2, Yanagi2, Yanagi Dr., Miyake}.
Three sheets of hole surfaces with a cylinder-type structure exist around the $\Gamma$-point of the zone center and two electron pockets exist around the X-points of the zone boundary.
We focus on the degenerate $y'z$ and $zx'$ bands occupied up to half filling, which are lifted by the transverse ultrasonic waves and acoustic phonons for $C_{66}$ with a small-wavenumber limit of $|\boldsymbol{q}| \rightarrow 0$.
 
We show  the electronic states of the degenerate  $y'z$ and $zx'$ orbitals in the crystalline electric field (CEF) Hamiltonian on an Fe$^{2+}$ ion with $D_{2d}$ site symmetry, which consists of an electric monopole, quadrupole, and hexadecapole expressed in terms of spherical harmonics as
\begin{align}
\label{HCEF}
H_\mathrm{CEF}
&= A_0^0
\nonumber \\
&\quad
+ A_2^0 \left( \frac{3z^2-r^2} {2r^2} \right)
\nonumber \\
&\qquad
+ A_4^0 \left( \frac{35z^4-30z^2r^2+3r^4} {8r^4} \right)
\nonumber \\
&\quad \qquad
+ A_4^4 \left( \frac{\sqrt{35} }{8} \frac{x'^4-6x'^2y'^2+y'^4} {r^4} \right)
.
\end{align}
Here, $A_l^m$ denotes CEF parameters for the electric multipole potentials.
The 3$d$ orbitals of the Fe$^{2+}$ ion in the CEF Hamiltonian of Eq. (\ref{HCEF}) splits into five orbital states with energy $E_\mathrm{CEF}^l$ as
\begin{align}
\label{ECEFl}
E_\mathrm{CEF}^l
= \int d\boldsymbol{r} \psi_{l}^{\ast} \left( \boldsymbol{r} \right) H_\mathrm{CEF} \psi_{l} \left( \boldsymbol{r} \right)
.
\end{align}
Here, $l$ denotes the 3$d$ orbital suffix.
The point charge model of the CEF Hamiltonian
gives the low-lying singlet states of $\psi_{3z^2-r^2}$ with $A_1$ symmetry and $\psi_{x'^2 - y'^2}$ with $B_1$ symmetry and the mid-lying doublet state of $\psi_{y'z}$ and $\psi_{zx'}$ with $E$ symmetry and the excited singlet of $\psi_{x'y'}$ with $B_2$ symmetry \cite{Kurihara Dr}.
Actually, the band calculations show that the three hole sheets around the $\Gamma$-point consist of the doublet state of $\psi_{y'z}$ and $\psi_{zx'}$ and the singlet state $\psi_{x'^2-y'^2}$.

In the present investigation, we suppose that the structural and superconducting transitions of the iron pnictide manifest themselves as a result of the spontaneous symmetry breaking associated with the degenerate $y'z$ and $zx'$ bands, which are mapped on the special unitary group SU(2).
In order to explore the quadrupole-strain interaction in the iron pnictide, we specially treat the degenerate $\psi_{y'z}$ and $\psi_{zx'}$ orbitals of the $E$ symmetry with half filling as \cite{Miyake, Inui Group}
\begin{align}
\label{wave function y'z}
\psi_{y'z} \left( \boldsymbol{r} \right)
&= \frac{i}{\sqrt{2} } f_d \left( r \right) \left[ Y_2^1 \left( \theta, \varphi \right) + Y_2^{-1} \left( \theta, \varphi \right) \right]
\nonumber \\
&= \sqrt{\frac{15}{4\pi} } f_d \left( r \right) \frac{y'z} {r^2}
, \\
\label{wave function zx'}
\psi_{zx'} \left( \boldsymbol{r} \right)
&= \frac{-1}{\sqrt{2} } f_d \left( r \right) \left[ Y_2^1 \left( \theta, \varphi \right) - Y_2^{-1} \left( \theta, \varphi \right) \right]
\nonumber \\
&= \sqrt{\frac{15}{4\pi} } f_d \left( r \right) \frac{zx'} {r^2}
.
\end{align}
Here, the polar coordinate $\left( r, \theta, \varphi \right)$ represents the position vector $\boldsymbol{r} = \left( x', y', z \right)$ of an electron.
$f_d \left( r \right)$ is the radius function of a 3$d$ electron, and $Y_2^{\pm1}\left( \theta, \varphi \right)$ is the spherical harmonics with angular momentum $l = 2$ and azimuthal quantum number $m = \pm1$. 

Since the direct product of the $E$ doublet is reduced as $E \otimes E = A_1 \oplus A_2 \oplus B_1 \oplus B_2$, we deduce that the degenerate $y'z$ and $zx'$ orbitals carry the electric quadrupoles
$O_{3z^2 - r^2} = \bigl( 3z^2 - r^2 \bigr) / \bigl( 2r^2 \bigr)$ with $A_1$ symmetry,
$O_{x'y'} = \sqrt{3} x'y'/r^2$ with $B_1$ symmetry,
and $O_{x'^2-y'^2} = \sqrt{3} \bigl(x'^2 - y'^2 \bigr)/ \bigl( 2r^2 \bigr)$ with $B_2$ symmetry
and the angular momentum $l_z = -i \bigl(x' \partial /  \partial y' - y'  \partial /  \partial x' \bigr) = -i  \partial /  \partial \varphi$ with $A_2$ symmetry.
The quadrupole $O_{x'^2-y'^2}$ couples to the strain $\varepsilon_{xy}$ of the transverse ultrasonic wave of $C_{66}$, while $O_{x'y'}$ couples to the strain $\varepsilon_{x^2 - y^2}$ of the transverse ultrasonic wave of $\left( C_{11} - C_{12} \right) /2$. 
The CEF Hamiltonian of Eq. (\ref{HCEF}) includes  $O_{3z^2 - r^2}$ with full symmetry.

Employing the identity matrix $\tau_0$ and the Pauli matrices $\tau_x$, $\tau_y$, and $\tau_z$ corresponding to the generator elements of the special unitary group SU(2), we present the quadrupoles $O_{3z^2-r^2}$, $O_{x'^2-y'^2}$, and $O_{x'y'}$ and the angular momentum $l_z$ in terms of the matrices for the orbital state $\psi_{y'z}$ of Eq. (\ref{wave function y'z}) and  $\psi_{zx'}$ of Eq. (\ref{wave function zx'}) as
\begin{align}
\label{Matrix Ou}
O_{3z^2-r^2}
&= \frac{1}{7}
\bordermatrix{
& \psi_{y'z} & \psi_{zx'} \cr
& 1 & 0 \cr
& 0 & 1 \cr
}
= \frac{1}{7} \tau_0
,
\\
\label{Matrix Ox'y'}
O_{x'y'}
&= \frac{\sqrt{3} }{7}
\left(
\begin{array}{cc}
0&1 \\
1&0 \\
\end{array}
\right)
= \frac{\sqrt{3} }{7} \tau_x
,
\\
\label{Matrix lz}
l_z
&= - \left(
\begin{array}{cc}
0&-i \\
i&0 \\
\end{array}
\right)
= - \tau_y
,
\\
\label{Matrix Ox'2-y'2}
O_{x'^2-y'^2}
&= - \frac{\sqrt{3} }{7}
\left(
\begin{array}{cc}
1&0 \\
0&-1 \\
\end{array}
\right)
= - \frac{\sqrt{3} }{7} \tau_z
.
\end{align}
The Pauli matrices $\tau_x$, $\tau_y$, and $\tau_z$ obey the commutation relation
\begin{align}
\label{Pauli matrix Commutation relation}
\left[ \tau_i, \tau_j \right]
= 2i \varepsilon_{ijk} \tau_k \ \ \ \left( i, j, k = x, y ,z\right)
.
\end{align}
Here, $\varepsilon_{ijk}$ is the Levi$-$Civita symbol.
As will be shown in Sects. 4.6 and 4.10, the commutation relation among the quadrupoles $O_{x'^2 - y'^2}$ and $O_{x'y'}$ and the angular momentum $l_z$ of Eq. (\ref{Pauli matrix Commutation relation}) brings about quantum fluctuations, which play a significant role in the manifestation of the specific superconductivity accompanying the hexadecapole ordering in the vicinity of the QCP.

\begin{figure}[h]
\begin{center}
\includegraphics[angle=0,width=0.46\textwidth, bb=0 15 595 371]{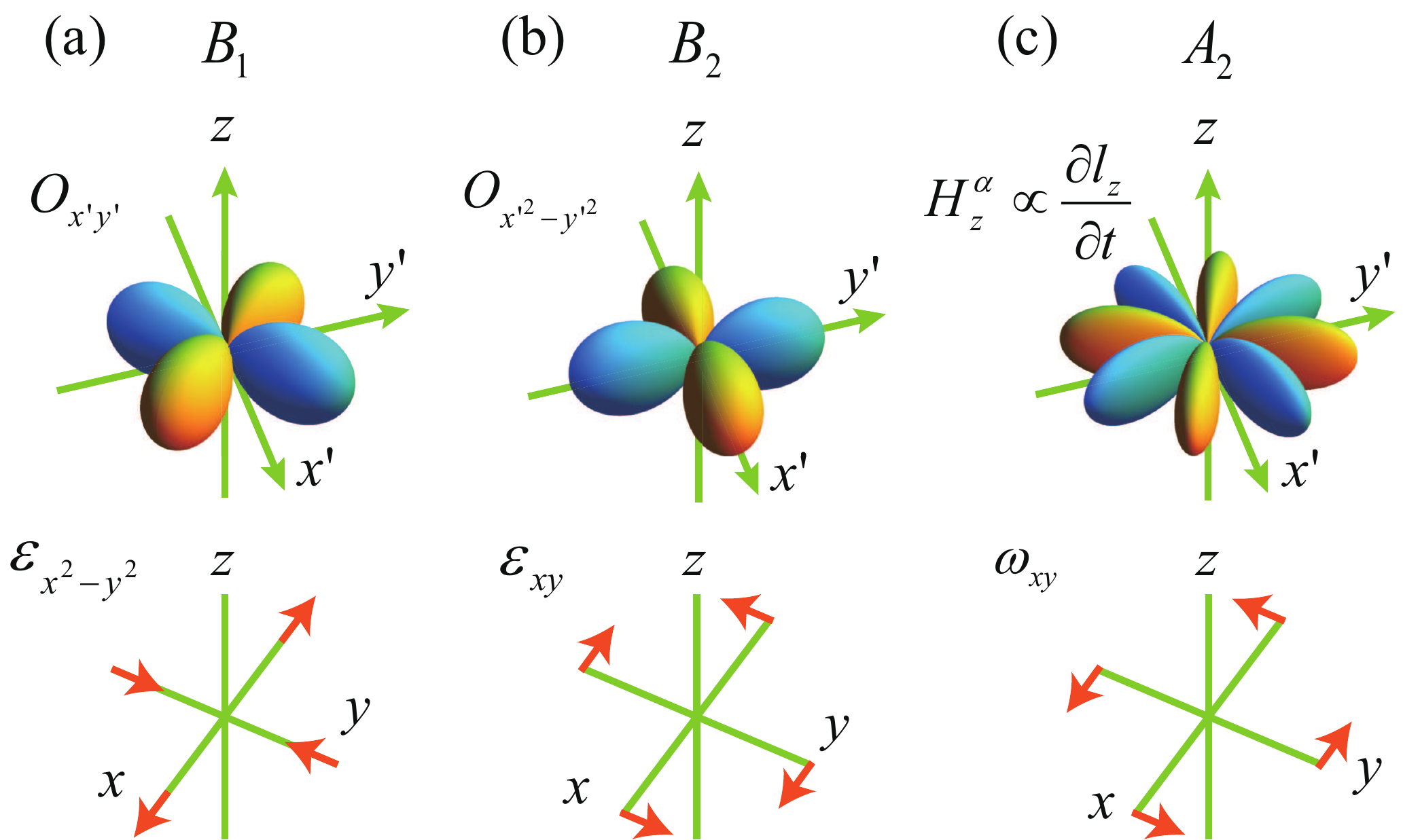}
\end{center}
\caption{
(Color online)
Schematic view of interactions between electric quadrupole $O_{x'y'}$ and strain $\varepsilon_{x^2 - y^2}$ with $B_1$ symmetry in (a), $O_{x'^2 - y'^2}$ and $\varepsilon_{xy}$ with $B_2$ symmetry in (b), and electric hexadecapole $H_z^\alpha \propto \partial l_z / \partial t$ and rotation $\omega_{xy}$ with $A_2$ symmetry in (c) of the space group $D_{4h}^{17}
$ in the present iron pnictide.
The coordinates of $x$ and $y$ ($x'$ and $y'$) are adopted for the neighboring Ba-Ba (Fe-Fe) directions.
}
\label{Fig5}
\end{figure}

The elastic strains $\varepsilon_{\mathrm{\Gamma}_\gamma}$ induced by the thermally excited transverse acoustic phonons or the experimentally generated transverse ultrasonic waves perturb the CEF Hamiltonian as
\begin{align}
\label{perturbation of CEF}
H_\mathrm{CEF} \left( \boldsymbol{r}, \varepsilon_{\mathrm{\Gamma}_\gamma} \right)
= H_\mathrm{CEF} \left( \boldsymbol{r} \right)
+ \sum_{\mathrm{\Gamma}_{\gamma}} 
\frac{\partial H_\mathrm{CEF} \left( \boldsymbol{r} \right) } 
{\partial \varepsilon_{\mathrm{\Gamma}_\gamma} } \varepsilon_{{\mathrm{\Gamma}}_\gamma}
.
\end{align}
Because electrons accommodated in the degenerate $y'z$ and $zx'$ orbitals bear the electric quadrupoles, the second term of Eq. (\ref{perturbation of CEF}) proportional to the strain $\varepsilon_{\mathrm{\Gamma}_\gamma}$ is expressed in terms of the quadrupole-strain interaction as \cite{Dohm, Thalmeier, Luthi Phys. Ac.}
\begin{align}
\label{HQS}
H_\mathrm{QS} = -g_{x'^2-y'^2}O_{x'^2-y'^2}\varepsilon_{xy} -g_{x'y'}O_{x'y'}\varepsilon_{x^2-y^2}.
\end{align}
As the present ultrasonic experiments give a large coupling constant of $g_{x'^2 - y'^2} \sim 1 \times 10^3$ K per electron, the first term $-g_{x'^2 - y'^2} O_{x'^2 - y'^2} \varepsilon_{xy}$ in Eq. (\ref{HQS}) is important for the critical divergence of the attenuation coefficient $\alpha_{66}$ and the softening of the elastic constant $C_{66}$ around the structural transition.
The little softening of $\left( C_{11} - C_{12} \right) /2$ in the present experiments implies that the second term $-g_{x'y'} O_{x'y'} \varepsilon_{x^2-y^2}$ in Eq. (\ref{HQS}) scarcely affects the system.
We schematically show the quadrupole $O_{x'y'}$ conjugate with the strain $\varepsilon_{x^2 - y^2}$ in Fig. \ref{Fig5}(a) and the quadrupole $O_{x'^2 - y'^2}$ conjugate with $\varepsilon_{xy}$ in Fig. \ref{Fig5}(b).

Next, we consider the interaction of the rotation $\omega_{xy}$ associated with the transverse ultrasonic wave with one-electron states having the angular momentum $l_z$ in the CEF Hamiltonian.
The rotation operator of $\exp \bigl[ -i l_z \omega_{xy} \bigr]$ twists the $\psi_{y'z}$ and $\psi_{zx'}$ orbital states in the CEF Hamiltonian given by Eq. (\ref{HCEF}) as \cite{HoVO4}
\begin{align}
\label{HCEF rot.}
\left\langle l \left| H_\mathrm{CEF} \bigl( \omega_{xy} \bigr) \right| l' \right\rangle
= \int d\boldsymbol{r} \psi_{l}^{\ast} \left( \boldsymbol{r} \right) e^{il_z\omega_{xy}}
H_\mathrm{CEF} e^{ -il_z\omega_{xy}} \psi_{l'} \left( \boldsymbol{r} \right)
.
\end{align}
Here, ket $ | l \rangle = \psi_{l} \left( \boldsymbol{r} \right)$ stands for the orbital states $l = y'z$ of Eq. (\ref{wave function y'z}) and $zx'$ of  Eq. (\ref{wave function zx'}).
Supposing infinitesimal rotation of $\omega_{xy} \ll 1$ associated with the ultrasound in experiments, the rotation-operated Hamiltonian of Eq. (\ref{HCEF rot.}) is expanded in terms of power series up to the second order of $\omega_{xy}$ as
\begin{align}
\label{Rotation of HCEF}
e^{ i l_z \omega_{xy} } H_\mathrm{CEF} e^{ - i l_z \omega_{xy} }
\approx &H_\mathrm{CEF}
+ i \left[ l_z, H_\mathrm{CEF} \right] \omega_{xy}
\nonumber \\
&- \frac{1}{2} \left[ l_z, \bigl[ l_z, H_\mathrm{CEF} \bigr] \right] \bigl( \omega_{xy} \bigr)^2
.
\end{align}
Here, $\left[ A, B \right] = AB - BA$ denotes the Poisson bracket.
Applying the differential operation of $l_z = -i \left( x' \partial / \partial y' - y' \partial / \partial x' \right) = -i \partial / \partial \varphi$ to the CEF Hamiltonian of Eq. (\ref{HCEF}), we obtain the perturbation Hamiltonian as 
\begin{align}
\label{Hrot from HCEF}
H_\mathrm{rot}^\mathrm{CEF} \bigl( \omega_{xy} \bigr)
= 4A_4^4 H_z^{\alpha} \omega_{xy} - 8A_4^4 H_4^4 \bigl( \omega_{xy} \bigr)^2
.
\end{align}
The first term of Eq. (\ref{Hrot from HCEF}) represents the linear coupling of the rotation $\omega_{xy}$ to the electric hexadecapole $H_z^\alpha =  \sqrt{35}x'y' \bigl( x'^2 - y'^2 \bigr) / \bigl( 2r^4 \bigr)$ with $A_2$ symmetry of the $D_{2d}$ point group \cite{Kuramoto}.
In Fig. \ref{Fig5}(c), we schematically show the hexadecapole $H_z^\alpha$ carried by the one-electron state and the rotation $\omega_{xy}$ in the interaction Hamiltonian $H_\mathrm{rot}^\mathrm{CEF} \bigl( \omega_{xy} \bigr)$ of Eq. (\ref{Hrot from HCEF}).
The second term in Eq. (\ref{Hrot from HCEF}) is the coupling of the square of the rotation $\bigl( \omega_{xy} \bigr)^2$ to the electric hexadecapole $H_4^4 = \sqrt{35} \bigl( x'^4 - 6x'^2y'^2 + y'^4 \bigr)/ \bigl( 8r^4 \bigr)$ with full symmetry.
The hexadecapole $H_4^4$ has already appeared in the CEF Hamiltonian of Eq. (\ref{HCEF}). 

 The coefficient proportional to the rotation $\omega_{xy}$ in Eq. (\ref{Rotation of HCEF}) is identified with the Heisenberg equation of the angular momentum $l_z$ for the CEF Hamiltonian of Eq. (\ref{HCEF}).
This gives the time derivative of the angular momentum $l_z$, which is identical to the torque $\tau_{xy}$ for a quantum system, as
\begin{align}\
\label{torque of HCEF}
i \hbar \frac{ \partial l_z} {\partial t} = \left[ l_z, H_\mathrm{CEF} \right] = -i g_\mathrm{H} H_z^{\alpha} = i \tau_{xy}
.
\end{align}
Note that the coupling parameter of $g_\mathrm{H} = 4A_4^4$ in Eq. (\ref{torque of HCEF}) has already been given in the CEF Hamiltonian of Eq. (\ref{HCEF}).
Since the angular momentum $l_z$ is constant for the motion of the one-electron states in the CEF Hamiltonian of Eq. (\ref{HCEF}), $l_z$ commutes with $H_\mathrm{CEF}$ as $\left[ l_z, H_\mathrm{CEF} \right] = 0$.
Consequently, the hexadecapole $H_z^{\alpha}$ proportional to the time derivative of the angular momentum $\partial l_z / \partial t$ of Eq. (\ref{torque of HCEF}) is expected to vanish.
This is well known as the rotational invariance for the electronic states in a central force potential such as the CEF potential. 
The rotation invariance can be naturally understood by the fact that the hexadecapole interaction of Eq. (\ref{Rotation of HCEF}) does not change the energy of an electron moving along the contour line of the CEF potential.

The rotation $\omega_{xy}$ due to a transverse ultrasonic wave should also affect the quadrupole-strain interaction of Eq. (\ref{HQS}) within the bilinear coupling between the strain and rotation as
\begin{align}
\label{Hrot from HQS}
H_\mathrm{rot}^\mathrm{QS} \bigl( \omega_{xy} \bigr)
&= i \left[ l_z, H_\mathrm{QS} \right] \omega_{xy}
\nonumber \\
&= 2g_{x'^2-y'^2}O_{x'y'}\varepsilon_{xy}\omega_{xy}
\nonumber \\
&\qquad
-2g_{x'y'}O_{x'^2-y'^2}\varepsilon_{x^2-y^2}\omega_{xy}
.
\end{align}
Here, the commutation relation among the Pauli matrices of Eq. (\ref{Pauli matrix Commutation relation}) is employed.
The perturbation energies for the orbital states $l = y'z$ and $zx'$ with energy $E_l^0$ due to the perturbation Hamiltonian of Eq. (\ref{Hrot from HQS}) are written as
\begin{align}
\label{Perturbation for l}
E_l \bigl( \omega_{xy} \bigr)
&= E_l^0
+ \left\langle l \left| H_\mathrm{rot}^\mathrm{QS} \bigl( \omega_{xy} \bigr) \right| l \right\rangle
\nonumber \\
&= E_l^0
+ 2g_{x'^2-y'^2} \left\langle l \left| O_{x'y'} \right| l \right\rangle \varepsilon_{xy}\omega_{xy}
\nonumber \\
&\qquad \qquad
-2g_{x'y'} \left\langle l \left| O_{x'^2-y'^2} \right| l \right\rangle \varepsilon_{x^2-y^2}\omega_{xy}
.
\end{align}
The null of the diagonal elements of the Pauli matrix of $\boldsymbol{O}_{x'y'}$ of Eq. (\ref{Matrix Ox'y'}) diminishes the second term $2g_{x'^2-y'^2} \left\langle l \left| O_{x'y'} \right| l \right\rangle$ in the second line in Eq. (\ref{Perturbation for l}).
Although the matrix of $\boldsymbol{O}_{x'^2-y'^2}$ of Eq. (\ref{Matrix Ox'2-y'2}) possesses diagonal elements, we may ignore the third term $- 2g_{x'y'} \left\langle l \left| O_{x'^2-y'^2} \right| l \right\rangle$ in the second line in Eq. (\ref{Perturbation for l}) because the little softening of $(C_{11}-C_{12})/2$ indicates a small quadrupole-strain coupling constant of $g_{x'y'} << 1$. 
Presumably, the strain-rotation bilinear coupling terms of Eq. (\ref{Hrot from HQS}) have little effect on the present system with degenerate $y'z$ and $zx'$ orbitals. 

 The applied magnetic field, however, breaks the rotational invariance for the one-electron states in the CEF potential.
This effect of rotation on the elastic properties in applied magnetic fields has been theoretically investigated \cite{Dohm, Thalmeier} and experimentally verified by ultrasonic measurements on the antiferro-magnetic compound MnF$_2$ \cite{Melcher} and the rare-earth compounds TmSb \cite{Wang and Luthi}, CeAl$_2$ \cite{Luthi and Ligner}, CeB$_6$ \cite{Luthi CeB6}, Ce$_{0.5}$La$_{0.5}$B$_6$ \cite{Goto Ce0.5La0.5B6}, and HoVO$_4$ \cite{HoVO4}.
Furthermore, the effect of rotation on the quantum oscillation of elastic constants of transverse ultrasonic waves for Fermi surfaces with an ellipsoidal shape has been discussed \cite{Kataoka and Goto}.
In Sect. 4.6, we show the coupling of the rotation associated with the transverse acoustic phonons to the hexadecapole carried by two-electron states.

\subsection{Quadrupole-strain interaction}

As shown in Eq. (20), the quadrupoles carried by electrons in the CEF Hamiltonian interact to the strains of transverse ultrasonic waves.
This quadrupole-strain interaction brings about the sizable softening of $C_{66}$ and the divergence of the attenuation coefficients $\alpha_{66}$ and $\alpha_{\mathrm{L}[110]}$ around the structural transition.
In order to properly describe the structural transition of the iron pnictide, we will consider the interaction of the quadrupole $O_{x'^2-y'^2} \left( \boldsymbol{r}_i \right)$ of band electrons to the transverse acoustic phonons carrying the strain $\varepsilon_{xy} \left( \boldsymbol{r}_i \right)$ for the position vector $\boldsymbol{r}_i$ per electron as
\begin{align}
\label{HQS by field operator}
H_\mathrm{QS}
= - g_{x'^2-y'^2}
\sum_{l_1, l_2 = y'z, zx'} \sum_{\sigma = \uparrow, \downarrow}
& \int d\boldsymbol{r}_i
\mathit{\Psi}_{l_1, \sigma}^\ast \left( \boldsymbol{r}_i \right)
O_{x'^2-y'^2} \left( \boldsymbol{r}_i \right)
\nonumber \\
& \times \mathit{\Psi}_{l_2, \sigma} \left( \boldsymbol{r}_i \right) 
\varepsilon_{x'^2-y'^2} \left( \boldsymbol{r}_i \right)
.
\end{align}
The electron field operators of $\mathit{\Psi}_{y'z, \sigma} \left( \boldsymbol{r}_i \right)$ and $\mathit{\Psi}_{zx', \sigma} \left( \boldsymbol{r}_i \right)$ at position $\boldsymbol{r}_i$ acting on the degenerate $y'z$ and $zx'$ bands with spin orientation $\sigma$ in Eq. (\ref{HQS by field operator}) are written as \cite{K. Yoshida}
\begin{align}
\label{Filed operator 1.1}
\mathit{\Psi}_{l, \sigma} \left( \boldsymbol{r}_i \right)
= d_{i, l, \sigma} \psi_l  \left( \boldsymbol{r}_i \right) v_\sigma \left( \boldsymbol{r}_i \right)
, \\
\label{Filed operator 1.2}
\mathit{\Psi}_{l, \sigma}^\ast \left( \boldsymbol{r}_i \right)
= d_{i, l, \sigma}^\dagger \psi_l^\ast  \left( \boldsymbol{r}_i \right) v_\sigma^\ast \left( \boldsymbol{r}_i \right)
.
\end{align}
Here, $l$ denotes the band suffix of $y'z$ and $zx'$, and $v_\sigma \left( \boldsymbol{r}_i \right)$ is the spin function of $\alpha \left( \boldsymbol{r}_i \right)$ for up-spin $\sigma = \uparrow$ or $\beta \left( \boldsymbol{r}_i \right)$ for down-spin $\sigma =  \downarrow$. 
The annihilation operator $d_{i, l, \sigma}$ and creation operator $d_{i, l, \sigma}^\dagger$ are expressed by Fourier transforms as below
\begin{align}
\label{Fourier transformation of d 1}
d_{i, l, \sigma} 
&= \sum_{ \boldsymbol{k}} 
d_{\boldsymbol{k}, l, \sigma} e^{i\boldsymbol{k} \cdot \boldsymbol{r}_i} 
, \\
\label{Fourier transformation of d 2}
d_{i, l, \sigma}^{\dagger}
&= \sum_{\boldsymbol{k}}
d_{\boldsymbol{k},  l, \sigma}^{\dagger}e^{-i\boldsymbol{k} \cdot \boldsymbol{r}_i }
.
\end{align}
Here, $\boldsymbol{k}$ is the wavevector of an electron.
We express the Hamiltonian $H_\mathrm{K}$ for the electrons accommodated in the degenerate $y'z$ and $zx'$ bands, which play a significant role in the appearance of the quadrupole ordering in the system as
\begin{align}
\label{HK}
H_\mathrm{K} 
&= \sum_{\boldsymbol{k}} \sum_{\sigma}
\left[
\varepsilon_{y'z} \left( \boldsymbol{k} \right) 
d_{\boldsymbol{k}, y'z,  \sigma}^{\dagger} d_{\boldsymbol{k}, y'z,  \sigma} \right.
\nonumber \\
&\left. \qquad \qquad \qquad
+\varepsilon_{zx'} \left( \boldsymbol{k} \right) 
d_{\boldsymbol{k}, zx',  \sigma}^{\dagger} d_{\boldsymbol{k}, zx',  \sigma}
\right]
.
\end{align}
Here, $\varepsilon_{l} \left( \boldsymbol{k} \right)$ for suffix $l = y'z$ or $zx'$ is the energy from the Fermi level.

The quadrupole-strain interaction of Eq. (\ref{HQS by field operator}) is rewritten in terms of Fourier transforms as
\begin{align}
\label{HQS 2nd quant.}
H_\mathrm{QS}
= - G_{x'^2-y'^2}
\sum_{\boldsymbol{k}, \boldsymbol{q} }
O_{x'^2-y'^2, \boldsymbol{k}, \boldsymbol{q}} \varepsilon_{xy} \left( \boldsymbol{q} \right)
.
\end{align}
Here, we use the coupling constant $G_{x'^2-y'^2} = \sqrt{3} g_{x'^2-y'^2} /7$. 
The interaction Hamiltonian of Eq. (\ref{HQS 2nd quant.}) means that the electrons with wavevector $\boldsymbol{k}$ bearing the quadrupole $O_{x'^2-y'^2, \boldsymbol{k}, \boldsymbol{q} }$ are scattered by the strain $\varepsilon_{xy} \left( \boldsymbol{q} \right)$ of the transverse acoustic phonons with wavevector $\boldsymbol{q}$. 
The quadrupole $O_{x'^2-y'^2, \boldsymbol{k}, \boldsymbol{q} }$ in Eq. (\ref{HQS 2nd quant.}) is described in terms of the annihilation and creation operators of Eqs. (\ref{Fourier transformation of d 1}) and (\ref{Fourier transformation of d 2}) as
\begin{align}
\label{Ox'2-y'2 2nd quant.}
O_{x'^2-y'^2, \boldsymbol{k}, \boldsymbol{q} }
= \sum_{\sigma}
\left( - d_{\boldsymbol{k}  + \boldsymbol{q}, y'z, \sigma}^{\dagger} d_{\boldsymbol{k}, y'z, \sigma}
+ d_{\boldsymbol{k}  + \boldsymbol{q}, zx', \sigma}^{\dagger} d_{\boldsymbol{k}, zx', \sigma}
\right)
.
\end{align}
Note that the quadrupole $O_{x'^2-y'^2, \boldsymbol{k},  \boldsymbol{q} }$ in Eq. (\ref{Ox'2-y'2 2nd quant.}) is simply described by the difference between the occupation numbers for the degenerate $y'z$ and $zx'$ bands at the long-wavelength limit of $ \left| \boldsymbol{q} \right| = 2 \pi / \lambda \rightarrow 0$ of the transverse acoustic phonons carrying the strain $\varepsilon_{xy} \left( \boldsymbol{q} \right)$ as
\begin{align}
\label{Ox'2-y'2 as number operator}
O_{x'^2-y'^2, \boldsymbol{k}, \boldsymbol{q} = 0}
&= \sum_{\sigma}
\left( - d_{\boldsymbol{k}, y'z, \sigma}^{\dagger} d_{\boldsymbol{k}, y'z, \sigma}
+ d_{\boldsymbol{k}, zx', \sigma}^{\dagger} d_{\boldsymbol{k}, zx', \sigma}
\right)
\nonumber \\
&= \sum_{\sigma}
\left( -n_{\boldsymbol{k}, y'z, \sigma} + n_{\boldsymbol{k}, zx', \sigma} \right)
.
\end{align}
In Sect. 4.6, we will use the quadrupole  $O_{x'y', \boldsymbol{k}, \boldsymbol{q} }$ expressed as
\begin{align}
\label{Ox'y' 2nd quant.}
O_{x'y', \boldsymbol{k}, \boldsymbol{q} }
= \sum_{\sigma}
\left (
d_{\boldsymbol{k}  + \boldsymbol{q}, y'z, \sigma}^{\dagger} d_{\boldsymbol{k}, zx', \sigma}
+ d_{\boldsymbol{k}  + \boldsymbol{q}, zx', \sigma}^{\dagger} d_{\boldsymbol{k}, y'z, \sigma}
\right).
\end{align}

The strain $\varepsilon_{xy} \left( \boldsymbol{q} \right)$ in Eq. (\ref{HQS 2nd quant.}) is expressed by the annihilation and creation operators of phonons defined in Eq. (\ref{Phonon energy}) as
\begin{align}
\label{Strain 2nd quant.}
\varepsilon_{xy} \left( \boldsymbol{q} \right)
=& \frac{i}{2} \sqrt{\mathstrut \frac{\hbar} {2V\rho_\mathrm{M} \omega_{y} \left( \boldsymbol{q} \right)}} q_{x}
\left( a_{y,\boldsymbol{q}} - a_{y, -\boldsymbol{q}}^{\dagger} \right)
\nonumber \\
&+ \frac{i}{2} \sqrt{\mathstrut \frac{\hbar} {2V\rho_\mathrm{M} \omega_{x} \left( \boldsymbol{q} \right)}} q_{y}
\left( a_{x, \boldsymbol{q}}  - a_{x, -\boldsymbol{q}}^{\dagger} \right)
.
\end{align}
The strain $\varepsilon_{xy}$ induced by a transverse ultrasonic wave with frequency as low as 100 MHz is identified with $\varepsilon_{xy} \left( \boldsymbol{q} \right)$ of Eq. (\ref{Strain 2nd quant.}) in the long-wavelength limit of $ \left| \boldsymbol{q} \right| = 2 \pi / \lambda \rightarrow 0$ as $\varepsilon_{xy} = \varepsilon_{xy} \left( \boldsymbol{q} = 0 \right)$.
In the case of the quadrupole-strain interaction of Eq. (\ref{HQS 2nd quant.}) mediated by thermally excited transverse acoustic phonons, the emission of phonons carrying the strain $\varepsilon_{xy} \left( \boldsymbol{q} \right)$ scatters the electron state of momentum $\hbar \boldsymbol{k}$ to the state of $\hbar \boldsymbol{k} + \hbar \boldsymbol{q}$ and the absorption of phonons scatters the electron state of $\hbar \boldsymbol{k}$ to the state of $\hbar \boldsymbol{k} - \hbar \boldsymbol{q}$. 

\begin{figure}[t]
\begin{center}
\includegraphics[angle=0,width=0.48\textwidth, bb=0 5 174 119]{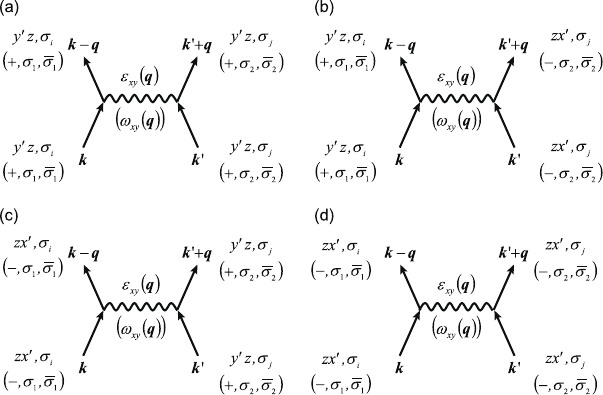}
\end{center}
\caption{
The indirect quadrupole interaction $H_\mathrm{ind}^\mathrm{QQ}$ of Eq. (\ref{HindQQ}) gives four scattering processes between the one-electron states through the strain $\varepsilon_{xy} \left( \boldsymbol{q} \right)$  of a transverse acoustic phonon, while indirect hexadecapole interaction $H_\mathrm{ind}^\mathrm{HH}$ of Eq. (\ref{HindHH1}) gives four scattering processes between the two-electron states through the rotation $\omega_{xy} \left( \boldsymbol{q} \right)$ of a transverse acoustic phonon.
}
\label{Fig6}
\end{figure}

The canonical transformation for the electron-phonon scattering processes by the quadrupole-strain interaction $H_\mathrm{QS}$ of Eq. (\ref{HQS 2nd quant.}) provides us with the electron interaction mediated by the transverse acoustic phonons as \cite{QTS}
\begin{align}
\label{HindQQ}
H_\mathrm{ind}^\mathrm{QQ}
= & -\sum_{\boldsymbol{k}, \boldsymbol{k}', \boldsymbol{q}} \sum_{\sigma, \sigma'}
D_{y'z}^\mathrm{QQ} \left( \boldsymbol{k}, \boldsymbol{q} \right)
d_{\boldsymbol{k} - \boldsymbol{q}, y'z, \sigma}^{\dagger} d_{\boldsymbol{k}, y'z, \sigma}
d_{\boldsymbol{k}' + \boldsymbol{q}, y'z, \sigma'}^{\dagger}  d_{\boldsymbol{k}', y'z, \sigma'}
\nonumber \\
& +\sum_{\boldsymbol{k}, \boldsymbol{k}', \boldsymbol{q}} \sum_{\sigma, \sigma'}
D_{y'z}^\mathrm{QQ} \left( \boldsymbol{k}, \boldsymbol{q} \right)
d_{\boldsymbol{k} - \boldsymbol{q}, y'z, \sigma}^{\dagger} d_{\boldsymbol{k}, y'z, \sigma}
d_{\boldsymbol{k}' + \boldsymbol{q}, zx', \sigma'}^{\dagger} d_{\boldsymbol{k}', zx', \sigma'}
\nonumber  \\
& +\sum_{\boldsymbol{k}, \boldsymbol{k}', \boldsymbol{q}} \sum_{\sigma, \sigma'}
D_{zx'}^\mathrm{QQ} \left( \boldsymbol{k}, \boldsymbol{q} \right)
d_{\boldsymbol{k} - \boldsymbol{q} , zx',  \sigma}^{\dagger} d_{\boldsymbol{k}, zx', \sigma} 
d_{\boldsymbol{k}' + \boldsymbol{q}, y'z, \sigma'}^{\dagger} d_{\boldsymbol{k}', y'z, \sigma'}
\nonumber  \\
&- \sum_{\boldsymbol{k}, \boldsymbol{k}', \boldsymbol{q}} \sum_{\sigma, \sigma'}
D_{zx'}^\mathrm{QQ} \left( \boldsymbol{k}, \boldsymbol{q} \right)
d_{\boldsymbol{k} - \boldsymbol{q}, zx', \sigma}^{\dagger} d_{\boldsymbol{k}, zx', \sigma}
d_{\boldsymbol{k}' + \boldsymbol{q}, zx',  \sigma'}^{\dagger} d_{\boldsymbol{k}', zx',  \sigma'}
.
\end{align}
Here, the coupling coefficient $D_{l}^\mathrm{QQ} \left( \boldsymbol{k}, \boldsymbol{q} \right)$ in Eq. (\ref{HindQQ}) for $l = y'z$ and $zx'$ is given by
\begin{align}
\label{DQQ}
&
D_l^\mathrm{QQ} \left( \boldsymbol{k}, \boldsymbol{q} \right)
= - \frac{1}{2} G_{x'^2-y'^2}^2
\nonumber \\
& \times
\left\{
\frac{\hbar} {2V\rho_\mathrm{M} \omega_{y} \left( \boldsymbol{q} \right)} q_{x}^2 
\frac{\hbar \omega_y \left( \boldsymbol{q} \right)}
{
\left[
\varepsilon_l \left( \boldsymbol{k} \right) -  \varepsilon_l \left( \boldsymbol{k} - \boldsymbol{q} \right) 
\right]^2 
- \hbar^2 \omega_y \left( \boldsymbol{q} \right) ^2
} \right.
\nonumber \\
&\left.
+ \frac{\hbar} {2V\rho_\mathrm{M} \omega_{x} \left( \boldsymbol{q} \right)} q_{y}^2 
\times \frac{\hbar \omega_x \left( \boldsymbol{q} \right)}
{\left[ \varepsilon_l \left( \boldsymbol{k} \right) -  \varepsilon_l \left( \boldsymbol{k} - \boldsymbol{q} \right) \right]^2 
- \hbar^2 \omega_x \left( \boldsymbol{q} \right)^2}
\right\}
.
\end{align}
The four independent scattering processes involving virtually excited one-phonon states due to the strain $\varepsilon_{xy} \left( \boldsymbol{q} \right)$ in Eq. (\ref{HQS 2nd quant.}) are schematically pictured in Fig. \ref{Fig6}.
The processes in Fig. \ref{Fig6}(a) indicate the scattering of the electrons accommodated in the same band $y'z$ and those in Fig. \ref{Fig6}(d) indicate the scattering in the same band $zx'$.
The processes in Figs. 6(b) and 6(c) indicate the scattering between two electrons in the different bands $y'z$ and $zx'$. 

Supposing the identity of the energy $\varepsilon_{y'z} \left( \boldsymbol{k} \right) = \varepsilon_{zx'} \left( \boldsymbol{k} \right)$ for electrons with a small wavevector $\boldsymbol{k}$ located near the Fermi level, we take the indirect quadrupole interaction coefficient $D_{y'z}^\mathrm{QQ} \left( \boldsymbol{k}, \boldsymbol{q} \right) = D_{zx'}^\mathrm{QQ} \left( \boldsymbol{k}, \boldsymbol{q} \right) = D^\mathrm{QQ} \left( \boldsymbol{k}, \boldsymbol{q} \right)$ in the tetragonal phase.
Using the quadrupole $O_{x'^2-y'^2, \boldsymbol{k}, \boldsymbol{q}}$ expressed in Eq. (\ref{Ox'2-y'2 2nd quant.}), we reduce the quadrupole interaction Hamiltonian of Eq. (\ref{HindQQ}) as
\begin{align}
\label{HindQQ_2}
H_\mathrm{ind}^\mathrm{QQ} 
= - \sum_{\boldsymbol{k}, \boldsymbol{k}', \boldsymbol{q}}
D^\mathrm{QQ} \left( \boldsymbol{k}, \boldsymbol{q} \right)
O_{x'^2-y'^2, \boldsymbol{k}, -\boldsymbol{q} } O_{x'^2-y'^2, \boldsymbol{k}', \boldsymbol{q} }
.
\end{align}
In order to understand the frequency dependence of the ultrasonic attenuation of $\alpha_{\mathrm{L} [110]}$ in Fig. \ref{Fig3}(a), we will examine the indirect quadrupole interaction coefficient $D^\mathrm{QQ} \left( \boldsymbol{k}, \boldsymbol{q} \right)$ in Eq. (\ref{HindQQ_2}), which strongly depends on the magnitude of wavevectors $\boldsymbol{k}$ and $\boldsymbol{k}'$ for the interacting electrons and $\boldsymbol{q}$ for the transverse acoustic phonons participating in the scattering.
Taking the electron energies of
$\varepsilon_{y'z} \left( \boldsymbol{k} \right)
= \varepsilon_{zx'} \left( \boldsymbol{k} \right)
= \varepsilon \left( \boldsymbol{k} \right)
= \hbar^2 \left| \boldsymbol{k} \right|^2 / 2m_e$
with effective electron mass $m_e$ and transverse acoustic phonon energies of
$\hbar \omega_x \left( \boldsymbol{q} \right)
= \hbar \omega_y \left( \boldsymbol{q} \right)
=\hbar \omega \left( \boldsymbol{q} \right)
= \hbar v_{66} |\boldsymbol{q}|$
with ultrasonic velocity $v_{66}$, we obtain the indirect quadrupole interaction coefficient $D^\mathrm{QQ} \left( \boldsymbol{k}, \boldsymbol{q} \right)$ of Eq. (\ref{HindQQ_2}) as
\begin{align}
\label{DQQ2}
D^\mathrm{QQ} \left( \boldsymbol{k}, \boldsymbol{q} \right)
&= - G_{x'^2-y'^2}^2
\frac{\hbar} {2V\rho_\mathrm{M} \omega \left( \boldsymbol{q} \right)} q^2 
\nonumber \\
&\qquad \quad \times
\frac{\hbar \omega \left( \boldsymbol{q} \right)}
{
\left[
\varepsilon \left( \boldsymbol{k} \right) -  \varepsilon \left( \boldsymbol{k} - \boldsymbol{q} \right) 
\right]^2 
- \hbar^2 \omega \left( \boldsymbol{q} \right) ^2
}
.
\end{align}

In the small-wavenumber limit of $| \boldsymbol{q} | \rightarrow 0$ for the transverse acoustic phonons, the indirect quadrupole interaction coefficient of Eq. (\ref{DQQ2}) is reduced to the simple formula
\begin{align}
\label{DQQ(k,0)}
D^\mathrm{QQ} \left( \boldsymbol{k}, \boldsymbol{q} = 0 \right)
= - \frac{G_{x'^2-y'^2}^2}{2 V \rho_\mathrm{M} } 
\frac{1} {\left( \displaystyle \frac{\hbar} {m_e}\right)^2 \left( k^2 - \displaystyle \frac{m_e^2v_{66}^2} {\hbar^2} \right)} 
.
\end{align}
The indirect quadrupole interaction coefficient of Eq. (\ref{DQQ(k,0)}) for electrons with a small wavenumber of $-\big| \boldsymbol{k}_\mathrm{b}^\mathrm{Q} \bigr| < k < \big| \boldsymbol{k}_\mathrm{b}^\mathrm{Q} \bigr| = m_e v_{66}/ \hbar$ has a positive sign, $D^\mathrm{QQ} \left( \boldsymbol{k}, \boldsymbol{q} = 0 \right) > 0$, showing the ferro-type quadrupole interaction, while the interaction coefficient for a relatively large wavenumber $k > \big| \boldsymbol{k}_\mathrm{b}^\mathrm{Q} \bigr| = m_e v_{66}/ \hbar$ possesses a negative sign, $D^\mathrm{QQ} \left( \boldsymbol{k}, \boldsymbol{q} = 0 \right) < 0$, indicating the antiferro-type quadrupole interaction.
The ferro-type interaction of $D^\mathrm{QQ} \left( \boldsymbol{k}, \boldsymbol{q} = 0 \right) > 0$ brings about the ferro-quadrupole ordering accompanying the structural transition, as actually observed in the under-doped compound $x = 0.036$.

It is worth estimating the boundary wavenumber $k_\mathrm{b}^\mathrm{Q} = \left| \boldsymbol{k}_\mathrm{b}^\mathrm{Q} \right|$ and the corresponding energy for the present iron pnictide.
In the compound with $x = 0.036$, the ultrasonic velocity $v_{66} = 772$ m/s at 65 K in the vicinity of the structural transition (1780 m/s at 100 K far above the transition) gives the boundary wavenumber $k_\mathrm{b}^\mathrm{Q} = 6.7 \times 10^6 $ m$^{-1}$ ($15 \times 10^6$ m$^{-1}$),
the boundary wavelength $\lambda_\mathrm{b}^\mathrm{Q} = 0.94 \times 10^{-6}$ m ($0.41 \times 10^{-6}$ m), and the corresponding frequency $f_\mathrm{b}^\mathrm{Q} = \varepsilon \left( k_\mathrm{b}^\mathrm{Q} \right)/h = 0.41$ GHz (2.2 GHz).
Here, we take the electron rest mass as $m_e$.
In the low-energy regime below 2 GHz $\approx$ 100 mK, therefore, the electrons bearing the quadrupoles $O_{x'^2-y'^2, \boldsymbol{k}, - \boldsymbol{q} }$ and $O_{x'^2-y'^2, \boldsymbol{k}', \boldsymbol{q} }$ in Eq. (\ref{HindQQ_2}) interact with each other through the strain $\varepsilon_{xy} \left( \boldsymbol{q} \right)$ in the manner of the ferro-type quadrupole interaction.
Actually, the relaxation rate $\tau^{-1}$ observed around the structural transition for $x = 0.036$ in Fig. \ref{Fig3}(b) is always smaller than the estimated crossover frequency of about 2 GHz.
This implies that the transverse acoustic waves with frequencies below 2 GHz exhibit the full extent of elastic softening and considerable damping in the vicinity of the structural transition point, while the acoustic waves with frequencies higher than 2 GHz exhibit less softening and damping.
As shown in Fig. \ref{Fig2}(e), the ultrasonic measurements with frequencies less than 260 MHz actually exhibited the full amount of softening in $C_{66}$, while the Raman scattering measurements of the transverse acoustic phonons with frequencies as high as 20 GHz in the end material $x = 0$ exhibited only 25\% of the full amount of softening \cite{Gallais}.

\subsection{Quadrupole interaction}

 The electrons accommodated in the degenerate $y'z$ and $zx'$ bands carry the quadrupoles.
In addition to the indirect quadrupole interaction $H_\mathrm{ind}^\mathrm{QQ}$ mediated by the strain $\varepsilon_{xy}$ presented in Sect. 4.3, we expect direct quadrupole interactions through the long-range Coulomb potential between electrons at positions $\boldsymbol{r}_i$ and $\boldsymbol{r}_j$, expressed as

\begin{align}
\label{H_Coulomb}
&\left\langle l_1 \ \sigma, l_2 \ \sigma' \left| H_\mathrm{Coulomb} \right| l_3 \ \sigma', l_4 \ \sigma \right\rangle
\nonumber \\
&  \ \ \ \ \ \ \ \ \ \ 
= \sum_{i \leq j} \iint d\boldsymbol{r}_i d\boldsymbol{r}_j
\psi_{l_1}^{\ast} \bigl( \boldsymbol{r}_i \bigr) v_\sigma^\ast \bigl( \boldsymbol{r}_i \bigr)
\psi_{l_2}^{\ast} \bigl( \boldsymbol{r}_j \bigr) v_{\sigma'}^\ast \bigl( \boldsymbol{r}_j \bigr)
\nonumber \\
& \ \ \ \ \ \ \ \ \ \ \ \ \ \ \ 
\times \frac{e^2} {\left| \boldsymbol{r}_i - \boldsymbol{r}_j \right|}
\psi_{l_3} \bigl( \boldsymbol{r}_j \bigr) v_{\sigma'} \bigl( \boldsymbol{r}_j \bigr)
\psi_{l_4} \bigl( \boldsymbol{r}_i \bigr)v_\sigma \bigl( \boldsymbol{r}_i \bigr)
.
\end{align}
Here, $l_n$ ($n = 1, 2, 3, 4$) denote suffixes of $y'z$ and $zx'$ orbital bands and $v_\sigma$ is the spin function in Eqs. (\ref{Filed operator 1.1}) and (\ref{Filed operator 1.2}).
The electric multipole expansion for the Coulomb interaction in Eq. (\ref{H_Coulomb}) for an isotropic space with $r_i < r_j$ provides terms consisting of the monopole, dipole, and quadrupole as \cite{Kuramoto, Arfken}
\begin{align}
\label{Multipole expansion}
\frac{e^2} {\left| \boldsymbol{r}_i - \boldsymbol{r}_j \right| }
&= \frac{e^2} {r_j}
\sum_{k=0}^{\infty}
\frac{4\pi} {2k+1} \left( \frac{r_i} {r_j} \right)^k
\sum_{m=-k}^{k}
 Y_k^m \bigl( \theta_i, \varphi_i \bigr) Y_k^{m\ast} \bigl( \theta_j, \varphi_j \bigr)
\nonumber \\
&= \frac{e^2} {r_j}
+ e^2 \frac{r_i} {r_j^2}
\left[
 \frac{x_i}{r_i} \frac{x_j}{r_j} + \frac{y_i}{r_i} \frac{y_j}{r_j} + \frac{z_i}{r_i} \frac{z_j}{r_j}
\right]
\nonumber \\
&\quad
+ e^2 \frac{r_i^2} {r_j^3}
\left[
O_{3z^2-r^2} \bigl( \boldsymbol{r}_i \bigr) O_{3z^2-r^2} \bigl( \boldsymbol{r}_j \bigr)
+ O_{x'^2-y'^2} \bigl( \boldsymbol{r}_i \bigr) O_{x'^2-y'^2} \bigl( \boldsymbol{r}_j \bigr)
\right.
\nonumber \\
& \qquad \qquad
+ O_{y'z} \bigl( \boldsymbol{r}_i \bigr) O_{y'z} \bigl( \boldsymbol{r}_j \bigr)
+ O_{zx'} \bigl( \boldsymbol{r}_i \bigr) O_{zx'} \bigl( \boldsymbol{r}_j \bigr)
\nonumber \\
& \qquad \qquad \quad
 \left.+ O_{x'y'} \bigl( \boldsymbol{r}_i \bigr) O_{x'y'} \bigl( \boldsymbol{r}_j \bigr)
\right]
.
\end{align}
According to the multipole expansion of Eq. (\ref{Multipole expansion}), we identify the electric quadrupole with the order parameter of the structural transition in the present iron pnictide. 
In our attempt to clarify the spontaneous symmetry breaking of the degenerate $y'z$ and $zx'$ orbital bands in the tetragonal lattice, we express the direct quadrupole interaction $H_\mathrm{C}^\mathrm{QQ}$ in terms of the quadrupole $O_{x'^2-y'^2}$ with the $B_2$ symmetry and $O_{x'y'}$ with $B_1$ symmetry as 
\begin{align}
\label{HCQQ}
H_\mathrm{C}^\mathrm{QQ}
= & - \sum_{i \leq j}
J_\mathrm{C} \bigl( \boldsymbol{r}_i - \boldsymbol{r}_j \bigr)
\nonumber \\
&\times
\left[
O_{x'^2-y'^2} \bigl( \boldsymbol{r}_i \bigr) O_{x'^2-y'^2} \bigl( \boldsymbol{r}_j \bigr)
+ O_{x'y'} \bigl( \boldsymbol{r}_i \bigr) O_{x'y'} \bigl( \boldsymbol{r}_j \bigr)
\right]
.
\end{align}
Here, we adopt the Coulomb interaction coefficient $J_\mathrm{C} \bigl( \boldsymbol{r}_i - \boldsymbol{r}_j \bigr)$, which  depends on the relative position $\boldsymbol{r}_i - \boldsymbol{r}_j$.
The quadrupole interaction of $O_{3z^2 - r^2}$ with $A_1$ symmetry is excluded because symmetry breaking is not expected.
The case of quadrupoles $O_{y'z}$ and $O_{zx'}$ being absent in the $y'z$ and $zx'$ bands is beyond the scope of this study. 
Note that the monopole interaction among the electrons in Eq. (\ref{Multipole expansion}) is effectively included in the band model concerned and that the dipole interaction is irrelevant for the present lattice bearing the inversion symmetry in the $xy$ plane.

Using the expressions $O_{x'y'} = \sqrt{3}\tau_x / 7$ in Eq. (\ref{Matrix Ox'y'}) and $O_{x'^2 - y'^2} = \sqrt{3}\tau_z / 7$ in Eq. (\ref{Matrix Ox'2-y'2}), we map the quadrupole interaction of Eq. (\ref{HCQQ}) on the ideal $xz$ model.
In the present real system, however, the deviation of the anisotropic quadrupole interaction from the ideal $xz$ model is expected.
From the viewpoint of the symmetry of the tetragonal lattice, the mutual quadrupole interactions of $O_{x'^2-y'^2}$ with $B_2$ symmetry and $O_{x'y'}$ with $B_1$ symmetry may be different.
Actually, it has been reported that the mixing between the 4$p$ states of As ions and the 3$d$ states of Fe ions brings about an anisotropic quadrupole interaction \cite{Yamada}.
The quadrupoles $O_{x'^2-y'^2}$ at positions $\boldsymbol{r}_i$ and $\boldsymbol{r}_j$ interact with each other through the strain $\varepsilon_{xy}$ of the transverse acoustic phonons for $C_{66}$, which exhibits the large amount of softening.
The quadrupoles $O_{x'y'}$ may also interact with each other mediated by the strain $\varepsilon_{x^2 - y^2}$ of the transverse acoustic phonons for $(C_{11} - C_{12})/2$, which monotonically increases.
The former quadrupole interaction of $O_{x'^2-y'^2}$ plays a significant role in the structural transition, while the latter quadrupole interaction of $O_{x'y'}$ has a minor effect on the phase transition.
Consequently, the indirect quadrupole interaction mediated by the transverse acoustic phonons also indicates the anisotropic nature of the quadrupole interaction. 

Combining the direct quadrupole interaction $H_\mathrm{C}^\mathrm{QQ}$ of Eq. (\ref{HCQQ}) followed by the Coulomb potential and the indirect quadrupole interaction $H_\mathrm{ind}^\mathrm{QQ}$ of Eq. (\ref{HindQQ_2}) mediated by the strain $\varepsilon_{xy}$ of the transverse acoustic phonons, we obtain the anisotropic quadrupole interaction $H_\mathrm{QQ}$ specified by parameter $\gamma$ as
\begin{align}
\label{Anisotropic HQQ}
&H_\mathrm{QQ} \bigl( \gamma \bigr)
\nonumber \\
& \quad 
= - \sum_{i \leq j} J_\mathrm{Q} \bigl( \boldsymbol{r}_i - \boldsymbol{r}_j \bigr)
\nonumber \\
& \qquad
\times \left[
O_{x'^2-y'^2} \bigl( \boldsymbol{r}_i \bigr) O_{x'^2-y'^2} \bigl( \boldsymbol{r}_j \bigr)
+ \gamma O_{x'y'} \bigl( \boldsymbol{r}_i \bigr) O_{x'y'} \bigl( \boldsymbol{r}_j \bigr)
\right]
.
\end{align}
The quadrupole interaction coefficient $J_\mathrm{Q} \bigl( \boldsymbol{r}_i - \boldsymbol{r}_j \bigr)$ in Eq. (\ref{Anisotropic HQQ}) is written as the sum of the Coulomb interaction coefficient $J_\mathrm{C} \bigl( \boldsymbol{r}_i - \boldsymbol{r}_j \bigr)$ in the direct interaction of Eq. (\ref{HCQQ}) and the indirect quadrupole interaction coefficient $D^\mathrm{QQ}\bigl( \boldsymbol{r}_i - \boldsymbol{r}_j \bigr)$ mediated by the strain $\varepsilon_{xy}$ of  the transverse acoustic phonons in Eq. (\ref{DQQ2}).
\begin{align}
\label{JQ = JC + DQQ}
J_\mathrm{Q} \bigl( \boldsymbol{r}_i - \boldsymbol{r}_j \bigr)
= J_\mathrm{C} \bigl( \boldsymbol{r}_i - \boldsymbol{r}_j \bigr) + D^\mathrm{QQ} \bigl( \boldsymbol{r}_i - \boldsymbol{r}_j \bigr)
.
\end{align}
The quadrupole interaction coefficient $J_\mathrm{Q} \bigl( \boldsymbol{r}_i - \boldsymbol{r}_j \bigr)$ and the anisotropic parameter $\gamma$ in Eq. (\ref{Anisotropic HQQ}) play a substantial role in the appearance of the ferro-quadrupole ordering associated with the structural transition and the hexadecapole ordering associated with the superconducting transition.

\subsection{Quadrupole ordering}

\begin{figure}[t]
\begin{center}
\includegraphics[angle=0,width=0.48\textwidth, bb=0 10 235 249]{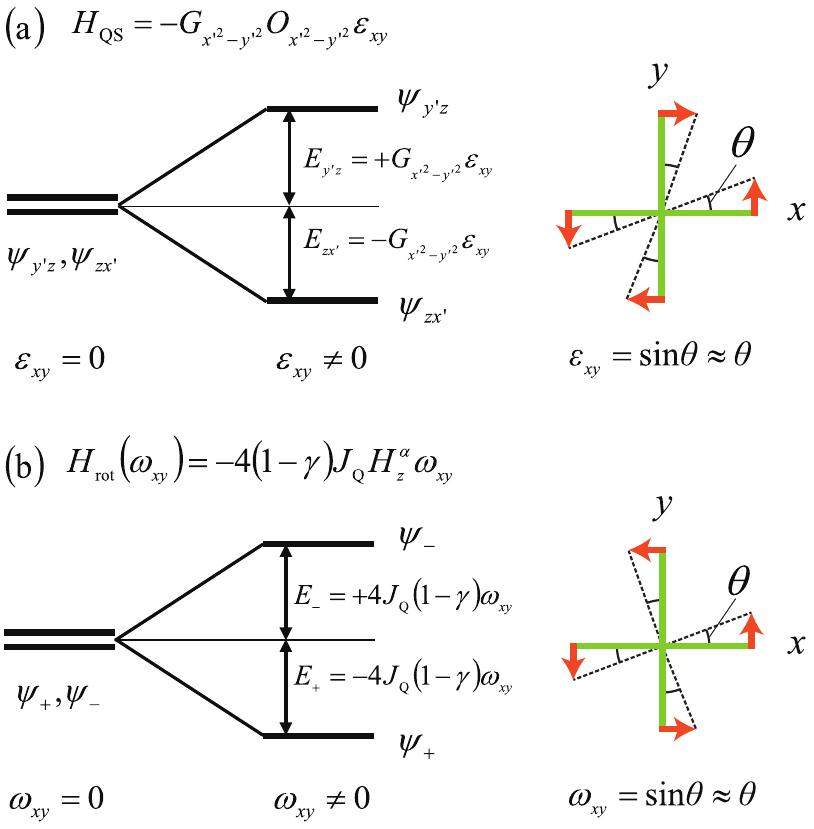}
\end{center}
\caption{
(Color online)
(a) Energy splitting of the degenerate $y'z$ and $zx'$ bands by the quadrupole-strain interaction $H_\mathrm{QS}$ of Eq. (\ref{HQS 2nd quant.}).
(b) Energy splitting of the two-electron states $\psi_\pm \bigl( \boldsymbol{r}_i, \boldsymbol{r}_j \bigr)$ of Eq. (\ref{psi+-}) by the hexadecapole-rotation interaction $H_\mathrm{rot} \bigl( \omega_{xy} \bigr)$ of Eq. (\ref{Hrot by Bij}).
The transverse ultrasonic wave with amplitude $\xi_y^0$ and frequency $\omega$ induces infinitesimal strain $\varepsilon_{xy} = \sin\theta$ and rotation $\omega_{xy} = \sin\theta$ with $\sin\theta = q_x \xi_y^0 = \left( \omega / v_x \right) \xi_y^0 = \left(2\pi / \lambda_x \right) \xi_y^0$ for a given phase velocity $v_x = v_{66}$.
}
\label{Fig7}
\end{figure}

In order to explain the elastic softening of $C_{66}$ and critical slowing down of the relaxation time $\tau$ around the structural transition in the iron pnictide, we consider the quadrupole interaction Hamiltonian $H_\mathrm{QQ} (\gamma)$ of Eq. (\ref{Anisotropic HQQ}) for the case of $\gamma = 0$ mapped on the Ising model.
A transverse ultrasonic wave with a small-wavenumber limit of $| \boldsymbol{q} | \rightarrow 0$ induces the strain $\varepsilon_{xy} = \varepsilon_{xy} \left( \boldsymbol{q} = 0 \right)$, which is treated as a classical quantity.
According to the quadrupole-strain interaction of Eq. (\ref{HQS 2nd quant.}), the strain $\varepsilon_{xy}$ generated by the ultrasonic wave lifts the degenerate $y'z$ and $zx'$ bands, as schematically shown in Fig. \ref{Fig7}(a).
This ultrasonic perturbation gives a quadrupole susceptibility proportional to the reciprocal temperature as 
\begin{align}
\label{chiQ}
\chi_\mathrm{Q}
=\frac{N_\mathrm{Q} G_{x'^2-y'^2}^2 / C_{66}^0} {T} = \frac{\mathit{\Delta}_\mathrm{Q} } {T}
.
\end{align}
Here, $N_\mathrm{Q}$ is the number of electrons carrying the ferro-type quadrupole interaction in the small-wavenumber regime of $-k_\mathrm{b}^\mathrm{Q} < k < k_\mathrm{b}^\mathrm{Q}$ for the boundary wavenumber $k_\mathrm{b}^\mathrm{Q}$ as discussed for Eq. (\ref{DQQ(k,0)}).
The quadrupole-strain interaction energy of $\mathit{\Delta}_\mathrm{Q} = N_\mathrm{Q} G_{x'^2-y'^2}^2 / C_{66}^0 $ in Eq. (\ref{chiQ}) is determined in terms of $D^\mathrm{QQ} \left( \boldsymbol{k}, \boldsymbol{q} \right)$ of Eq. (\ref{DQQ(k,0)}) for the small-wavenumber regime of $|\boldsymbol{q}| \rightarrow 0$ as
\begin{align}
\label{DQQ and deltaQ}
\frac{\mathit{\Delta}_\mathrm{Q} } {2VN_\mathrm{Q} } \frac{ C_{66}^0 } { \rho_\mathrm{M} v_{66}^2 } 
&=  \frac{G_{x'^2-y'^2}^2 } {2V \rho_\mathrm{M} v_{66}^2 }
\nonumber \\
&= \frac{1}{N_\mathrm{Q} } \sum_{ |\boldsymbol{k}| < k_\mathrm{b}^\mathrm{Q} }
D^\mathrm{QQ} \left( \boldsymbol{k}, \boldsymbol{q} = 0 \right)
\nonumber \\
&= \widetilde{D}^\mathrm{QQ}
.
\end{align}
Here, $\sum_{ |\boldsymbol{k}| < k_\mathrm{b}^\mathrm{Q} }$ means the sum over the electron states with wavenumber $|\boldsymbol{k}| < k_\mathrm{b}^\mathrm{Q}$, that participate in the ferro-type quadrupole interaction.
$\widetilde{D}^\mathrm{QQ}$ in Eq. (\ref{DQQ and deltaQ}) stands for the effective indirect quadrupole interaction of Eq. (\ref{DQQ2}).
As shown in Fig. \ref{Fig4}, the ultrasonic experiments give a quadrupole-strain interaction energy of $\mathit{\Delta}_\mathrm{Q} \sim 20$ K that is almost independent of the Co concentration $x$.
The indirect quadrupole interaction energy $D^\mathrm{QQ}$ in Eq. (\ref{DQQ and deltaQ}) is enhanced by the softening of $\rho_\mathrm{M} v_{66}^2 = C_{66}$ due to the quadrupole-strain interaction of Eq. (\ref{HQS 2nd quant.}).

The softening of $C_{66}$ as a precursor of the structural transition is expressed as 
\begin{align}
\label{Elastic with chiQ tilde}
C_{66}
=C_{66}^0 \left( 1 - \widetilde{\chi}_\mathrm{Q} \right)
&= C_{66}^0\left(\frac{T - T_\mathrm{s}^0} {T - \mathit{\Theta}_\mathrm{Q} } \right)
\nonumber \\
& = C_{66}^0\left( 1- \frac{\mathit{\Delta}_\mathrm{Q} } {T - \mathit{\Theta}_\mathrm{Q} } \right)
.
\end{align}
Here, we adopt the renormalized susceptibility $\widetilde{\chi}_\mathrm{Q}$ expressed by the quadrupole interaction energy $\mathit{\Theta}_\mathrm{Q}$ and the quadrupole-strain interaction energy  $\mathit{\Delta}_\mathrm{Q}$ as
\begin{align}
\label{chiQ tilde}
\widetilde{\chi}_\mathrm{Q}
= \frac{\mathit{\Delta}_\mathrm{Q} } {T-\mathit{\Theta}_\mathrm{Q} }
.
\end{align}
The elastic instability point due to the full softening $C_{66} \rightarrow 0$ gives the structural transition temperature as $T_\mathrm{s}^0 = \mathit{\Theta}_\mathrm{Q} + \mathit{\Delta}_\mathrm{Q}$.
This has already been used in Eq. (\ref{Temp. dep. C66}) for the analysis of the experimentally observed softening of $C_{66}$.

By using the mean field approximation, we obtain the following self-consistent equation giving the temperature dependence of the quadrupole order parameter $O_{x'^2 - y'^2}$:
\begin{align}
\label{Self-consistent of Ox'2-y'2}
\left\langle O_{x'^2 - y'^2} \right\rangle
= \tanh \left[ 
\widetilde{J}_\mathrm{Q}  \left\langle O_{x'^2 - y'^2}  \right\rangle /k_\mathrm{B}T  \right]
.
\end{align}
Here, $\langle A \rangle$ stands for the thermal average.
We use the effective coupling constant $\widetilde{J}_\mathrm{Q}$, which is the average of the Fourier transformed  quadrupole interaction of
$
J_\mathrm{Q} \bigl( \boldsymbol{r}_i - \boldsymbol{r}_j \bigr)
= J_\mathrm{C} \bigl( \boldsymbol{r}_i - \boldsymbol{r}_j \bigr) + D^\mathrm{QQ} \bigl( \boldsymbol{r}_i - \boldsymbol{r}_j \bigr)
$
in Eq. (\ref{JQ = JC + DQQ}) over the electron states with wavenumber $|\boldsymbol{k}| < k_\mathrm{b}^\mathrm{Q}$.
$\widetilde{J}_\mathrm{Q}$ is expressed as
\begin{align}
\label{JQ'}
\widetilde{J}_\mathrm{Q}
&= \widetilde{J}_\mathrm{C} + \widetilde{D}^\mathrm{QQ}
\nonumber \\
&= \frac{1}{N_\mathrm{Q} } \sum_{ |\boldsymbol{k}| < k_\mathrm{b}^\mathrm{Q} } J_\mathrm{C} \bigl( \boldsymbol{k}, \boldsymbol{q} = 0 \bigr)
+ \frac{1}{N_\mathrm{Q} } \sum_{ |\boldsymbol{k}| < k_\mathrm{b}^\mathrm{Q} } D^\mathrm{QQ} \bigl( \boldsymbol{k}, \boldsymbol{q} = 0 \bigr)
.
\end{align}
Here, $\widetilde{J}_\mathrm{C}$ is the effective energy of the direct Coulomb interaction of Eq. (\ref{HCQQ}). 
$\widetilde{D}^\mathrm{QQ}$ is the effective indirect quadrupole interaction coefficient in Eq. (\ref{DQQ and deltaQ}).
We suppose that the ferro-quadrupole ordering is caused by the quadrupole interaction with the positive coupling constant $\widetilde{J}_\mathrm{Q} > 0$.
We expect there to be a non vanishing solution $\langle O_{x'^2 - y'^2} \rangle \neq 0$ at low temperatures below the ferro-quadrupole ordering point of $T_\mathrm{c} = \widetilde{J}_\mathrm{Q} / k_\mathrm{B}$.

The elastic instability due to the full softening $C_{66} \rightarrow 0$  brings about spontaneous lattice distortion accompanying the ferro-quadrupole ordering as
\begin{align}
\label{Mean strain}
\left\langle \varepsilon_{xy} \right\rangle
= \frac{ N_\mathrm{Q} G_{x'^2 - y'^2} \left\langle O_{x'^2 - y'^2} \right\rangle } { C_{66} }
= \sqrt{ \frac{ N_\mathrm{Q} \mathit{\Delta}_\mathrm{Q} } { C_{66} } } \left\langle O_{x'^2 - y'^2} \right\rangle
.
\end{align}
Here, we use the relation between the coupling constant $G_{x'^2 - y'^2}$ and the quadrupole-strain interaction energy $\mathit{\Delta}_\mathrm{Q}$ of $G_{x'^2 - y'^2} = \sqrt{ \mathit{\Delta}_\mathrm{Q} C_{66}^0 / N_\mathrm{Q} }$.
The ferro-quadrupole ordering of  $\bigl\langle O_{x'^2 - y'^2} \bigr\rangle \neq 0$ associated with the spontaneous strain of $\bigl\langle \varepsilon_{xy} \bigr\rangle \neq 0$ in Eq. (\ref{Mean strain}) is caused by the structural transition from the tetragonal phase with the high-symmetry space group $D_{4h}^{17}$ to the orthorhombic phase with the low-symmetry space group $D_{2h}^{23}$.
The generators of the symmorphic space group $D_{4h}^{17}$ consist of the rotation, reflection, and inversion operations but do not involve the screw and glide operations.
The quenching of the $B_{2g}$ symmetry in the mother phase of $D_{4h}^{17}$ associated with the ferro-quadrupole ordering of $\langle O_{x'^2-y'^2} \rangle \neq 0$ and the spontaneous strain $\langle \varepsilon_{xy} \rangle \neq 0$ across the structural transition loses the symmetry operations, which consist of rotation through $\pm \pi / 2$ about the vertical axis of $C_4$, rotations through $\pi$ about the horizontal $x$- and $y$-axes of $2C_2'$, rotation through $\pm \pi /2$ about the vertical axis followed by inversion of $2IC_4$, and mirror reflection in the vertical plane of $2\sigma_\mathrm{v}$ \cite{Inui Group, International Table}.
Note that the ferro-type quadrupole ordering denoted in the present paper has been frequently referred to as cooperative Jahn-Teller effects \cite{Gehring and Gehring}, orbital orderings, or electronic nematic orders \cite{Murakami, Takata, Yi, Zhang, Nakayama, Shimojima, Kruger, Lv, Kontani_2, Yamakawa, Kasahara}.

The neutron scattering experiments on the end material BaFe$_2$As$_2$ have shown a spontaneous strain $\bigl\langle \varepsilon_{xy} \bigr\rangle$ of $5.08 \times 10^{-3}$ at 5 K in the distorted orthorhombic phase \cite{Huang}, where the order parameter is fully polarized as $\bigl\langle O_{x'^2 - y'^2} \bigr\rangle = 1$.
Adopting the quadrupole-strain interaction energy $\mathit{\Delta}_\mathrm{Q} = 21$ K and the elastic constant $C_{66}^0 = 3.19 \times 10^{10}$ J/m$^3$ of BaFe$_2$As$_2$ \cite{Kurihara Dr}, we estimate the number of 3$d$ electrons $N_\mathrm{Q}$ in Eq. (\ref{Mean strain}) to be $2.84 \times 10^{21}$ cm$^{-3}$, which is approximately one-fourteenth of the total number of electrons of $2N_\mathrm{Fe} = 3.92 \times 10^{22}$ cm$^{-3}$. 
Furthermore, we deduce the coupling energy $G_{x'^2 - y'^2}$ to be $\sim 4 \times 10^3$ K per electron, which is four times larger than the tentatively estimated $G_{x'^2 - y'^2}$ of $\sim 1 \times 10^3$ K for the localized electron picture in Sect. 3.1. 
The large quadrupole-strain coupling energy of $G_{x'^2 - y'^2} \sim 4 \times 10^3$ K is caused by the enhanced quadrupole due to the extended orbital radius compatible with the itinerant feature of the  3$d$ electron bands. 
Taking the number of 3$d$ electrons $N_\mathrm{Q} = 2.84 \times 10^{21}$ cm$^{-3}$ into account, we may roughly estimate the indirect quadrupole interaction energy in BaFe$_2$As$_2$ as $D^\mathrm{QQ} \sim 220$ K at $T = T_\mathrm{s} = 135$ K for $\mathit{\Delta}_\mathrm{Q} = 21$ K, $C_{66}^0 = 3.19 \times 10^{10}$ J/m$^3$, and $C_{66} = \rho_\mathrm{M} v_{66}^2 = 0.281 \times 10^{10}$ J/m$^3$.
This value is comparable with the structural transition temperature $T_\mathrm{s}$ of the under-doped compounds.
It is worth referring to the extremely enhanced quadrupole-strain interaction energy of $2.8 \times 10^5$ K for the vacancy orbital of a silicon wafer with a large orbital radius, which was verified by means of bulk ultrasonic waves and surface acoustic waves  \cite{Okabe, Mitsumoto}.

The divergence of the relaxation time $\tau$ observed via the ultrasonic attenuation coefficient $\alpha_{66}$ for $x = 0.036$ in Fig. \ref{Fig3}(b) is expressed in terms of the critical slowing down due to freezing of the ferro-quadrupole order parameter $O_{x'^2 - y'^2}$ at the structural transition temperature $T_\mathrm{s}^0 \cong T_\mathrm{c}^0 = 65$ K. 
The relaxation time $\tau_\mathrm{Q}$ is written as
\begin{align}
\label{tauQ}
\tau_\mathrm{Q}
= \tau_0 \left| \frac{ T - T_\mathrm{c}^0 } {T_\mathrm{c}^0} \right|^{-1}
\propto 
\frac{1} { 1 - \widetilde{\chi}_\mathrm{Q} }.
\end{align}
This expression for $\tau_\mathrm{Q}$ based on the degenerate $y'z$ and $zx'$ bands well reproduces the experimental results analyzed in terms of Eq. (\ref{Relaxation time}) for the critical index $z\nu =$ 1 for both tetragonal and orthorhombic phases.
The critical slowing down of the quadrupole $O_{x'^2 - y'^2}$ in the present iron pnictide agrees with the critical dynamics of the kinetic Ising model \cite{Suzuki}.
According to the dissipation fluctuation theorem, the critical slowing down of the relaxation time $\tau_\mathrm{Q}$ is expressed by the divergence of the susceptibility for the order parameter of the quadrupole $O_{x'^2-y'^2}$ due to the infinite increase in the correlation length $\zeta$ in the vicinity of the structural transition. 
By analogy with the ferro-magnetic spin system \cite{Mori, Halperin and Hohenberg}, the development of the correlation function of $\langle O_{x'^2-y'^2}(\boldsymbol{k}, t) O_{x'^2-y'^2}(-\boldsymbol{k}, 0) \rangle$ obeying the time decay factor of $\exp[-t/\tau_\mathrm{Q}(\boldsymbol{k})]$ should cause the relaxation time $\tau_\mathrm{Q}(\boldsymbol{k})$ to diverge at the ferro-type ordering point.
Knowledge of the diffusion coefficient is necessary for the numerical estimation of $\tau_0$ in Eq. (\ref{tauQ}).

\subsection{Hexadecapole-rotation interaction of two-electron states}

In the over-doped compound $x = 0.071$ exhibiting superconductivity, the antiferro-type quadrupole interaction with the negative quadrupole interaction energy $\mathit{\Theta}_\mathrm{Q} = -47$ K is verified by analysis of the softening of $C_{66}$ in the normal phase.
The softening of $C_{66}$ tends to zero at the fictitious structural instability point at the negative temperature $T_\mathrm{s}^0 = -26.5$ K.
This means that the critical slowing down of the relaxation time $\tau$ around the superconducting transition at $T_\mathrm{c}^0 = T_\mathrm{SC} = 23$ K in Fig. \ref{Fig1}(b) is caused by an appropriate order parameter, which is strictly distinguished from the quadrupole.
In the present section, we introduce two-electron states bound by the quadrupole interaction and consider the coupling of the hexadecapole carried by two-electron states to the rotation $\omega_{xy}$ of transverse ultrasonic waves.

Taking the Pauli exclusion principle into account, we express the energy of the anisotropic quadrupole interaction of Eq. (\ref{Anisotropic HQQ}) by using the two-electron state $\psi_{\Gamma_{\gamma} }^{S, S_z} \bigl( \boldsymbol{r}_i, \boldsymbol{r}_j \bigr)$ of the Slater determinant with the irreducible representation $\Gamma_\gamma$ of the orbital part, and the spin state denoted by the total spin $S$ and $z$ component $S_z$ as
\begin{align}
\label{Matrix elements of HQQ}
E_\mathrm{QQ}^{\Gamma_\gamma, S, S_z}
&=  \bigl\langle \Gamma_\gamma \ S \ S_z | H_\mathrm{QQ}\bigl(\gamma \bigr) | \Gamma_\gamma \ S \ S_z \bigr\rangle
\nonumber \\
&= \iint d\boldsymbol{r}_i d\boldsymbol{r}_j
\psi_{\Gamma_\gamma}^{S, S_z \ast} \bigl( \boldsymbol{r}_i, \boldsymbol{r}_j \bigr)
H_\mathrm{QQ}\bigl( \gamma \bigr)
\psi_{\Gamma_\gamma }^{S, S_z} \bigl( \boldsymbol{r}_i, \boldsymbol{r}_j \bigr)
.
\end{align}
From the viewpoint of symmetry for the orbital state, the direct product of the $E$-doublet in the point group symmetry $D_{2d}$ reduces as $E \otimes E = A_1 \oplus A_2 \oplus B_1 \oplus B_2$.
Thus, the symmetric $A_1$, $B_1$, and $B_2$ orbital states exist upon exchanging $\boldsymbol{r}_i$ for $\boldsymbol{r}_j$ as well as the antisymmetric $A_2$ state.
Furthermore, there are two kinds of spin states consisting of a spin singlet of total spin $S = 0$ that is symmetric upon exchanging $\boldsymbol{r}_i$ for $\boldsymbol{r}_j$ and the antisymmetric spin triplet with $S = 1$.
Consequently, we deduce the two-electron state of $\psi_{\Gamma_{\gamma} }^{S, S_z} \bigl( \boldsymbol{r}_i, \boldsymbol{r}_j \bigr)$ in Eq. (\ref{Matrix elements of HQQ}), which is denoted by the orbital state of the symmetry $\Gamma_\gamma = A_1$, $B_2$, and $B_1$ with the spin singlet of $S = 0$ and the orbital state of the $A_2$ symmetry with the spin triplet of $S = 1$ for $S_z = 1$, $0$, and $-1$.
As a result, we obtain the Slater determinant in terms of the one-electron orbital states of $\psi_{y'z} \bigl( \boldsymbol{r}_i \bigr)$ and $\psi_{zx'} \bigl( \boldsymbol{r}_i \bigr)$ and the spin states of $\alpha (\boldsymbol{r}_i) $ and $\beta (\boldsymbol{r}_i)$ as
\begin{align}
\label{Two-electron states A1}
\psi_{\Gamma_\gamma = A_1}^{S = 0, S_z = 0} \bigl( \boldsymbol{r}_i, \boldsymbol{r}_j \bigr)
&= \frac{1} {\sqrt{2} }
\left[
\psi_{y'z} \bigl( \boldsymbol{r}_i \bigr) \psi_{y'z} \bigl( \boldsymbol{r}_j \bigr) 
+ \psi_{zx'} \bigl( \boldsymbol{r}_i \bigr) \psi_{zx'} \bigl( \boldsymbol{r}_j \bigr) 
\right]
\nonumber \\
& \quad \times  \frac{1} {\sqrt{2} }
\left[
\alpha \bigl( \boldsymbol{r}_i \bigr) \beta \bigl( \boldsymbol{r}_j \bigr)
- \beta \bigl( \boldsymbol{r}_i \bigr) \alpha \bigl( \boldsymbol{r}_j \bigr)
\right]
, \\
\label{Two-electron states B2}
\psi_{B_2}^{0, 0} \bigl( \boldsymbol{r}_i, \boldsymbol{r}_j \bigr)
&= \frac{1} {\sqrt{2} }
\left[
\psi_{y'z} \bigl( \boldsymbol{r}_i \bigr) \psi_{y'z} \bigl( \boldsymbol{r}_j \bigr) 
- \psi_{zx'} \bigl( \boldsymbol{r}_i \bigr) \psi_{zx'} \bigl( \boldsymbol{r}_j \bigr) 
\right]
\nonumber \\
& \quad \times 
 \frac{1} {\sqrt{2} }
\left[
\alpha \bigl( \boldsymbol{r}_i \bigr) \beta \bigl( \boldsymbol{r}_j \bigr)
- \beta \bigl( \boldsymbol{r}_i \bigr) \alpha \bigl( \boldsymbol{r}_j \bigr)
\right]
, \\
\label{Two-electron states B1}
\psi_{B_1}^{0, 0} \bigl( \boldsymbol{r}_i, \boldsymbol{r}_j \bigr)
&= \frac{1} {\sqrt{2} }
\left[
\psi_{y'z} \bigl( \boldsymbol{r}_i \bigr) \psi_{zx'} \bigl( \boldsymbol{r}_j \bigr) 
+ \psi_{zx'} \bigl( \boldsymbol{r}_i \bigr) \psi_{y'z} \bigl( \boldsymbol{r}_j \bigr) 
\right]
\nonumber \\
& \quad \times 
 \frac{1} {\sqrt{2} }
\left[
\alpha \bigl( \boldsymbol{r}_i \bigr) \beta \bigl( \boldsymbol{r}_j \bigr)
- \beta \bigl( \boldsymbol{r}_i \bigr) \alpha \bigl( \boldsymbol{r}_j \bigr)
\right]
, \\
\label{Two-electron states S = 1}
\psi_{A_2}^{1, 1} \bigl( \boldsymbol{r}_i, \boldsymbol{r}_j \bigr)
&= \frac{1} {\sqrt{2} }
\left[
\psi_{y'z} \bigl( \boldsymbol{r}_i \bigr) \psi_{zx'} \bigl( \boldsymbol{r}_j \bigr) 
- \psi_{zx'} \bigl( \boldsymbol{r}_i \bigr) \psi_{y'z} \bigl( \boldsymbol{r}_j \bigr) 
\right]
\nonumber \\
&\quad \times 
\alpha \bigl( \boldsymbol{r}_i \bigr) \alpha \bigl( \boldsymbol{r}_j \bigr)
, \\
\label{Two-electron states S = 0}
\psi_{A_2}^{1, 0} \bigl( \boldsymbol{r}_i, \boldsymbol{r}_j \bigr)
&= \frac{1} {\sqrt{2} }
\left[
\psi_{y'z} \bigl( \boldsymbol{r}_i \bigr) \psi_{zx'} \bigl( \boldsymbol{r}_j \bigr) 
- \psi_{zx'} \bigl( \boldsymbol{r}_i \bigr) \psi_{y'z} \bigl( \boldsymbol{r}_j \bigr) 
\right]
\nonumber \\
& \quad \times 
 \frac{1} {\sqrt{2} }
\left[
\alpha \bigl( \boldsymbol{r}_i \bigr) \beta \bigl( \boldsymbol{r}_j \bigr)
+ \beta \bigl( \boldsymbol{r}_i \bigr) \alpha \bigl( \boldsymbol{r}_j \bigr)
\right]
, \\
\label{Two-electron states S = -1}
\psi_{A_2}^{1, -1} \bigl( \boldsymbol{r}_i, \boldsymbol{r}_j \bigr)
&= \frac{1} {\sqrt{2} }
\left[
\psi_{y'z} \bigl( \boldsymbol{r}_i \bigr) \psi_{zx'} \bigl( \boldsymbol{r}_j \bigr) 
- \psi_{zx'} \bigl( \boldsymbol{r}_i \bigr) \psi_{y'z} \bigl( \boldsymbol{r}_j \bigr) 
\right]
\nonumber \\
& \quad \times 
\beta \bigl( \boldsymbol{r}_i \bigr) \beta \bigl( \boldsymbol{r}_j \bigr)
.
\end{align}

In order to properly describe the energy of the anisotropic quadrupole interaction of Eq. (\ref{Matrix elements of HQQ}), we calculate the quadrupole interaction energy of the Hamiltonian $H_\mathrm{QQ} \bigl(\gamma \bigr)$ of Eq. (\ref{Anisotropic HQQ}) in terms of the matrix representation for the two-electron states $\psi_{\Gamma_{\gamma} }^{S, S_z} \bigl( \boldsymbol{r}_i, \boldsymbol{r}_j \bigr)$ of Eqs. (\ref{Two-electron states A1})-(\ref{Two-electron states S = -1}).
\begin{align}
\label{Matrix of HQQ}
&\boldsymbol{H}_\mathrm{QQ} \bigl(\gamma \bigr)
= - \sum_{i \leq j} J_\mathrm{Q}^{ij}
\nonumber \\
\times &\bordermatrix{
& \psi_{A_1}^{0, 0} & \psi_{B_2}^{0, 0} & \psi_{B_1}^{0, 0} & \psi_{A_2}^{1, 1} & \psi_{A_2}^{1, 0}  & \psi_{A_2}^{1, -1} \cr
& 1 + \gamma & 0& 0 & 0 & 0 & 0 \cr
& 0 & 1 - \gamma & 0 & 0 & 0 & 0 \cr
& 0 & 0 & - 1 + \gamma & 0 & 0 & 0 \cr
& 0 & 0 & 0 & - 1 - \gamma & 0 & 0 \cr
& 0 & 0 & 0 & 0 & -  1 - \gamma & 0 \cr
& 0 & 0 & 0 & 0 & 0 &  -  1 - \gamma
}
.
\end{align}
Here, the quadrupole interaction energy of $J_\mathrm{Q}^{ij}$ between electrons at positions $\boldsymbol{r}_i$ and $\boldsymbol{r}_j$ is calculated in terms of the radius function $f_d \bigl( r \bigr)$ for the 3$d$ electron in Eqs. (\ref{wave function y'z}) and (\ref{wave function zx'}) as 
\begin{align}
\label{JQij}
J_\mathrm{Q}^{ij} 
&= \left( \frac{ \sqrt{3} }{7} \right)^2 \iint dr_i dr_j
J_\mathrm{Q} \bigl( \boldsymbol{r}_i - \boldsymbol{r}_j \bigr) 
r_i^2 f_d \bigl( r_i \bigr)^2
r_j^2 f_d \bigl( r_j \bigr)^2
\nonumber \\
&= \frac{3}{49} \iint dr dr'
J_\mathrm{Q} \bigl( \boldsymbol{r} - \boldsymbol{R}_i - \boldsymbol{r}' + \boldsymbol{R}_j \bigr) 
|\boldsymbol{r} - \boldsymbol{R}_i|^2 f_d \bigl( |\boldsymbol{r} - \boldsymbol{R}_i| \bigr)^2
\nonumber \\
& \qquad\qquad\qquad \times |\boldsymbol{r}' - \boldsymbol{R}_j|^2 f_d \bigl( |\boldsymbol{r}' - \boldsymbol{R}_j| \bigr)^2
.
\end{align}
In the integral in the lower line in Eq. (\ref{JQij}), we take coordinates $\boldsymbol{r}$ and $\boldsymbol{r}'$ instead of $\boldsymbol{r}_i = \boldsymbol{r} - \boldsymbol{R}_i$ and $\boldsymbol{r}_j = \boldsymbol{r}' - \boldsymbol{R}_j$ for the positions $\boldsymbol{R}_i$ and $\boldsymbol{R}_j$ of Fe$^{2+}$ ion, respectively. 
The intra-atomic quadrupole interaction of $J_\mathrm{Q}^{ij}$ for the electrons accommodated in the orbital at the same Fe$^{2+}$ ion, $i$. $e$., $\boldsymbol{R}_i = \boldsymbol{R}_j$, is much stronger than the inter-atomic quadrupole interaction for electrons located at different Fe$^{2+}$ ions, $i$. $e$., $\boldsymbol{R}_i \neq \boldsymbol{R}_j$.
It is expected that the intra-atomic quadrupole interaction $J_\mathrm{Q}^{i = j}$ for $\boldsymbol{R}_i = \boldsymbol{R}_j$, which is mostly independent of the lattice structure, will show isotropic azimuth angle dependence.
This almost isotropic feature of the quadrupole interaction of $J_\mathrm{Q}^{ij}$ dominated by the intra-atomic coupling favors the s-like superconducting energy gap of the present iron pnictide.
This will be discussed in Sect. 4.9.

Ultrasonic measurements under pulsed magnetic fields of up to 62 T reveal increases of 7.4\% for the softened $C_{66}$ at 84 K in the tetragonal phase of $x = 0.036$ and 3.3\% for the softened $C_{66}$ at 30 K in the normal phase of $x = 0.071$ \cite{M. Akatsu_Pulse}.
Taking into account the robustness of the softening of $C_{66}$ against magnetic fields in the normal phase above the structural and superconducting transition temperatures,  we adopt the two-electron states of the Slater determinant with the spin-singlet states of Eqs. (\ref{Two-electron states A1})-(\ref{Two-electron states B1}), but disregard those with the spin-triplet states of Eqs. (\ref{Two-electron states S = 1})-(\ref{Two-electron states S = -1}).
We abbreviate the two-electron state with the spin singlet of
$\psi_{\Gamma_{\gamma} }^{0, 0} \bigl( \boldsymbol{r}_i, \boldsymbol{r}_j \bigr)$
in Eqs. (\ref{Two-electron states A1})-(\ref{Two-electron states B1}) to
$\psi_{\Gamma_{\gamma} } \bigl( \boldsymbol{r}_i, \boldsymbol{r}_j \bigr)$
in the following discussion.

In the case of the highly anisotropic quadrupole interaction with $\gamma = 0$, the quadrupole interaction energies of the $\psi_{A_1}\bigl( \boldsymbol{r}_i, \boldsymbol{r}_j \bigr)$ and $\psi_{B_2}\bigl( \boldsymbol{r}_i, \boldsymbol{r}_j \bigr)$ states in Eq. (\ref{Matrix of HQQ}) causes the degeneration of each other and the energy of the $\psi_{B_1}\bigl( \boldsymbol{r}_i, \boldsymbol{r}_j \bigr)$ state deviates from them.
 This highly anisotropic case of $\gamma \approx 0$ in Eq. (\ref{Anisotropic HQQ}) leads to the ferro-quadrupole ordering of $O_{x'^2-y'^2}$ accompanying the structural transition.
With increasing the Co concentration $x$ to the QCP of $x_\mathrm{QCP} = 0.061$, the quadrupole interaction of Eq. (\ref{Anisotropic HQQ}) develops an almost isotropic feature with $\gamma \stackrel{<}{_\sim} 1$, where the energies of the $\psi_{B_2}\bigl( \boldsymbol{r}_i, \boldsymbol{r}_j \bigr)$ and $\psi_{B_1}\bigl( \boldsymbol{r}_i, \boldsymbol{r}_j \bigr)$ states get closer to each other and the energy of the $\psi_{A_1}\bigl( \boldsymbol{r}_i, \boldsymbol{r}_j \bigr)$ state is different.
Since the quadrupoles $O_{x'^2-y'^2}$ of Eq. (\ref{Matrix Ox'2-y'2}) and $O_{x'y'}$ of Eq. (\ref{Matrix Ox'y'}) and the angular momentum $l_z$ of Eq. (\ref{Matrix lz}) obey the commutation relation of Eq. (\ref{Pauli matrix Commutation relation}) among the Pauli matrices, we expect quantum fluctuation between the quadrupoles $O_{x'^2-y'^2}$ and $O_{x'y'}$ in the vicinity of the QCP, which is particularly important for explaining the superconductivity accompanying the hexadecapole ordering.
The special case of $\gamma = 1$ mapped on the ideal $xz$ model leads to the fully degenerate $\psi_{B_2}\bigl( \boldsymbol{r}_i, \boldsymbol{r}_j \bigr)$ and $\psi_{B_1}\bigl( \boldsymbol{r}_i, \boldsymbol{r}_j \bigr)$ states, which causes the disappearance of the hexadecapole.
This is inconsistent with the experimental observation of the critical slowing down around the superconducting transition temperature.

When the rotation $\omega_{xy}$ is induced by either the thermally excited  transverse acoustic phonons or the experimentally generated transverse ultrasonic waves, the phases of the two-electron state of $\psi_{{\Gamma}_\gamma} \bigl( \boldsymbol{r}_i, \boldsymbol{r}_j \bigr)$ at positions $\boldsymbol{r}_i$ and $\boldsymbol{r}_j$ in Eq. (\ref{Matrix elements of HQQ}) bound by the quadrupole interaction of Eq. (\ref{Anisotropic HQQ}) are simultaneously changed by the rotation operator as follows:
\begin{align}
\label{Rotation effect for two-electron state}
\psi_{\Gamma_\gamma} \bigl( \boldsymbol{r}_i, \boldsymbol{r}_j \bigr)
& \rightarrow
\exp \left[ -i l_z \bigl( \boldsymbol{r}_i \bigr) \omega_{xy} \right]
\exp \left[ -i l_z \bigl( \boldsymbol{r}_j \bigr) \omega_{xy} \right]
\nonumber \\
& \qquad \qquad \qquad \qquad \qquad \times
\psi_{\Gamma_\gamma} \bigl( \boldsymbol{r}_i, \boldsymbol{r}_j \bigr)
.
\end{align}
Thus, an infinitesimal amount of rotation $\omega_{xy}$ perturbs the quadrupole interaction Hamiltonian of Eq. (\ref{Anisotropic HQQ}) as
\begin{align}
\label{Rotation of Anisotropic HQQ}
&
\left\langle \Gamma_\gamma \left| H_\mathrm{QQ} \bigl( \omega_{xy} \bigr) \right| \Gamma_{\gamma '} \right\rangle
\nonumber \\
&
= \iint d\boldsymbol{r}_i d\boldsymbol{r}_j
\psi_{\Gamma_\gamma}^\ast \bigl( \boldsymbol{r}_i, \boldsymbol{r}_j \bigr)
\exp \left\{ i \left[ l_z \bigl( \boldsymbol{r}_i \bigr) + l_z \bigl( \boldsymbol{r}_j \bigr) \right] \omega_{xy} \right\}
\nonumber \\
& \qquad \qquad 
\times H_\mathrm{QQ} \bigl( \gamma \bigr)
\exp \left\{ -i \left[ l_z \bigl( \boldsymbol{r}_i \bigr) + l_z \bigl( \boldsymbol{r}_j \bigr) \right] \omega_{xy} \right\}
\psi_{\Gamma_{\gamma'}} \bigl( \boldsymbol{r}_i, \boldsymbol{r}_j \bigr)
\nonumber \\
&
\approx \iint d\boldsymbol{r}_i d\boldsymbol{r}_j
\psi_{\Gamma_\gamma}^\ast \bigl( \boldsymbol{r}_i, \boldsymbol{r}_j \bigr)
H_\mathrm{QQ} \bigl( \gamma \bigr)
\psi_{\Gamma_{\gamma'} } \bigl( \boldsymbol{r}_i, \boldsymbol{r}_j \bigr)
\nonumber \\ 
&
+ \iint d\boldsymbol{r}_i d\boldsymbol{r}_j
\psi_{\Gamma_\gamma}^\ast \bigl( \boldsymbol{r}_i, \boldsymbol{r}_j \bigr)
i \left[
l_z \bigl( \boldsymbol{r}_i \bigr) + l_z \bigl( \boldsymbol{r}_j \bigr),  H_\mathrm{QQ} \bigl( \gamma \bigr)
\right] 
\nonumber \\
&\qquad \qquad \qquad \qquad \qquad \qquad \qquad \times 
\psi_{\Gamma_{\gamma'}} \bigl( \boldsymbol{r}_i, \boldsymbol{r}_j \bigr)  \omega_{xy}
\nonumber \\
&
+ \iint d\boldsymbol{r}_i d\boldsymbol{r}_j
\psi_{\Gamma_\gamma}^\ast \bigl( \boldsymbol{r}_i, \boldsymbol{r}_j \bigr)
\nonumber \\
& \qquad \qquad \times \left( - \frac{1}{2} \right) 
 \left[
l_z \bigl( \boldsymbol{r}_i \bigr) + l_z \bigl( \boldsymbol{r}_j \bigr), 
\left[
l_z \bigl( \boldsymbol{r}_i \bigr) + l_z \bigl( \boldsymbol{r}_j \bigr),  H_\mathrm{QQ} \bigl( \gamma \bigr)
\right] \right] 
\nonumber \\
& \qquad \qquad \qquad \qquad \times 
 \psi_{\Gamma_{\gamma'}} \bigl( \boldsymbol{r}_i, \boldsymbol{r}_j \bigr) \bigl( \omega_{xy} \bigr)^2
.
\end{align}
Using the commutation relation of Eq. (\ref{Pauli matrix Commutation relation}) for the Pauli matrices, we reduce the perturbation Hamiltonian $H_\mathrm{rot} \bigl( \omega_{xy} \bigr)$ which depends on the rotation $\omega_{xy}$ as follows:
\begin{align}
\label{Rotation Hamiltonian 1}
&H_\mathrm{rot} \bigl( \omega_{xy} \bigr)
\nonumber \\
&= 2\bigl( 1 - \gamma \bigr) \sum_{i \leq j} J_\mathrm{Q} \bigl( \boldsymbol{r}_i - \boldsymbol{r}_j \bigr)
\nonumber \\
& \qquad \times
\left[
O_{x'y'} \bigl( \boldsymbol{r}_i \bigr) O_{x'^2-y'^2} \bigl( \boldsymbol{r}_j \bigr)
+ O_{x'^2-y'^2} \bigl( \boldsymbol{r}_i \bigr) O_{x'y'} \bigl( \boldsymbol{r}_j \bigr)
\right] \omega_{xy}
\nonumber \\
&+ 4\bigl( 1 - \gamma \bigr) \sum_{i \leq j} J_\mathrm{Q} \bigl( \boldsymbol{r}_i - \boldsymbol{r}_j \bigr)
\nonumber \\
& \qquad \times 
\left[
O_{x'^2-y'^2} \bigl( \boldsymbol{r}_i \bigr) O_{x'^2-y'^2} \bigl( \boldsymbol{r}_j \bigr)
- O_{x'y'} \bigl( \boldsymbol{r}_i \bigr) O_{x'y'} \bigl( \boldsymbol{r}_j \bigr)
\right] \bigl( \omega_{xy} \bigr)^2
.
\end{align}
From the first term in Eq. (\ref{Rotation Hamiltonian 1}), which linearly depends on the rotation $\omega_{xy}$, we identify the electric hexadecapole as
\begin{align}
\label{Hexadecapole}
H_z^\alpha \bigl( \boldsymbol{r}_i, \boldsymbol{r}_j \bigr)
= O_{x'y'} \bigl( \boldsymbol{r}_i \bigr) O_{x'^2-y'^2} \bigl( \boldsymbol{r}_j \bigr)
+ O_{x'^2-y'^2} \bigl( \boldsymbol{r}_i \bigr) O_{x'y'} \bigl( \boldsymbol{r}_j \bigr)
.
\end{align}
The hexadecapole $H_z^\alpha \bigl( \boldsymbol{r}_i, \boldsymbol{r}_j \bigr)$ of Eq. (\ref{Hexadecapole}) is invariant upon the exchange of positions $\boldsymbol{r}_i$ and $\boldsymbol{r}_j$ but is antisymmetric upon the exchange of coordinates $x'$ and $y'$.  
In Fig. \ref{Fig5}(c), we schematically show the interaction between the hexadecapole $H_z^\alpha \bigl( \boldsymbol{r}_i, \boldsymbol{r}_j \bigr)$ of Eq. (\ref{Hexadecapole}) carried by the two-electron states and the rotation $\omega_{xy}$ of the transverse acoustic phonons. 
The hexadecapole $H_z^\alpha \bigl( \boldsymbol{r}_i, \boldsymbol{r}_j \bigr)$ operates on the two-electron state that is bound by the quadrupole interaction of Eq. (\ref{Anisotropic HQQ}), while the hexadecapole $H_z^\alpha \bigl( \boldsymbol{r}' \bigr)$ in Eq. (\ref{Hrot from HCEF}) acts on the one-electron state that is trapped in the CEF Hamiltonian of the central force.
Note that the hexadecapole $H_z^\alpha \bigl( \boldsymbol{r}_i, \boldsymbol{r}_j \bigr)$ and rotation $\omega_{xy}$ commonly belong to the $A_2$ symmetry of the point group symmetry $D_{2d}$.
The coefficient of the second term proportional to $\bigl( \omega_{xy} \bigr)^2$ in Eq. (\ref{Rotation Hamiltonian 1}) represents the energy modulation of the anisotropic quadrupole interaction. 

Comparing the term proportional to the rotation $\omega_{xy}$ in the expressions of Eq. (\ref{Rotation of Anisotropic HQQ}) with that in Eq. (\ref{Rotation Hamiltonian 1}), we write the Heisenberg equation of the time derivative for the total angular momentum $l_z \bigl( \boldsymbol{r}_i, \boldsymbol{r}_j \bigr) = l_z \bigl( \boldsymbol{r}_i \bigr) + l_z \bigl( \boldsymbol{r}_j \bigr)$ which is proportional to the hexadecapole $H_z^\alpha \bigl( \boldsymbol{r}_i, \boldsymbol{r}_j \bigr)$ as follows:
\begin{align}
\label{Torque for two-electron state}
i\hbar \frac{\partial} {\partial t} l_z \bigl( \boldsymbol{r}_i, \boldsymbol{r}_j \bigr) 
&= \left[ l_z \bigl( \boldsymbol{r}_i, \boldsymbol{r}_j \bigr),  H_\mathrm{QQ} \bigl( \gamma \bigr)  \right]
\nonumber \\
&= -2i \bigl( 1 - \gamma \bigr) J_\mathrm{Q} \bigl( \boldsymbol{r}_i - \boldsymbol{r}_j \bigr)
H_z^\alpha \bigl( \boldsymbol{r}_i, \boldsymbol{r}_j \bigr)
.
\end{align}
Furthermore, the time derivative of the angular momentum in Eq. (\ref{Torque for two-electron state}) is identified with the torque $\tau_{xy} \bigl( \boldsymbol{r}_i, \boldsymbol{r}_j \bigr)$ for the two-electron state bound by the anisotropic quadrupole interaction of Eq. (\ref{Anisotropic HQQ}) as
\begin{align}
\label{Define torque}
\hbar \frac{\partial} {\partial t} l_z \bigl( \boldsymbol{r}_i, \boldsymbol{r}_j \bigr) 
= \tau_{xy} \bigl( \boldsymbol{r}_i, \boldsymbol{r}_j \bigr)
.
\end{align}
Note that the torque $\tau_{xy} \bigl( \boldsymbol{r}_i, \boldsymbol{r}_j \bigr)$ in Eq. (\ref{Define torque}) vanishes for the special case of the ideal isotropic quadrupole interaction with $\gamma = 1$ of Eq. (\ref{Anisotropic HQQ}) identified with the ideal $xz$ model.
The hexadecapole $H_z^\alpha \bigl( \boldsymbol{r}_i, \boldsymbol{r}_j \bigr)$ proportional to the time derivative of the angular momentum in Eq. (\ref{Torque for two-electron state}) conserves the time-reversal symmetry, while the angular momentum $l_z \bigl( \boldsymbol{r}_i, \boldsymbol{r}_j \bigr)$ itself breaks the time-reversal symmetry.

The hexadecapole-rotation interaction linearly coupled to the rotation $\omega_{xy}$ in Eq. (\ref{Rotation Hamiltonian 1}) has the following matrix elements for the two-electron states of $\psi_{\Gamma_\gamma} \bigl( \boldsymbol{r}_i, \boldsymbol{r}_j \bigr)$:
\begin{align}
\label{Matrix elements of Hrot}
&\left\langle \Gamma_\gamma \left| H_\mathrm{rot}^0 \bigl( \omega_{xy} \bigr) \right| \Gamma_{\gamma '} \right\rangle
\nonumber \\ 
&=  2(1-\gamma)  \sum_{i \leq j} \iint d\boldsymbol{r}_i d\boldsymbol{r}_j
\psi_{\Gamma_\gamma}^\ast \bigl( \boldsymbol{r}_i, \boldsymbol{r}_j \bigr)
J_\mathrm{Q} \bigl( \boldsymbol{r}_i - \boldsymbol{r}_j \bigr) H_z^\alpha \bigl( \boldsymbol{r}_i, \boldsymbol{r}_j \bigr)
\nonumber \\
& \qquad \qquad \times \psi_{\Gamma_{\gamma'} } \bigl( \boldsymbol{r}_i, \boldsymbol{r}_j \bigr)
\omega_{xy}
\nonumber \\
&= 2(1-\gamma) \sum_{i \leq j} \iint d\boldsymbol{r} d\boldsymbol{r}'
\psi_{\Gamma_\gamma}^\ast \bigl( \boldsymbol{r} - \boldsymbol{R}_i, \boldsymbol{r}' - \boldsymbol{R}_j \bigr)
\nonumber \\
&\qquad \qquad  \times J_\mathrm{Q} \bigl( \boldsymbol{r}_i - \boldsymbol{R}_i - \boldsymbol{r}_j + \boldsymbol{R}_j \bigr)
H_z^\alpha \bigl( \boldsymbol{r} - \boldsymbol{R}_i, \boldsymbol{r}' - \boldsymbol{R}_j \bigr)
\nonumber \\
&\qquad \qquad \qquad \qquad \times
\psi_{\Gamma_{\gamma'} } \bigl( \boldsymbol{r} - \boldsymbol{R}_i, \boldsymbol{r}' - \boldsymbol{R}_j \bigr)
\omega_{xy}
.
\end{align}
In the integral of Eq. (\ref{Matrix elements of Hrot}), we respectively take coordinates $\boldsymbol{r}$ and $\boldsymbol{r}'$ instead of $\boldsymbol{r}_i = \boldsymbol{r} - \boldsymbol{R}_i$ and $\boldsymbol{r}_j = \boldsymbol{r}' - \boldsymbol{R}_j$ for the positions $\boldsymbol{R}_i$ and $\boldsymbol{R}_j$ of Fe$^{2+}$ ions.
As mentioned for Eq. (\ref{JQij}), the intra-atomic quadrupole interaction between electrons accommodated in the same Fe$^{2+}$ ion with $\boldsymbol{R}_i = \boldsymbol{R}_j$ is dominant over the inter-atomic quadrupole interaction with $\boldsymbol{R}_i \neq \boldsymbol{R}_j$.
The hexadecapole formed by the intra-atomic quadrupole interaction plays an important role in the appearance of the hexadecapole ordering in the superconducting state.

Among the two-electron states listed in Eqs. (\ref{Two-electron states A1})-(\ref{Two-electron states S = -1}), we take the spin-singlet states consisting of the two-electron states $\psi_{\Gamma_\gamma}^{0,0} \bigl( \boldsymbol{r}_i, \boldsymbol{r}_j \bigr) = \psi_{\Gamma_\gamma} \bigl( \boldsymbol{r}_i, \boldsymbol{r}_j \bigr)$ for the orbital symmetries of $\Gamma_\gamma = A_1$, $B_2$, and $B_1$ compatible with the magnetic robustness of the present system, while we disregard the spin-triplet state of $\psi_{A_2}^{S = 1, \ S_z} \bigl( \boldsymbol{r}_i, \boldsymbol{r}_j \bigr)$ with $S_z = 1$, $0$, and $-1$.
We deduce the hexadecapole-rotation interaction of Eq. (71) for the spin-singlet states as 
\begin{align}
\label{Matrix of Hrot off-diagonal}
\boldsymbol{H}_\mathrm{rot}^0 \bigl( \omega_{xy} \bigr)
= - 4 \bigl(1 - \gamma \bigr) \sum_{i \leq j} J_\mathrm{Q}^{ij}
\bordermatrix{
& \psi_{A_1} & \psi_{B_2} & \psi_{B_1} \cr
& 0 & 0 & 0 \cr
& 0 & 0 & 1 \cr
& 0 & 1 & 0
}
\omega_{xy}
.
\end{align}
Here, the matrix elements
$\bigl\langle \Gamma_\gamma | H_\mathrm{rot}^0 \bigl( \omega_{xy} \bigr) | \Gamma_{\gamma '} \bigr\rangle$
in Eq. (\ref{Matrix of Hrot off-diagonal}) for the two-electron state $\psi_{\Gamma_\gamma} \bigl( \boldsymbol{r}_i, \boldsymbol{r}_j \bigr)$ with the symmetries of $\Gamma_\gamma = A_1$, $B_2$, and $B_1$ and the spin singlet of S = 0 of Eqs. (\ref{Two-electron states A1})-(\ref{Two-electron states B1}) are calculated.
The hexadecapole $H_z^\alpha \bigl( \boldsymbol{r}_i, \boldsymbol{r}_j \bigr)$ with the $A_2$ symmetry possesses the off-diagonal elements between the two-electron states $\psi_{B_2} \bigl( \boldsymbol{r}_i, \boldsymbol{r}_j \bigr)$ and $\psi_{B_1} \bigl( \boldsymbol{r}_i, \boldsymbol{r}_j \bigr)$, but these matrix elements vanish for the state $\psi_{A_1} \bigl( \boldsymbol{r}_i, \boldsymbol{r}_j \bigr)$.
This is confirmed by the symmetry property of $B_2 \otimes B_1 = A_2$.

We have particular interest in the interplay of the rotation $\omega_{xy}$ with the appearance of the hexadecapole ordering and the superconductivity.
Employing the unitary transformation for Eq. (\ref{Matrix of Hrot off-diagonal}), we obtain the diagonal representation of the hexadecapole-rotation interaction as
\begin{align}
\label{Matrix of Hrot diagonal}
\boldsymbol{H}_\mathrm{rot}\bigl( \omega_{xy} \bigr)
= - 4 \left(1 - \gamma \right) \sum_{i \leq j} J_\mathrm{Q}^{ij}  
\bordermatrix{
& \psi_+  & \psi_- \cr
& 1 & 0 \cr
& 0 & -1
}
\omega_{xy}
.
\end{align}
Here, we adopt the wave function $\psi_h \bigl( \boldsymbol{r}_i, \boldsymbol{r}_j \bigr)$ with $h= \pm$ for the two-electron states as
\begin{align}
\label{psi+- by y'z and zx'}
\psi_\pm \bigl( \boldsymbol{r}_i, \boldsymbol{r}_j \bigr)
&= \frac{1} {\sqrt{2} }
\left[
\psi_{B_2} \bigl( \boldsymbol{r}_i, \boldsymbol{r}_j \bigr) \pm \psi_{B_1} \bigl( \boldsymbol{r}_i, \boldsymbol{r}_j \bigr)
\right]
\nonumber \\
&= \frac{1} {\sqrt{2} }
\left[
\frac{ \psi_{y'z} \bigl( \boldsymbol{r}_i \bigr) \pm \psi_{zx'} \bigl( \boldsymbol{r}_i \bigr) } { \sqrt{2} }
\psi_{y'z} \bigl( \boldsymbol{r}_j \bigr) \right.
\nonumber \\
& \left. \qquad \qquad \pm
\frac{ \psi_{y'z} \bigl( \boldsymbol{r}_i \bigr) \mp \psi_{zx'} \bigl( \boldsymbol{r}_i \bigr) } { \sqrt{2} }
\psi_{zx'} \bigl( \boldsymbol{r}_j \bigr)
\right]
\nonumber \\
&\qquad \times 
\frac{1}{\sqrt{2} } 
\left[
\alpha \bigl( \boldsymbol{r}_i \bigr) \beta \bigl( \boldsymbol{r}_j \bigr)
- \beta \bigl( \boldsymbol{r}_i \bigr) \alpha \bigl( \boldsymbol{r}_j \bigr)
\right]
.
\end{align}
By analogy with the Zeeman term for the magnetic dipole moments in a magnetic field, we identify the wave function $\psi_+ \bigl( \boldsymbol{r}_i, \boldsymbol{r}_j \bigr)$ with the eigenstate for the hexadecapole corresponding to the right-hand rotation of $h = +$ and the wave function $\psi_- \bigl( \boldsymbol{r}_i, \boldsymbol{r}_j \bigr)$ with eigenstate for the hexadecapole corresponding to the left-hand rotation of $h = -$. 
The hexadecapole-rotation interaction of Eq. (\ref{Matrix of Hrot diagonal}) indicates that the rotation $\omega_{xy}$ acts as a symmetry-breaking field on the hexadecapole carried by the two-electron states $\psi_+ \bigl( \boldsymbol{r}_i, \boldsymbol{r}_j \bigr)$ and $\psi_- \bigl( \boldsymbol{r}_i, \boldsymbol{r}_j \bigr)$ of Eq. (\ref{psi+- by y'z and zx'}).

 In order to properly describe the hexadecapole ordering, it is convenient to use the one-electron states $\lambda_\pm \left( \boldsymbol{r}_i \right)$ described in terms of the spherical harmonics of $Y_l^m \left( \theta_i, \varphi_i \right)$ with the orbital quantum number $l = 2$ and the azimuthal quantum number $m = \pm1$ and the radial part of $f_d \left( r_i \right)$ as
\begin{align}
\label{wave function lambda}
\lambda_\pm \left( \boldsymbol{r}_i \right)
&= f_d \left( r_i \right) Y_2^{\pm1} \left( \theta_i, \varphi_i \right)
\nonumber \\
&= -\frac{1}{\sqrt{2} } 
\left[
i \psi_{y'z} \left( \boldsymbol{r}_i \right) \pm \psi_{zx'} \left( \boldsymbol{r}_i \right)
\right]
.
\end{align}
We rewrite the two-electron states $\psi_\pm \bigl( \boldsymbol{r}_i, \boldsymbol{r}_j \bigr)$ of Eq. (\ref{psi+- by y'z and zx'}) in terms of the one-electron wave function $\lambda_\pm \left( \boldsymbol{r}_i \right)$ of Eq. (\ref{wave function lambda}) as
\begin{align}
\label{psi+-}
\psi_\pm \bigl( \boldsymbol{r}_i, \boldsymbol{r}_j \bigr)
&= \frac{1} {\sqrt{2} }
\left[
e^{\mp i \frac{3\pi}{4} } \lambda_{+1} \bigl( \boldsymbol{r}_i \bigr) \lambda_{+1} \bigl( \boldsymbol{r}_j \bigr)
+ e^{\pm i \frac{3\pi}{4} } \lambda_{-1} \bigl( \boldsymbol{r}_i \bigr) \lambda_{-1} \bigl( \boldsymbol{r}_j \bigr)
\right]
\nonumber \\
& \qquad \qquad \times 
\frac{1} {\sqrt{2}}
\left[
\alpha \bigl( \boldsymbol{r}_i \bigr) \beta \bigl( \boldsymbol{r}_j \bigr)
- \beta \bigl( \boldsymbol{r}_i \bigr) \alpha \bigl( \boldsymbol{r}_j \bigr)
\right]
.
\end{align}
Note that the eigenstates of $\psi_\pm \bigl( \boldsymbol{r}_i, \boldsymbol{r}_j \bigr)$ of Eq. (\ref{psi+-}) consist of the two-electron states of $\lambda_{+1} \bigl( \boldsymbol{r}_i \bigr) \lambda_{+1} \bigl( \boldsymbol{r}_j \bigr)$ and $\lambda_{-1} \bigl( \boldsymbol{r}_i \bigr) \lambda_{-1} \bigl( \boldsymbol{r}_j \bigr)$, which are superposed on each other while maintaining an orthogonal relation with a phase difference of $\pm 3\pi / 2$.

Taking the spin-singlet state in Eq. (\ref{psi+-}) into account, we introduce annihilation operators of $B_{ i, j, \pm, \sigma, \overline{\sigma} }$ and creation operators of $B_{ i, j, \pm, \sigma, \overline{\sigma} }^\dagger$ for the two-electron eigenstates $\psi_\pm \bigl( \boldsymbol{r}_i, \boldsymbol{r}_j \bigr)$ of the hexadecapole as
\begin{align}
\label{Operator B by l 1} 
B_{i, j, \pm, \sigma, \overline{\sigma} } 
= \frac{1}{\sqrt{2}}
\left( 
e^{\mp i \frac{3\pi}{4} } l_{ j, +1, \overline{\sigma} } l_{ i, +1, \sigma}
+ e^{\pm i \frac{3\pi}{4} } l_{ j, -1, \overline{\sigma} } l_{ i, -1, \sigma}
 \right)
, \\
\label{Operator B by l 2}
B_{i, j, \pm, \sigma, \overline{\sigma} }^\dagger
= \frac{1}{\sqrt{2}}
\left( 
e^{\pm i \frac{3\pi}{4} } l_{ i, +1, \sigma}^\dagger l_{ j, +1, \overline{\sigma} }^\dagger
+ e^{\mp i \frac{3\pi}{4} } l_{ i, -1, \sigma}^\dagger l_{ j, -1, \overline{\sigma} }^\dagger
 \right)
.
\end{align}
Here, we use annihilation operators $l_{ i, \pm1, \sigma }$ and creation operators $l_{ i, \pm1, \sigma }^\dagger$ for the one-electron eigenstates $\lambda_{\pm1} \left( \boldsymbol{r}_i \right) v_\sigma \left( \boldsymbol{r}_i \right)$ of Eq. (\ref{wave function lambda}), which obey the following anticommutation relation for fermions:
\begin{align}
\label{Commutation relation of l}
\left\{
l_{ i, m, \sigma }, l_{ j, m', \sigma'}^\dagger
\right\}
= \delta_{i, j} \delta_{m,m'} \delta_{\sigma, \sigma'}
\ \  \left( m, m' = \pm 1, \ \sigma, \sigma' = \uparrow \mathrm{or} \downarrow \right) 
.
\end{align}
These electron operators are written in terms of  $d_{ i, l, \sigma }$  and  $d_{ i, l, \sigma }^\dagger$  of Eqs. (\ref{Fourier transformation of d 1}) and (\ref{Fourier transformation of d 2}) as
\begin{align}
\label{Operator l}
l_{ i, \pm1, \sigma }
= - \frac{1}{\sqrt{2}} 
\left( i d_{ i, y'z, \sigma } \pm  d_{ i, zx', \sigma } \right)
.
\end{align}

By using the two-electron operators of $B_{i, j, \pm, \sigma, \overline{\sigma} }$ of Eq. (\ref{Operator B by l 1}) and $B_{i, j, \pm, \sigma, \overline{\sigma} }^\dagger$ of Eq. (\ref{Operator B by l 2}), we rewrite the hexadecapole-rotation interaction $H_\mathrm{rot} \bigl( \omega_{xy} \bigr)$ of Eq. (\ref{Matrix of Hrot diagonal}) as
\begin{align}
\label{Hrot by Bij}
H_\mathrm{rot} \bigl( \omega_{xy} \bigr)
= -4 \left( 1 - \gamma \right) \sum_{i \leq j} J_\mathrm{Q}^{ij} H_{z, i, j}^\alpha \omega_{xy}
.
\end{align}
Here, we describe the hexadecapole operator $H_{z, i, j}^\alpha$ in terms of the difference between the occupation numbers
$N_{i, j, +, \sigma, \overline{\sigma} } = B_{i, j, +, \sigma, \overline{\sigma} }^\dagger B_{i, j, +, \sigma, \overline{\sigma} }$
of the two-electron state $\psi_+ \bigl( \boldsymbol{r}_i, \boldsymbol{r}_j \bigr)$ and
$N_{i, j, -, \sigma, \overline{\sigma} } = B_{i, j, -, \sigma, \overline{\sigma} }^\dagger B_{i, j, -, \sigma, \overline{\sigma} }$
of $\psi_- \bigl( \boldsymbol{r}_i, \boldsymbol{r}_j \bigr)$ as
\begin{align}
\label{Hexadecapole by  Bij}
H_{z, i, j}^\alpha
&= \frac{1}{2} \sum_{ \sigma \neq \overline{\sigma} }
\left(
B_{i, j, +, \sigma, \overline{\sigma} }^\dagger B_{i, j, +, \sigma, \overline{\sigma} }
-B_{i, j, -, \sigma, \overline{\sigma} }^\dagger B_{i, j, -, \sigma, \overline{\sigma} }
\right)
\nonumber \\
&= \frac{1}{2} \sum_{ \sigma \neq \overline{\sigma} }
\left( N_{i, j, +, \sigma, \overline{\sigma} } - N_{i, j, -, \sigma, \overline{\sigma} } \right)
.
\end{align}
The two-electron wave function $\psi_\pm \bigl( \boldsymbol{r}_i, \boldsymbol{r}_j \bigr)$ of Eq. (\ref{psi+- by y'z and zx'}), consisting of a linear combination of  $\psi_{B_2} \bigl( \boldsymbol{r}_i, \boldsymbol{r}_j \bigr)$ with the $B_2$ symmetry and $\psi_{B_1} \bigl( \boldsymbol{r}_i, \boldsymbol{r}_j \bigr)$ with the $B_1$ symmetry, possesses the hexadecapole with the $A_2$ symmetry of Eq. (\ref{Hexadecapole by  Bij}) in the diagonal elements of Eq. (\ref{Hrot by Bij}).
This is confirmed by the fact that the direct product of $B_1 \oplus B_2$ for the two-electron wave functions $\psi_\pm \bigl( \boldsymbol{r}_i, \boldsymbol{r}_j \bigr)$ of Eq. (\ref{psi+- by y'z and zx'}) is reduced as $(B_1 \oplus B_2) \otimes (B_1 \oplus B_2) = 2A_1 \oplus 2A_2$ for the point group of the $D_{2d}$ symmetry of the Fe$^{2+}$ ion site.

Fourier transforms of the two-electron operators of $B_{i, j, \pm, \sigma, \overline{\sigma} }$ of Eq. (\ref{Operator B by l 1}) and $ B_{i, j, \pm, \sigma, \overline{\sigma} }^\dagger$ of Eq. (\ref{Operator B by l 2}) give their momentum representations.
For example, the Fourier transform of $B_{i, j, +, \sigma, \overline{\sigma} }$ is
\begin{align}
\label{Fourier transformation of B+}
&
B_{i, j, +, \sigma, \overline{\sigma} }
= \frac{1} {\sqrt{2} } \sum_{\boldsymbol{k}_\mathrm{G}, \boldsymbol{k}_\mathrm{R} }
\left(
e^{- i \frac{3\pi}{4} } l_{\frac{1}{2} \boldsymbol{k}_\mathrm{G} - \boldsymbol{k}_\mathrm{R}, +1, \overline{\sigma} }
l_{\frac{1}{2} \boldsymbol{k}_\mathrm{G} + \boldsymbol{k}_\mathrm{R}, +1, \sigma } 
\right.
\nonumber \\
&\left. \qquad \qquad \qquad \qquad 
+
e^{i \frac{3\pi}{4} } l_{\frac{1}{2} \boldsymbol{k}_\mathrm{G} - \boldsymbol{k}_\mathrm{R}, -1, \overline{\sigma} }
l_{\frac{1}{2} \boldsymbol{k}_\mathrm{G} + \boldsymbol{k}_\mathrm{R}, -1, \sigma } 
\right)
\nonumber \\
& \qquad  \qquad  \qquad \qquad \qquad  \qquad \times
e^{ i \boldsymbol{k}_\mathrm{G} \cdot \boldsymbol{r}_\mathrm{G} }
e^{ i \boldsymbol{k}_\mathrm{R} \cdot \boldsymbol{r}_\mathrm{R} }
.
\end{align}
Here, we use the gravity position $\boldsymbol{r}_\mathrm{G} = \bigl( \boldsymbol{r}_i + \boldsymbol{r}_j \bigr)/2$ and gravity momentum $\hbar \boldsymbol{k}_\mathrm{G} = \hbar \boldsymbol{k}_i + \hbar \boldsymbol{k}_j$.
It is supposed that the two-electron states are described in terms of the relative coordinate $\boldsymbol{r}_\mathrm{R} = \boldsymbol{r}_i - \boldsymbol{r}_j$ and relative momentum $\hbar \boldsymbol{k}_\mathrm{R} = \bigl( \hbar \boldsymbol{k}_i - \hbar \boldsymbol{k}_j \bigr)/2$ under the constraint that the gravity momentum vanishes as $\hbar \boldsymbol{k}_\mathrm{G} = \hbar \boldsymbol{k}_i + \hbar \boldsymbol{k}_j = 0$.
Under this constraint, the motion of the two-electron state is explained only in terms of bound states with the relative momentum $\hbar \boldsymbol{k}_\mathrm{R} = \hbar \boldsymbol{k}$.
Consequently, the annihilation operators of  $B_{\boldsymbol{k}, \pm, \sigma, \overline{\sigma} }$ for the eigenstates of the hexadecapole with the right-hand rotation of $h = +$ and the left-hand rotation of $h = -$ and the creation operators of $B_{\boldsymbol{k}, \pm, \sigma, \overline{\sigma} }^\dagger$ for the conjugate eigenstates are written as
\begin{align}
\label{Fourier transformation of  B 1}
B_{\boldsymbol{k}, \pm, \sigma, \overline{\sigma} }
= \frac{1} {\sqrt{2} }
\left(
e^{\mp i \frac{3\pi}{4} } l_{-\boldsymbol{k}, +1, \overline{\sigma} }  l_{\boldsymbol{k}, +1, \sigma} 
+ e^{\pm i \frac{3\pi}{4} }  l_{-\boldsymbol{k}, -1, \overline{\sigma} }  l_{\boldsymbol{k}, -1, \sigma} 
\right)
, \\
\label{Fourier transformation of  B 2}
B_{\boldsymbol{k}, \pm, \sigma, \overline{\sigma} }^\dagger
= \frac{1} {\sqrt{2} }
\left(
e^{\pm i \frac{3\pi}{4} } l_{\boldsymbol{k}, +1, \sigma}^\dagger l_{-\boldsymbol{k}, +1, \overline{\sigma} }^\dagger
+ e^{\mp i \frac{3\pi}{4} } l_{\boldsymbol{k}, -1, \sigma}^\dagger l_{-\boldsymbol{k}, -1, \overline{\sigma} }^\dagger
\right)
.
\end{align}
The two-electron operators of $B_{\boldsymbol{k}, \pm, \sigma, \overline{\sigma} }$ and $B_{\boldsymbol{k}, \pm, \sigma, \overline{\sigma} }^\dagger$ given by Eqs. (\ref{Fourier transformation of  B 1}) and (\ref{Fourier transformation of  B 2}) satisfy the mixed commutation relations \cite{QTS}
\begin{align}
\label{Commutation relation of B}
&\left[
B_{\boldsymbol{k}, \pm, \sigma, \overline{\sigma} }, B_{\boldsymbol{k}', \pm, \sigma, \overline{\sigma} }^\dagger
\right]
\nonumber \\
&= \left( 1 -
\frac{
n_{\boldsymbol{k}, +1, \sigma} + n_{-\boldsymbol{k}, +1, \overline{\sigma} }
+ n_{\boldsymbol{k}, -1, \sigma} + n_{-\boldsymbol{k}, -1, \overline{\sigma} }
}{2}
\right)
 \delta_{\boldsymbol{k}, \boldsymbol{k}' }
, \\
\label{Commutation relation of B 2}
&\left[
B_{\boldsymbol{k}, \pm, \sigma, \overline{\sigma} }, B_{\boldsymbol{k}', \mp, \sigma, \overline{\sigma} }^\dagger
\right]
\nonumber \\
&= \mp i 
\frac{
n_{\boldsymbol{k}, +1, \sigma} + n_{-\boldsymbol{k}, +1, \overline{\sigma} }
- n_{\boldsymbol{k}, -1, \sigma} - n_{-\boldsymbol{k}, -1, \overline{\sigma} }
}{2}
 \delta_{\boldsymbol{k}, \boldsymbol{k}' }
.
\end{align}
Here, we denote the one-electron number operator as $n_{\boldsymbol{k}, \pm1, \sigma} = l_{\boldsymbol{k}, \pm1, \sigma}^\dagger l_{\boldsymbol{k}, \pm1, \sigma}$.
The mixed commutation relations of Eqs. (\ref{Commutation relation of B}) and (\ref{Commutation relation of B 2}) are required to calculate the hexadecapole interaction in the normal phase and describe the hexadecapole ordering associated with the superconducting transition at $T_\mathrm{SC} = 23$ K for $x = 0.071$.

\subsection{Hexadecapole interaction}
 
In actual crystals of the iron pnictide, the thermally excited transverse acoustic phonon with wavevector $\boldsymbol{q}$ induces the rotation $\omega_{xy} \left( \boldsymbol{q} \right)$ as
\begin{align}
\label{Rotation 2nd quant.}
\omega_{xy} \left( \boldsymbol{q} \right) 
&= \frac{i}{2} \sqrt{
\mathstrut \frac{\hbar} {2V \rho_\mathrm{M} \omega_{y} \left( \boldsymbol{q} \right) } 
} q_{x}
\left( a_{y, \boldsymbol{q}} - a_{y, -\boldsymbol{q}}^{\dagger} \right)
\nonumber \\
& \qquad - \frac{i}{2} \sqrt{
\mathstrut \frac{\hbar} {2V\rho_\mathrm{M} \omega_{x} \left( \boldsymbol{q} \right)} 
} q_{y}
\left( a_{x, \boldsymbol{q}} - a_{x, -\boldsymbol{q}}^{\dagger} \right)
.
\end{align}
The two-electron states  $\psi_\pm \bigl( \boldsymbol{r}_i, \boldsymbol{r}_j \bigr)$ of Eq. (\ref{psi+-}) bearing the hexadecapole are scattered by the transverse acoustic phonons carrying the rotation $\omega_{xy} \left( \boldsymbol{q} \right)$.
These scattering are expressed in terms of the hexadecapole-rotation interaction in momentum space as 
\begin{align}
\label{Hrot by Bk}
&H_\mathrm{rot} \bigl( \omega_{xy} \bigr)
\nonumber \\
& \qquad
= -2 \bigl( 1 - \gamma \bigr)
\sum_{\boldsymbol{k}, \boldsymbol{q}} \sum_{\sigma, \neq \overline{\sigma}}
J_\mathrm{Q} \bigl( \boldsymbol{k}, \boldsymbol{q} \bigr)
\nonumber \\
&\qquad \qquad 
\times \left(
B_{\boldsymbol{k} + \boldsymbol{q}, +, \sigma, \overline{\sigma} }^\dagger B_{\boldsymbol{k}, +, \sigma, \overline{\sigma} }
- B_{\boldsymbol{k} + \boldsymbol{q}, -, \sigma, \overline{\sigma} }^\dagger B_{\boldsymbol{k}, -, \sigma, \overline{\sigma} }
\right) \omega_{xy} \bigl( \boldsymbol{q} \bigr)
\nonumber \\
&\qquad
= -4 \bigl( 1 - \gamma \bigr)
\sum_{\boldsymbol{k}, \boldsymbol{q}}
J_\mathrm{Q} \bigl( \boldsymbol{k}, \boldsymbol{q} \bigr)
H_{z,  \boldsymbol{k}, \boldsymbol{q} }^\alpha \omega_{xy} \bigl( \boldsymbol{q} \bigr)
.
\end{align}
The hexadecapole $H_{z,  \boldsymbol{k}, \boldsymbol{q} }^\alpha$ in Eq. (\ref{Hrot by Bk}) is written in terms of the two-electron state with wavevector $\boldsymbol{k}$ involving virtually excited one-phonon states with wavevector $\boldsymbol{q}$  as
\begin{align}
\label{Hexadecapole by Bk}
H_{z,  \boldsymbol{k}, \boldsymbol{q} }^\alpha
&= \frac{1}{2} \sum_{\sigma \neq \overline{\sigma} } H_{z,  \boldsymbol{k}, \boldsymbol{q}, \sigma, \overline{\sigma} }^\alpha
\nonumber \\
&= \frac{1}{2} \sum_{\sigma \neq \overline{\sigma} }
\left(
B_{\boldsymbol{k} + \boldsymbol{q}, +, \sigma, \overline{\sigma} }^\dagger B_{\boldsymbol{k}, +, \sigma, \overline{\sigma} }
- B_{\boldsymbol{k} + \boldsymbol{q}, -, \sigma, \overline{\sigma} }^\dagger B_{\boldsymbol{k}, -, \sigma, \overline{\sigma} }
\right)
.
\end{align}
We identify the hexadecapole of Eq. (\ref{Hexadecapole by Bk}) with the difference between the occupation numbers
$N_{\boldsymbol{k}, +, \sigma, \overline{\sigma} } = B_{\boldsymbol{k}, +, \sigma, \overline{\sigma} }^\dagger B_{\boldsymbol{k}, +, \sigma, \overline{\sigma} }$
and 
$N_{\boldsymbol{k}, -, \sigma, \overline{\sigma} } = B_{\boldsymbol{k}, -, \sigma, \overline{\sigma} }^\dagger B_{\boldsymbol{k}, -, \sigma, \overline{\sigma} }$
for the two-electron states with the small-wavenumber limit of $| \boldsymbol{q}| \rightarrow 0$ as follows:
\begin{align}
\label{Hexadecapole as number operator}
H_{z,  \boldsymbol{k}, \boldsymbol{q} = 0 }^\alpha
&= \frac{1}{2} \sum_{\sigma \neq \overline{\sigma} } H_{z,  \boldsymbol{k}, \boldsymbol{q} = 0, \sigma, \overline{\sigma} }^\alpha 
\nonumber \\
&= \frac{1}{2} \sum_{\sigma \neq \overline{\sigma} }
\left(  N_{\boldsymbol{k}, +, \sigma, \overline{\sigma} } - N_{\boldsymbol{k}, -, \sigma, \overline{\sigma} } \right)
.
\end{align}

The canonical transformation involving the virtual one-phonon processes gives the indirect interactions between the two-electron states carrying the hexadecapole as \cite{QTS}
\begin{align}
\label{HindHH1}
H_\mathrm{ind}^\mathrm{HH}
= & - \frac{1}{4} \sum_{\boldsymbol{k}, \boldsymbol{k}', \boldsymbol{q} } \sum_{\sigma_1 \neq \overline{\sigma}_1} \sum_{\sigma_2 \neq \overline{\sigma}_2}
D_+^\mathrm{HH} \left( \boldsymbol{k}, \boldsymbol{q} \right)
\nonumber \\
& \qquad \times 
B_{\boldsymbol{k} - \boldsymbol{q}, +, \sigma_1, \overline{\sigma}_1}^\dagger B_{\boldsymbol{k}, +, \sigma_1, \overline{\sigma}_1}
B_{\boldsymbol{k}' + \boldsymbol{q}, +, \sigma_2, \overline{\sigma}_2}^\dagger B_{\boldsymbol{k}', +, \sigma_2, \overline{\sigma}_2}
\nonumber \\
&+ \frac{1}{4} \sum_{\boldsymbol{k}, \boldsymbol{k}', \boldsymbol{q} } \sum_{\sigma_1 \neq \overline{\sigma}_1} \sum_{\sigma_2 \neq \overline{\sigma}_2}
D_+^\mathrm{HH} \left( \boldsymbol{k}, \boldsymbol{q} \right)
\nonumber \\
& \qquad \times 
B_{\boldsymbol{k} - \boldsymbol{q}, +, \sigma_1, \overline{\sigma}_1}^\dagger B_{\boldsymbol{k}, +, \sigma_1, \overline{\sigma}_1}
B_{\boldsymbol{k}' + \boldsymbol{q}, -, \sigma_2, \overline{\sigma}_2}^\dagger B_{\boldsymbol{k}', -, \sigma_2, \overline{\sigma}_2}
\nonumber \\
&+ \frac{1}{4} \sum_{\boldsymbol{k}, \boldsymbol{k}', \boldsymbol{q} } \sum_{\sigma_1 \neq \overline{\sigma}_1} \sum_{\sigma_2 \neq \overline{\sigma}_2} 
D_-^\mathrm{HH} \left( \boldsymbol{k}, \boldsymbol{q} \right)
\nonumber \\
& \qquad \times 
B_{\boldsymbol{k} - \boldsymbol{q}, -, \sigma_1, \overline{\sigma}_1} B_{\boldsymbol{k}, -, \sigma_1, \overline{\sigma}_1}
B_{\boldsymbol{k}' + \boldsymbol{q}, +, \sigma_2, \overline{\sigma}_2}^\dagger B_{\boldsymbol{k}', +, \sigma_2, \overline{\sigma}_2}
\nonumber \\
&- \frac{1}{4} \sum_{\boldsymbol{k}, \boldsymbol{k}', \boldsymbol{q} } \sum_{\sigma_1 \neq \overline{\sigma}_1} \sum_{\sigma_2 \neq \overline{\sigma}_2}
D_-^\mathrm{HH} \left( \boldsymbol{k}, \boldsymbol{q} \right)
\nonumber \\
& \qquad \times 
B_{\boldsymbol{k} - \boldsymbol{q}, -, \sigma_1, \overline{\sigma}_1} B_{\boldsymbol{k}, -, \sigma_1, \overline{\sigma}_1}
B_{\boldsymbol{k}' + \boldsymbol{q}, -, \sigma_2, \overline{\sigma}_2}^\dagger B_{\boldsymbol{k}', -, \sigma_2, \overline{\sigma}_2}
.
\end{align}
The indirect hexadecapole interaction Hamiltonian of Eq. (\ref{HindHH1}) mediated by the rotation $\omega_{xy} \bigl( \boldsymbol{q} \bigr)$  of the transverse acoustic phonon consists of the four scattering processes shown in Fig. \ref{Fig6}.
The scattering process of the two-electron states between the same rotation direction of $h = +$ is shown in Fig. \ref{Fig6}(a) and the scattering between the same rotation direction of $h = -$ is in Fig. \ref{Fig6}(d).
The scattering between opposite rotation directions of $h = +$ and $-$ are shown in Figs. \ref{Fig6}(b) and \ref{Fig6}(c). 

The coupling coefficient $D_h^\mathrm{HH} \left( \boldsymbol{k}, \boldsymbol{q} \right)$ of the indirect hexadecapole interaction in Eq. (\ref{HindHH1}) for the right-hand rotation direction of $h = +$ and the left-hand rotation of $h = -$ is expressed as
\begin{align}
\label{DHHh}
&D_h^\mathrm{HH} \left( \boldsymbol{k}, \boldsymbol{q} \right)
\nonumber \\
&= - \frac{1}{2} 
\left[
-4 \left( 1 - \gamma \right) J_\mathrm{Q} \left( \boldsymbol{k}, \boldsymbol{q} \right)
\right]^2
\nonumber \\
& \times \left\{
\frac{\hbar} { 2V\rho_\mathrm{M}\omega_{y} \left( \boldsymbol{q} \right) } q_{x}^2 
\frac{ 4\hbar \omega_y \left( \boldsymbol{q} \right) }
{
\left[
\varepsilon_h \left( \boldsymbol{k} \right) -  \varepsilon_h \left( \boldsymbol{k} - \boldsymbol{q} \right) 
\right]^2 
- \hbar^2 \omega_y \left( \boldsymbol{q} \right) ^2
} \right.
\nonumber \\
&\left. \qquad
+\frac{\hbar} { 2V\rho_\mathrm{M} \omega_{x} \left( \boldsymbol{q} \right) } q_{y}^2 
\frac{ 4\hbar \omega_x \left( \boldsymbol{q} \right) }
{
\left[ \varepsilon_h \left( \boldsymbol{k} \right) -  \varepsilon_h \left( \boldsymbol{k} - \boldsymbol{q} \right) \right]^2 
- \hbar^2 \omega_x \left( \boldsymbol{q} \right)^2 
} \right\}
.
\end{align}
Here, $\varepsilon_h \left( \boldsymbol{k} \right) = \hbar^2 \left| \boldsymbol{k} \right| ^2/2m^\ast$ with effective mass $m^\ast$ for wavevector $\boldsymbol{k}$ is the excitation energy of the two-electron states with the rotation direction of $h = \pm$ of Eq. (\ref{psi+-}).
The transverse acoustic phonon energy of $\hbar \omega_i \left( \boldsymbol{q} \right) = \hbar v_{66} q_i$ for $i = x$ or $y$ is given by the ultrasonic velocity $v_{66}$. 

In the normal phase without long-range ordering, the energies of the two-electron states $\psi_+ \bigl( \boldsymbol{r}_i, \boldsymbol{r}_j \bigr)$ and $\psi_- \bigl( \boldsymbol{r}_i, \boldsymbol{r}_j \bigr)$ cause the degeneration of each other as $\varepsilon_+ \left( \boldsymbol{k} \right) = \varepsilon_- \left( \boldsymbol{k} \right) = \varepsilon \left( \boldsymbol{k} \right)$.
By adopting the equality $D_+^\mathrm{HH} \left( \boldsymbol{k}, \boldsymbol{q} \right) = D_-^\mathrm{HH} \left( \boldsymbol{k}, \boldsymbol{q} \right) = D^\mathrm{HH} \left( \boldsymbol{k}, \boldsymbol{q} \right)$, we deduce the indirect hexadecapole interaction between the two-electron states of Eq. (\ref{HindHH1}) as
\begin{align}
\label{HindHH}
H_\mathrm{ind}^\mathrm{HH}
=& - \frac{1}{4} \sum_{\boldsymbol{k}, \boldsymbol{k}', \boldsymbol{q} } \sum_{\sigma_1 \neq \overline{\sigma}_1} \sum_{\sigma_2 \neq \overline{\sigma}_2}
D^\mathrm{HH} \left( \boldsymbol{k}, \boldsymbol{q} \right)
\nonumber \\
& \times 
\left(
B_{\boldsymbol{k} - \boldsymbol{q}, +, \sigma_1, \overline{\sigma}_1}^\dagger B_{\boldsymbol{k}, +, \sigma_1, \overline{\sigma}_1}
- B_{\boldsymbol{k} - \boldsymbol{q}, -, \sigma_1, \overline{\sigma}_1}^\dagger B_{\boldsymbol{k}, -, \sigma_1, \overline{\sigma}_1}
\right)
\nonumber \\
& \qquad \times \left(
B_{\boldsymbol{k}' + \boldsymbol{q}, +, \sigma_2, \overline{\sigma}_2}^\dagger B_{\boldsymbol{k}', +, \sigma_2, \overline{\sigma}_2}
- B_{\boldsymbol{k}' + \boldsymbol{q}, -, \sigma_2, \overline{\sigma}_2}^\dagger B_{\boldsymbol{k}', -, \sigma_2, \overline{\sigma}_2}
\right)
\nonumber \\
=& - \sum_{\boldsymbol{k}, \boldsymbol{k}', \boldsymbol{q} }
D^\mathrm{HH} \left( \boldsymbol{k}, \boldsymbol{q} \right)
H_{z,  \boldsymbol{k}, -\boldsymbol{q} }^\alpha  H_{z,  \boldsymbol{k}', \boldsymbol{q} }^\alpha 
.
\end{align}
The indirect hexadecapole interaction coefficient $D^\mathrm{HH} \left( \boldsymbol{k}, \boldsymbol{q} \right)$ in Eq. (\ref{HindHH}) is written as
\begin{align}
\label{DHH}
D^\mathrm{HH} \left( \boldsymbol{k}, \boldsymbol{q} \right)
&= - 8 \left( 1 - \gamma \right)^2 J_\mathrm{Q} \left( \boldsymbol{k}, \boldsymbol{q} \right)^2
\nonumber \\
& \times \frac{\hbar} { 2V\rho_\mathrm{M}\omega \left( \boldsymbol{q} \right) } q^2 
\frac{ 8\hbar \omega \left( \boldsymbol{q} \right) }
{
\left[
\varepsilon \left( \boldsymbol{k} \right) -  \varepsilon \left( \boldsymbol{k} - \boldsymbol{q} \right) 
\right]^2 
- \hbar^2 \omega \left( \boldsymbol{q} \right) ^2
} .
\end{align}
Here, we take the equivalence of the acoustic phonon energy $\hbar \omega_x \left( \boldsymbol{q} \right) = \hbar \omega_y \left( \boldsymbol{q} \right) = \hbar \omega \left( \boldsymbol{q} \right)$ for the tetragonal lattice.
The sign of the interaction coefficient $D^\mathrm{HH} \left( \boldsymbol{k}, \boldsymbol{q} \right)$ of Eq. (\ref{DHH}) dominates the character of the hexadecapole interaction.
The positive-sign regime of $D^\mathrm{HH} \left( \boldsymbol{k}, \boldsymbol{q} \right) > 0$ corresponds to the ferro-type hexadecapole interaction, while the negative-sign regime of $D^\mathrm{HH} \left( \boldsymbol{k}, \boldsymbol{q} \right) < 0$ corresponds to the antiferro-type hexadecapole interaction.
Note that the coefficient $ \left( 1 - \gamma \right)^2 J_\mathrm{Q} \left( \boldsymbol{k}, \boldsymbol{q} \right)^2$ in the indirect hexadecapole interaction of Eqs. (\ref{DHHh}) and (\ref{DHH}) is always positive for both the antiferro-type quadrupole interaction of $J_\mathrm{Q} \left( \boldsymbol{k}, \boldsymbol{q} \right) < 0$ and the ferro-type quadrupole interaction of $J_\mathrm{Q} \left( \boldsymbol{k}, \boldsymbol{q} \right) > 0$.
Consequently, the sign of $D^\mathrm{HH} \left( \boldsymbol{k}, \boldsymbol{q} \right)$ is determined by the sign of the denominator $\left[ \varepsilon \left( \boldsymbol{k} \right) -  \varepsilon \left( \boldsymbol{k} - \boldsymbol{q} \right) \right]^2 - \hbar^2 \omega \left( \boldsymbol{q} \right) ^2$ in Eq. (\ref{DHH}). 
Since the two-electron state of Eq. (\ref{psi+- by y'z and zx'}) with the spin-singlet state interacts with the rotation of the transverse acoustic phonons, the hexadecapole-rotation interaction of Eq. (\ref{Hrot by Bk}) and the hexadecapole interaction of Eq. (\ref{HindHH}) are independent of whether the spin orientation is $\sigma = \uparrow$ or $\downarrow$.

In the small-wavenumber regime of $\left| \boldsymbol{q} \right| \rightarrow 0$ for the transverse acoustic phonons, we rewrite the indirect hexadecapole interaction coefficient $D^\mathrm{HH} \left( \boldsymbol{k}, \boldsymbol{q} \right)$ in  Eq. (\ref{DHH}) as follows by disregarding high powers of $O \bigl( \left| \boldsymbol{q} \right| ^4 \bigr)$:
\begin{align}
\label{DHH (q = 0)}
&D^\mathrm{HH} \left( \boldsymbol{k}, \boldsymbol{q} = 0 \right)
\nonumber \\
& \qquad = - \frac{ 32 \left( 1 - \gamma \right)^2 J_\mathrm{Q} \left( \boldsymbol{k}, 0 \right)^2} {V \rho_\mathrm{M} }
\frac{1}
{
\left( \displaystyle \frac{\hbar} {m^\ast} \right)^2 \left( k^2 - \displaystyle \frac{m^{\ast 2} v_{66}^2} {\hbar^2} \right)
}.
\end{align}
This gives the boundary wavenumber $k_\mathrm{b}^\mathrm{H} = \left| \boldsymbol{k}_\mathrm{b}^\mathrm{H} \right| = m^\ast v_{66} / \hbar$, where the sign of $D^\mathrm{HH} \left( \boldsymbol{k}, \boldsymbol{q}= 0 \right)$ changes from positive to negative with increasing $\left| \boldsymbol{k} \right|$.
The experimental results in Fig. \ref{Fig1}(a) for $x = 0.071$ give the ultrasonic velocity $v_{66} = 1750$ m/s in the vicinity of the superconducting transition point $T_\mathrm{SC} = 23$ K ($v_{66} = 1970$ m/s at $T = 80$ K in the normal phase).
From these results, we estimate the excitation energy of
 $\varepsilon \bigl( k_\mathrm{b}^\mathrm{H} \bigr) = \hbar^2 k_\mathrm{b}^{\mathrm{H}2} / 2m^\ast$ to be $0.201$ K (0.256 K)
for the boundary wavenumber
$k_\mathrm{b}^\mathrm{H} = 34.0\times 10^6$ m$^{-1}$ ($30.2 \times 10^6$ m$^{-1}$)
and boundary wavelength
$\lambda_\mathrm{b}^\mathrm{H} = 0.185 \times 10^{-6}$ m ($0.208 \times 10^{-6}$ m).
Here, we suppose that the two-electron states have an energy of $\varepsilon \left( \boldsymbol{k} \right) = \hbar^2 \left| \boldsymbol{k} \right| ^2/2m^\ast$ with the effective mass being twice the rest electron mass $m^\ast = 2m_e$.

The low-lying two-electron state with the excitation energy $\varepsilon_h \left( \boldsymbol{k} \right) < \varepsilon \bigl( k_\mathrm{b}^\mathrm{H} \bigr) \sim 0.25$ K (5 GHz) contributes to the ferro-type hexadecapole interaction with the positive $D^\mathrm{HH} \left( \boldsymbol{k}, \boldsymbol{q} = 0 \right) > 0$.
This brings about the ferro-hexadecapole ordering, which is actually confirmed by the critical slowing down around the superconducting transition temperature for $x = 0.071$ in Fig. \ref{Fig1}(b).
The hexadecapole interaction of $D^\mathrm{HH} \left( \boldsymbol{k}, \boldsymbol{q}  = 0 \right)$ proportional to $(1 - \gamma)^2J_\mathrm{Q} \left( \boldsymbol{k}, \boldsymbol{q}  = 0 \right)^2$ in Eqs. (\ref{DHH}) and (\ref{DHH (q = 0)}) determines the ferro-type hexadecapole transition temperature, as will be shown by Eq. (\ref{DHH written by DeltaH}) of Sect. 4.8.

\subsection{Hexadecapole ordering}

We use the hexadecapole susceptibility to describe the critical slowing down associated with the ferro-type hexadecapole ordering.
According to the matrix representation of the hexadecapole-rotation interaction of Eq. (\ref{Matrix of Hrot diagonal}), the right-hand rotation of $\omega_{xy} > 0$ splits the two-electron states of Eq. (\ref{psi+-}) into a lower level with $E_+ = -4 \left( 1 - \gamma \right)J_\mathrm{Q}\omega_{xy}$ for $\psi_+ \bigl( \boldsymbol{r}_i, \boldsymbol{r}_j \bigr)$ with the right-hand rotation direction $h = +$ and an upper level with $E_- = 4 \left( 1 - \gamma \right)J_\mathrm{Q}\omega_{xy}$ for $\psi_- \bigl( \boldsymbol{r}_i, \boldsymbol{r}_j \bigr)$ with the left-hand rotation direction $h = -$.
The splitting of the two-electron states of $\psi_+ \bigl( \boldsymbol{r}_i, \boldsymbol{r}_j \bigr)$ and $\psi_- \bigl( \boldsymbol{r}_i, \boldsymbol{r}_j \bigr)$ by the rotation $\omega_{xy}$ with the symmetry-breaking character illustrated in Fig. \ref{Fig7}(b) gives the hexadecapole susceptibility of $\chi_\mathrm{H}$ for the two-electron state obeying the Curie law as
\begin{align}
\label{chiH}
\chi_\mathrm{H} 
= N_\mathrm{H}
\frac{
16 \left( 1 - \gamma \right)^2
J_\mathrm{Q}^2 / C_{66}^0 
} {T}
=\frac{  \mathit{\Delta}_\mathrm{H} }{T}
.
\end{align}
Here, $N_\mathrm{H}$ is the number of two-electron states participating in the ferro-type hexadecapole interaction.
The indirect hexadecapole-rotation interaction energy $\mathit{\Delta}_\mathrm{H}$ in Eq. (\ref{chiH}) is given by the hexadecapole interaction coefficient $D^\mathrm{HH} \left( \boldsymbol{k}, \boldsymbol{q}  = 0 \right)$ of Eq. (\ref{DHH (q = 0)}) as
\begin{align}
\label{DHH written by DeltaH}
\frac{2 \mathit{\Delta}_\mathrm{H} } {VN_\mathrm{H} } \frac{ C_{66}^0 }{ \rho_\mathrm{M} v_{66}^2 }
&= \frac{32 \left( 1 - \gamma \right)^2 J_\mathrm{Q}^2 } {V \rho_\mathrm{M} v_{66}^2 }
\nonumber \\
&=  \frac{1}{N_\mathrm{H} } \sum_{ |\boldsymbol{k}| < k_\mathrm{b}^\mathrm{H} }
D^\mathrm{HH} \left( \boldsymbol{k}, \boldsymbol{q} = 0 \right)
\nonumber \\
&=\widetilde{D}^\mathrm{HH}
.
\end{align}
The positive sign in the indirect hexadecapole interaction coefficient $\widetilde{D}^\mathrm{HH}$ due to $\mathit{\Delta}_\mathrm{H} > 0$ leads to the ferro-type hexadecapole ordering.
Note that the softening of $\rho_\mathrm{M} v_{66}^2 = C_{66}$ due to the quadrupole-strain interaction of Eq. (\ref{HQS 2nd quant.}) enhances the indirect hexadecapole interaction coefficient $\widetilde{D}^\mathrm{HH}$ in Eq. (\ref{DHH written by DeltaH}). 

The relaxation time $\tau$ around the superconducting transition for $x = 0.071$ diverges at the critical temperature $T_\mathrm{c}^0$ as shown in Fig. \ref{Fig4}.
Taking this experimental finding into account, we introduce the renormalized hexadecapole susceptibility $\widetilde{\chi}_\mathrm{H}$ as
\begin{align}
\label{chiH tilde}
\widetilde{\chi}_\mathrm{H} 
= \frac{ \mathit{\Delta}_\mathrm{H} } { T-\mathit{\Theta}_\mathrm{C} }
.
\end{align}
Here, we introduce the critical temperature $\mathit{\Theta}_\mathrm{C}$ corresponding to the experimentally observed critical temperature $T_\mathrm{c}^0$ for $\tau_\mathrm{H}$.
In the following Sects. 4.9 and 4.10, we present plausible model, where the hexadecapole ordering appears accompanying the superconductivity.
Consequently, the critical temperature $\mathit{\Theta}_\mathrm{C}$ of Eq. (\ref{chiH tilde}) consists of the indirect hexadecapole interaction energy $\widetilde{D}^\mathrm{HH}$ of Eq. (\ref{DHH written by DeltaH}) and the superconducting transition temperature $T_\mathrm{SC}$ as
\begin{align}
\label{ThetaC}
\mathit{\Theta}_\mathrm{C}
= \widetilde{D}^\mathrm{HH} + T_\mathrm{SC}
.
\end{align}
Note that the superconducting transition temperature $T_\mathrm{SC}$ will be given later by the self-consistent equation for the superconducting energy gap of Eq. (\ref{Gap eq}).
The internal energy based on the phenomenological theory given by Eq. (\ref{UH}) in Sect. 4.10 also indicates the ground state for the simultaneous ordering of the hexadecapole and superconductivity in the present iron pnictide.

The critical slowing down of the relaxation time $\tau_H$ is caused by the divergence of the correlation length associated with the ferro-type ordering of the hexadecapole $H_z^\alpha \bigl( \boldsymbol{r}_i, \boldsymbol{r}_j \bigr)$.
This is expressed in terms of the renormalized hexadecapole susceptibility of Eq. (\ref{chiH tilde}) as
\begin{align}
\label{tauH}
\tau_\mathrm{H}
= \tau_0 \left| \frac{ T - \mathit{\Theta}_\mathrm{C} } { \mathit{\Theta}_\mathrm{C} } \right| ^{-z\nu}
= \tau_0 \left| \frac{ T - \mathit{\Theta}_\mathrm{C} } { \mathit{\Theta}_\mathrm{C} } \right| ^{-1}
\propto  \widetilde{\chi}_\mathrm{H}
.
\end{align}
The critical index $z\nu = 1$  of Eq. (\ref{tauH}) based on mean field theory well reproduces the experimental results of the relaxation time $\tau$ above the superconducting transition temperature $T_\mathrm{SC}$ in Fig. \ref{Fig1}(b).
Because the indirect hexadecapole interaction of Eq. (\ref{HindHH1}) is mediated by the rotation of the transverse acoustic phonons with a long wavelength, the critical phenomena above the transition point is well described by mean field theory. 
The experimentally observed $z\nu = 1/3$ in the superconducting phase below $T_\mathrm{SC}$ in Fig. \ref{Fig1}(b), however,  distinctly deviates from $z\nu = 1$ of Eq. (\ref{tauH}) obtained from mean field theory.  
This discrepancy is accounted for by the fact that the hexadecapole correlation due to the two-electron states develops in both the normal and superconducting phases near the transition temperature $T_\mathrm{SC}$, while the hexadecapole correlation due to the Cooper pairs develops only in the superconducting phase as will be shown in Sect. 4.10.
The diffusion processes of the hexadecapole $H_z^\alpha \bigl( \boldsymbol{r}_i, \boldsymbol{r}_j \bigr)$ in the vicinity of the critical temperature $\mathit{\Theta}_\mathrm{C}$ determine the attempt relaxation time $\tau_0$ in Eq.  (\ref{tauH}) \cite{Mori, Halperin and Hohenberg}.
The calculation based on renormalization group theory for the inherent system exhibiting the hexadecapole ordering associated with the superconductivity may explains the experimental result of the critical index $z\nu = 1/3$ and the relative ratio $\tau_{0+} / \tau_{0-} = 1.03$ for $x = 0.071$.

\begin{figure}[t]
\begin{center}
\includegraphics[angle=0,width=0.48\textwidth, bb=0 10 412 395]{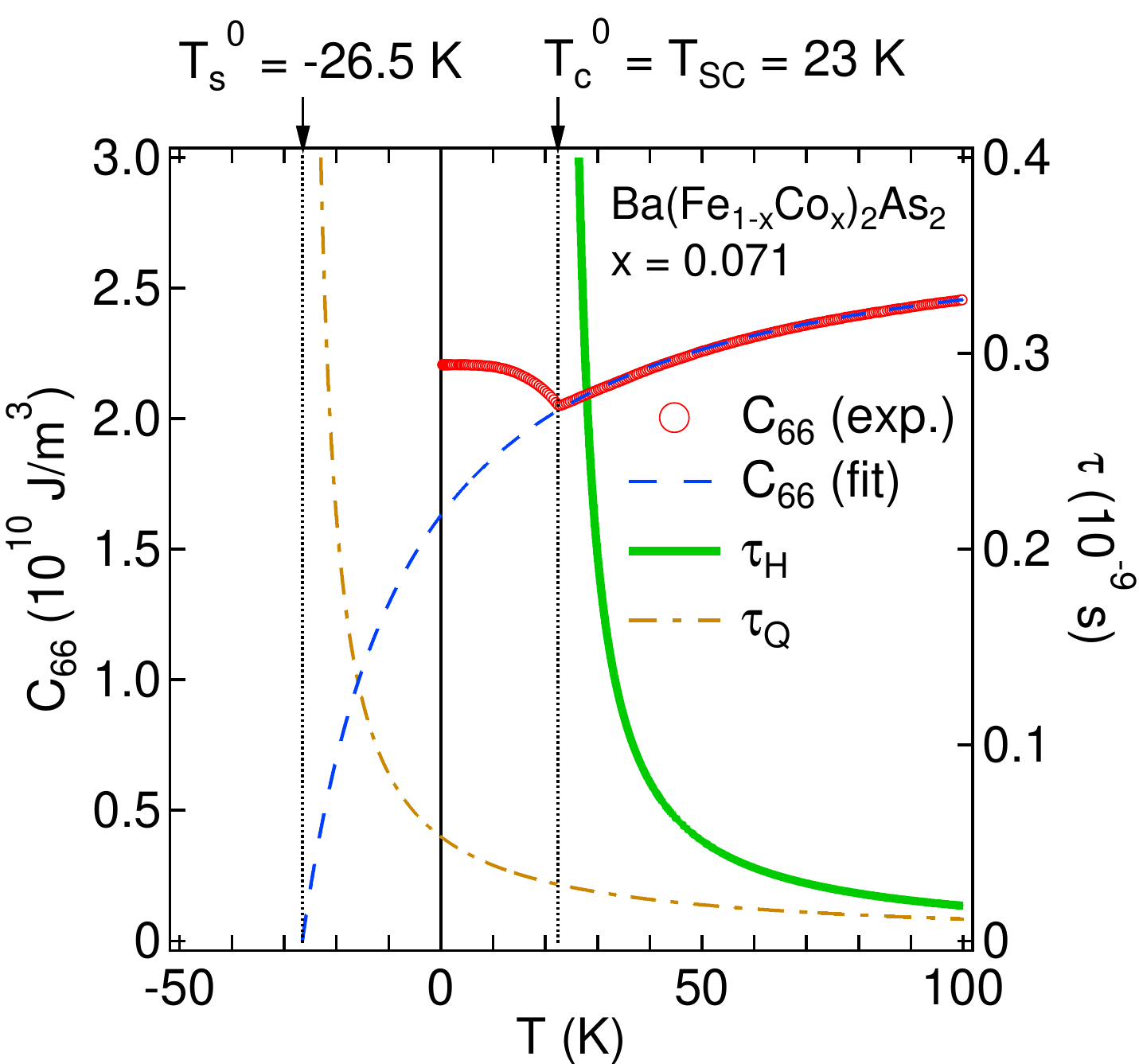}
\end{center}
\caption{
(Color online)
Divergence of the relaxation time $\tau_\mathrm{H}$ (green solid line) fit by Eq. (\ref{tauH}) and the softening of the elastic constant $C_{66}$ (blue dashed line) fit by Eq. (\ref{Elastic with chiQ tilde}) in the normal phase above the superconducting transition at $T_\mathrm{SC} = 23$ K for $x = 0.071$.
The temperature $T_\mathrm{c}^0 = 23$ K for the divergence of $\tau_\mathrm{H}$ indicates the hexadecapole ordering point of $\mathit{\Theta}_\mathrm{C}$, while the critical temperature $T_\mathrm{s}^0 = \mathit{\Theta}_\mathrm{Q} + \mathit{\Delta}_\mathrm{Q} = -26.5$ K of $C_{66}$ (red open circles) and $\tau_\mathrm{Q}$ (orange dashed and dotted line) give the fictitious lattice instability point for $C_{66} \rightarrow 0$.
}
\label{Fig8}
\end{figure}

It is worth presenting the hexadecapole susceptibility responsible for the attenuation coefficient $\alpha_{66}$ by comparing it with the quadrupole susceptibility responsible for the elastic constant $C_{66}$ in Fig. \ref{Fig8}.
The critical slowing down due to the ferro-type hexadecapole ordering brings about the divergence of the relaxation time $\tau_\mathrm{H}$, which is expressed in terms of the renormalized hexadecapole susceptibility $\widetilde{\chi}_\mathrm{H}$ of Eq. (\ref{chiH tilde}).
As shown by the solid green line in Fig. \ref{Fig8}, the divergence of the relaxation time $\tau_\mathrm{H}$ approaching the superconducting transition ensures the hexadecapole ordering at $\mathit{\Theta}_\mathrm{C} = T_\mathrm{c}^0 = T_\mathrm{SC} = 23$ K, consisting with the superconducting transition temperature.
On the other hand, the elastic constant $C_{66}$ shown by the red solid line in Fig. \ref{Fig8} exhibits softening obeying the renormalized quadrupole susceptibility $\widetilde{\chi}_\mathrm{Q}$ in Eq. (\ref{chiQ tilde}).
The critical temperature $T_\mathrm{s}^0$ is $\mathit{\Theta}_\mathrm{Q} + \mathit{\Delta}_\mathrm{Q} = -26.5$ K, implying that the fictitious lattice instability temperature where $C_{66} \rightarrow 0$ is strictly distinguished from the critical temperature $T_\mathrm{c}^0 = T_\mathrm{SC} = 23$ K for the divergence of the relaxation time $\tau_\mathrm{H}$.
 The critical slowing down due to the fictitious lattice instability may be possible, as indicated by the orange dashed and dotted line in Fig. \ref{Fig8}.
The ultrasonic attenuation due to the relaxation time $\tau_\mathrm{Q}$ at temperatures far above the fictitious instability point, however, is too small to detect.

The softening of $C_{66}$ in Fig. \ref{Fig8} is expressed in terms of the renormalized quadrupole susceptibility $\widetilde{\chi}_\mathrm{Q}$ in Eq. (\ref{Elastic with chiQ tilde}), which is caused by the quadrupole-strain interaction of Eq. (\ref{HQS 2nd quant.}).
The hexadecapole-rotation interaction of Eq. (\ref{Matrix of Hrot diagonal}) might affect the softening of $C_{66}^\mathrm{H}$ as
\begin{align}
\label{C66H}
C_{66}^\mathrm{H}
= C_{66} \left( 1- \widetilde{\chi}_\mathrm{H} \right)
= C_{66} \left( 1-  \frac{ \mathit{\Delta}_\mathrm{H} } { T-\mathit{\Theta}_\mathrm{C} } \right)
.
\end{align}
Here, $C_{66}$ is given by Eq. (\ref{Temp. dep. C66}) or Eq. (\ref{Elastic with chiQ tilde}) and represents the softening due to the quadrupole-strain interaction.
Because two electrons are accommodated in the degenerate $y'z$ and $zx'$ bands, we take the number of two-electron states as $N_\mathrm{H} \sim \left( 2N_\mathrm{Fe} \right) / 2 = 2\times10^{22}$ cm$^{-3}$ as an upper limit.
Adopting the quadrupole interaction energy $J_\mathrm{Q} = \mathit{\Theta}_\mathrm{Q} =  -47$ K for $x = 0.071$ and the anisotropic parameter $\gamma = 0.9$ as a tentative value, we deduce that the hexadecapole-rotation interaction energy  in Eq. (\ref{chiH}) is as small as
$\mathit{\Delta}_\mathrm{H}
= 16N_\mathrm{H} \left( 1 - \gamma \right)^2 J_\mathrm{Q}^2 / C_{66}^0 \sim 10^{-3}$ K.
This is too small to sizably affect the softening of the elastic constant $C_{66}$.
This small $\mathit{\Delta}_\mathrm{H}$ of $\sim 10^{-3}$ K is in strongly contrast to the considerable quadrupole-strain interaction energy of $\mathit{\Delta}_\mathrm{Q} \sim 20$ K, which brings about appreciable softening of $C_{66}$ with decreasing temperature.

\subsection{Superconductivity due to quadrupole interaction}

In out attempt to show the superconducting state compatible with the hexadecapole ordering, we will solve the superconducting Hamiltonian for a pair of electrons coupled to each other through the quadrupole interaction $H_\mathrm{QQ} \left( \gamma \right)$ of Eq. (\ref{Anisotropic HQQ}).
To this end, we treat band electrons of the orbital state $\lambda_m \left( \boldsymbol{r} \right)$ in Eq. (\ref{wave function lambda}) with the angular momentum $l = 2$ and the azimuthal quantum number $m = \pm1$.
The corresponding bare electron Hamiltonian is expressed in terms of the electron operators of $l_{\boldsymbol{k}, m, \sigma }$ and $l_{\boldsymbol{k}, m, \sigma }^\dagger$ for $m = \pm1$ as
\begin{align}
\label{HK'}
H_\mathrm{K}'
& = \sum_{\boldsymbol{k} } \sum_{\sigma}
\left[
\varepsilon_{+1, \sigma} \left( \boldsymbol{k} \right)
l_{\boldsymbol{k}, +1, \sigma}^\dagger l_{\boldsymbol{k}, +1, \sigma} \right.
\nonumber \\
&\left. \qquad \qquad \qquad
+ \varepsilon_{-1, \sigma} \left( \boldsymbol{k} \right)
l_{\boldsymbol{k}, -1, \sigma}^\dagger  l_{\boldsymbol{k}, -1, \sigma}
\right]
.
\end{align}
Here, the electron energy $\varepsilon_{m, \sigma} \left( \boldsymbol{k} \right)$ is measured from the Fermi energy. 

The quadrupoles expressed by the electron operators of $d_{i, l, \sigma}$ and $d_{i, l, \sigma}^\dagger$ $( l = y'z $ and $zx')$ in Eqs. (\ref{Ox'2-y'2 2nd quant.}) and (\ref{Ox'y' 2nd quant.}) are rewritten in terms of the electron operators of $l_{\boldsymbol{k}, m, \sigma }$ and $l_{\boldsymbol{k}, m, \sigma }^\dagger$ with $m = \pm1$ as
\begin{align}
\label{quadrupole Ov' by l}
O_{x'^2-y'^2, \boldsymbol{k}, \boldsymbol{q} }
&= - \sum_{\sigma}
\left(
 l_{\boldsymbol{k} + \boldsymbol{q}, +1, \sigma}^\dagger l_{\boldsymbol{k}, -1, \sigma}
+ l_{\boldsymbol{k} + \boldsymbol{q}, -1, \sigma}^\dagger l_{\boldsymbol{k}, +1, \sigma}
\right)
, \\ 
\label{quadrupole Ox'y' by l}
O_{x'y', \boldsymbol{k}, \boldsymbol{q} }
&= i \sum_{\sigma}
\left(
l_{\boldsymbol{k} + \boldsymbol{q}, +1, \sigma}^\dagger l_{\boldsymbol{k}, -1, \sigma}
- l_{\boldsymbol{k} + \boldsymbol{q}, -1, \sigma}^\dagger l_{\boldsymbol{k}, +1, \sigma}
\right)
.
\end{align}
Thus, we obtain an alternative expression for the quadrupole interaction Hamiltonian $H_\mathrm{QQ} \left( \gamma \right)$ with the anisotropic parameter $\gamma$  of Eq. (\ref{Anisotropic HQQ}) as
\begin{align}
\label{HQQ by l}
H_\mathrm{QQ} \left( \gamma \right)
=& -\frac{1}{2} \sum_{ \boldsymbol{k}, \boldsymbol{q} } \sum_{\sigma \neq \overline{\sigma} }
J_\mathrm{Q} \left( \boldsymbol{k}, \boldsymbol{q} \right)
\nonumber \\
& \quad \times
\left(
l_{\boldsymbol{k} + \boldsymbol{q}, +1, \sigma}^\dagger l_{-\boldsymbol{k} - \boldsymbol{q}, +1, \overline{\sigma} }^\dagger l_{-\boldsymbol{k}, -1, \overline{\sigma} } l_{\boldsymbol{k}, -1, \sigma } \right.
\nonumber \\
& \qquad + l_{\boldsymbol{k} + \boldsymbol{q}, +1, \sigma }^\dagger l_{-\boldsymbol{k} - \boldsymbol{q}, -1, \overline{\sigma} }^\dagger l_{-\boldsymbol{k}, +1, \overline{\sigma} } l_{\boldsymbol{k}, -1, \sigma} 
\nonumber \\ 
&\quad \qquad +l_{\boldsymbol{k} + \boldsymbol{q}, -1, \sigma}^\dagger l_{-\boldsymbol{k} - \boldsymbol{q}, +1, \overline{\sigma} }^\dagger l_{-\boldsymbol{k}, -1, \overline{\sigma} }  l_{\boldsymbol{k}, +1, \sigma}
\nonumber \\
&\qquad \qquad \left. 
+l_{\boldsymbol{k} + \boldsymbol{q}, -1, \sigma}^\dagger l_{-\boldsymbol{k} - \boldsymbol{q}, -1, \overline{\sigma} }^\dagger l_{-\boldsymbol{k}, +1, \overline{\sigma} } l_{\boldsymbol{k}, +1, \sigma}
\right)
\nonumber  \\
& -\frac{1}{2} \gamma \sum_{ \boldsymbol{k}, \boldsymbol{q} } \sum_{\sigma \neq \overline{\sigma} }
J_\mathrm{Q} \left( \boldsymbol{k}, \boldsymbol{q} \right)
\nonumber \\
& \quad \times 
\left(
- l_{\boldsymbol{k} + \boldsymbol{q}, +1, \sigma}^\dagger l_{-\boldsymbol{k} - \boldsymbol{q}, +1, \overline{\sigma} }^\dagger l_{-\boldsymbol{k}, -1, \overline{\sigma} } l_{\boldsymbol{k}, -1, \sigma} \right.
\nonumber \\
&\qquad + l_{\boldsymbol{k} + \boldsymbol{q}, +1, \sigma}^\dagger l_{-\boldsymbol{k} - \boldsymbol{q}, -1, \overline{\sigma} }^\dagger l_{-\boldsymbol{k}, +1, \overline{\sigma} } l_{\boldsymbol{k}, -1,\sigma}
\nonumber \\ 
&\quad \qquad  +l_{\boldsymbol{k} + \boldsymbol{q}, -1, \sigma}^\dagger l_{-\boldsymbol{k} - \boldsymbol{q}, +1, \overline{\sigma} }^\dagger l_{-\boldsymbol{k}, -1, \overline{\sigma} }  l_{\boldsymbol{k}, +1, \sigma}
\nonumber \\
&\qquad \qquad \left. 
- l_{\boldsymbol{k} + \boldsymbol{q}, -1, \sigma}^\dagger l_{-\boldsymbol{k} - \boldsymbol{q}, -1, \overline{\sigma} }^\dagger l_{-\boldsymbol{k}, +1, \overline{\sigma} } l_{\boldsymbol{k}, +1, \sigma}
\right)
.
\end{align}
The quadrupole interaction $H_\mathrm{QQ} \left( \gamma \right)$ of Eq. (\ref{HQQ by l}) gives four independent scattering processes between electrons with azimuthal quantum numbers of $m = +1$ and $-1$, wavevectors $\boldsymbol{k}$ of an electron bearing the electric quadrupoles and $\boldsymbol{q}$ of a transverse acoustic phonon, and spin orientations of $\sigma$ and $\overline{\sigma}$.
In order to account for the critical slowing down around the superconducting transition point, we notice that the two scattering processes consisting of 
$l_{\boldsymbol{k} + \boldsymbol{q}, +1, \sigma}^\dagger l_{-\boldsymbol{k} - \boldsymbol{q}, +1, \overline{\sigma} }^\dagger l_{-\boldsymbol{k}, -1, \overline{\sigma} } l_{\boldsymbol{k}, -1, \sigma} $
and
$l_{\boldsymbol{k} + \boldsymbol{q}, -1, \sigma}^\dagger l_{-\boldsymbol{k} - \boldsymbol{q}, -1, \overline{\sigma} }^\dagger l_{-\boldsymbol{k}, +1, \overline{\sigma} } l_{\boldsymbol{k}, +1, \sigma}$
in Eq. (\ref{HQQ by l}) give the superconducting ground state bearing the hexadecapole.
The former term of
$l_{\boldsymbol{k} + \boldsymbol{q}, +1, \sigma}^\dagger l_{-\boldsymbol{k} - \boldsymbol{q}, +1, \overline{\sigma} }^\dagger l_{-\boldsymbol{k}, -1, \overline{\sigma} } l_{\boldsymbol{k}, -1, \sigma} $
annihilates two electrons with the same azimuthal quantum number of $m = -1$ and creates two electrons with the same number of $m = +1$,
while the latter term of
$l_{\boldsymbol{k} + \boldsymbol{q}, -1, \sigma}^\dagger l_{-\boldsymbol{k} - \boldsymbol{q}, -1, \overline{\sigma} }^\dagger l_{-\boldsymbol{k}, +1, \overline{\sigma} } l_{\boldsymbol{k}, +1, \sigma}$
causes the reverse process.
On the other hand, we disregard the scattering of
$l_{\boldsymbol{k} + \boldsymbol{q}, -1, \sigma}^\dagger l_{-\boldsymbol{k} - \boldsymbol{q}, -1, \overline{\sigma} }^\dagger l_{-\boldsymbol{k}, +1, \overline{\sigma} } l_{\boldsymbol{k}, +1,\sigma}$
involved in exchange process from the electron state with $m = +1$ to the opposite state with $m = -1$, and that of
$l_{\boldsymbol{k} + \boldsymbol{q}, +1, \sigma}^\dagger l_{-\boldsymbol{k} - \boldsymbol{q}, +1, \overline{\sigma} }^\dagger l_{-\boldsymbol{k}, -1, \overline{\sigma} } l_{\boldsymbol{k}, -1,\sigma}$ for the exchange from $m = -1$ to $m = +1$.
These latter scattering lead to a superconducting ground state that does not carry the hexadecapole, which is incompatible with the critical slowing down due to the hexadecapole ordering.
Consequently, we adopt the restricted quadrupole interaction Hamiltonian consisting of the former process to properly describe the superconductivity accompanying the hexadecapole ordering:
\begin{align}
\label{Restricted HQQ}
&H_\mathrm{QQ}' \left( \gamma \right)
\nonumber \\
&\quad
= -\frac{1}{2} \left( 1 - \gamma \right) 
\sum_{ \boldsymbol{k}, \boldsymbol{q} } \sum_{\sigma \neq \overline{\sigma} }
J_\mathrm{Q} \left( \boldsymbol{k}, \boldsymbol{q} \right)
\nonumber \\
&\qquad \quad \times
\left(
l_{\boldsymbol{k} + \boldsymbol{q}, +1, \sigma}^\dagger l_{-\boldsymbol{k} - \boldsymbol{q}, +1, \overline{\sigma} }^\dagger l_{-\boldsymbol{k}, -1, \overline{\sigma} } l_{\boldsymbol{k}, -1, \sigma } \right.
\nonumber \\
&\left. \qquad \qquad \quad
+l_{\boldsymbol{k} + \boldsymbol{q}, -1, \sigma}^\dagger l_{-\boldsymbol{k} - \boldsymbol{q}, -1, \overline{\sigma} }^\dagger l_{-\boldsymbol{k}, +1, \overline{\sigma} } l_{\boldsymbol{k}, +1, \sigma}
\right)
.
\end{align}
Here, the gravity momentum for two-electron states bound by the restricted quadrupole interaction of Eq. (\ref{Restricted HQQ}) is constrained to vanish as $\hbar \boldsymbol{k}_\mathrm{G} = 0$.
Note that the restricted Hamiltonian $H_\mathrm{QQ}' \left( \gamma \right)$ of Eq. (\ref{Restricted HQQ}) satisfies the criterion of the $A_1$ symmetry of point group $D_{2d}$. 

We solve the superconducting Hamiltonian $H_\mathrm{SC}$ consisting of the bare electron Hamiltonian $H_\mathrm{K}'$ of Eq. (\ref{HK'}) and the restricted quadrupole interaction Hamiltonian $H_\mathrm{QQ}' \left( \gamma \right)$ of Eq. (\ref{Restricted HQQ}):
\begin{align}
\label{HSC}
&H_\mathrm{SC} 
\nonumber \\
&= H_\mathrm{K}' + H_\mathrm{QQ}' \left(\gamma \right)
\nonumber \\
&= \sum_{\boldsymbol{k} } \sum_{\sigma \neq \overline{\sigma} }
\left(
\begin{array}{c}
l_{\boldsymbol{k}, +1, \sigma}^\dagger \\
l_{-\boldsymbol{k}, +1, \overline{\sigma} } \\
l_{\boldsymbol{k}, -1, \sigma}^\dagger \\
l_{-\boldsymbol{k}, -1, \overline{\sigma} }
\end{array}
\right)^\mathrm{T} \times
\nonumber \\
&\left(
\begin{array}{cccc}
\varepsilon \left( \boldsymbol{k} \right) & - \mathit{\Delta}_{-1, -1}^{\sigma, \overline{\sigma} } \left( \boldsymbol{k} \right) & 0 & 0 \\
- \mathit{\Delta}_{-1, -1}^{\sigma, \overline{\sigma} \ast} \left( \boldsymbol{k} \right) & - \varepsilon \left( \boldsymbol{k} \right) & 0 & 0 \\
0 & 0 & \varepsilon \left( \boldsymbol{k} \right) & - \mathit{\Delta}_{+1, +1}^{\sigma, \overline{\sigma} } \left( \boldsymbol{k} \right) \\
0 & 0 & - \mathit{\Delta}_{+1, +1}^{\sigma, \overline{\sigma} \ast} \left( \boldsymbol{k} \right) & - \varepsilon \left( \boldsymbol{k} \right)  \\
\end{array}
\right)
\nonumber \\
&\qquad \qquad \qquad \qquad \qquad \qquad \times \left(
\begin{array}{c}
l_{\boldsymbol{k}, +1, \sigma} \\
l_{-\boldsymbol{k}, +1, \overline{\sigma} }^\dagger \\
l_{\boldsymbol{k}, -1, \sigma} \\
l_{-\boldsymbol{k}, -1, \overline{\sigma} }^\dagger 
\end{array}
\right)
.
\end{align}
Here, the electron state vector of
$\boldsymbol{l}_{\boldsymbol{k} } \equiv \left(
l_{\boldsymbol{k}, +1, \sigma},
l_{-\boldsymbol{k}, +1, \overline{\sigma} }^\dagger,
l_{\boldsymbol{k}, -1, \sigma},
l_{-\boldsymbol{k}, -1, \overline{\sigma} }^\dagger
\right)^\mathrm{T}$
is employed.
We use the equivalence of the energy
$\varepsilon_{+1, \sigma} \left( \boldsymbol{k} \right)
= \varepsilon_{-1, \sigma} \left( \boldsymbol{k} \right)
= \varepsilon \left( \boldsymbol{k} \right)$
for the electrons with the azimuthal quantum numbers of $m = +1$ and $-1$.
There are two different superconducting energy gaps of
$\mathit{\Delta}_{+1, +1}^{\sigma, \overline{\sigma} } \left( \boldsymbol{k} \right)$ denoted by the Cooper pairs with the right-hand azimuthal quantum number $m = +1$ and
$\mathit{\Delta}_{-1, -1}^{\sigma, \overline{\sigma} } \left( \boldsymbol{k} \right)$ for the left-hand azimuthal quantum number of $m = -1$ in Eq. (\ref{HSC}) as
\begin{align}
\label{Energy gap 1}
\mathit{\Delta}_{\pm 1, \pm 1}^{\sigma, \overline{\sigma} } \left( \boldsymbol{k} \right)
= \frac{1}{2} \left( 1-\gamma \right)
\sum_{\boldsymbol{q} }
J_\mathrm{Q} \left( \boldsymbol{k}, \boldsymbol{q} \right)
\bigl\langle
l_{-\boldsymbol{k} - \boldsymbol{q}, \pm1, \overline{\sigma} } l_{\boldsymbol{k} + \boldsymbol{q}, \pm1, \sigma }
\bigr\rangle
, \\
\label{Energy gap 2}
\mathit{\Delta}_{\pm 1, \pm 1}^{\sigma, \overline{\sigma} \ast} \left( \boldsymbol{k} \right)
= \frac{1}{2}  \left( 1-\gamma \right)
\sum_{\boldsymbol{q} }
J_\mathrm{Q} \left( \boldsymbol{k}, \boldsymbol{q} \right)
\bigl\langle
l_{\boldsymbol{k} + \boldsymbol{q}, \pm1, \sigma }^\dagger l_{-\boldsymbol{k} - \boldsymbol{q}, \pm1, \overline{\sigma} }^\dagger
\bigr\rangle 
.
\end{align}
Here, $\bigl\langle l_{-\boldsymbol{k}, \pm1, \overline{\sigma} } l_{\boldsymbol{k}, \pm1, \sigma } \bigr\rangle$
and
$\bigl\langle l_{\boldsymbol{k}, \pm1, \sigma }^\dagger l_{-\boldsymbol{k}, \pm1, \overline{\sigma} }^\dagger \bigr\rangle$
stand for the mean-field values of the Cooper pair indicating the off-diagonal long-range order parameter of the superconducting phase.
The quadrupole interaction coefficient $J_\mathrm{Q} \left( \boldsymbol{k}, \boldsymbol{q} \right)$ and the anisotropic parameter $\gamma$ in Eq. (\ref{Restricted HQQ}) dominate the two energy gaps of
$\mathit{\Delta}_{\pm1, \pm1}^{\sigma, \overline{\sigma} } \left( \boldsymbol{k} \right)$
in Eq. (\ref{Energy gap 1}) and
$\mathit{\Delta}_{\pm1, \pm1}^{\sigma, \overline{\sigma} \ast} \left( \boldsymbol{k} \right)$
in Eq. (\ref{Energy gap 2}), which characterize the inherent superconductivity of the system. 
The appearance of the energy gap $\mathit{\Delta}_{\pm 1, \pm 1}^{\sigma, \overline{\sigma} } \neq 0$ indicates the symmetry breaking of the U(1) gauge, where the electron number is not conserved across the superconducting transition \cite{P. W. Anderson}.

Supposing a Cooper pair consisting of two electrons with opposite spin orientations, we set the energy gap to 
$\mathit{\Delta}_{\pm1, \pm1}^{\uparrow, \downarrow} \left( \boldsymbol{k} \right)
= \mathit{\Delta}_{\pm1, \pm1}^{\downarrow, \uparrow} \left( \boldsymbol{k} \right)
= \mathit{\Delta}_{\pm1, \pm1} \left( \boldsymbol{k} \right) $
while omitting the spin orientations.
The superconducting Hamiltonian $H_\mathrm{SC}$ of Eq. (\ref{HSC}) is expressed by Bogoliubov quasiparticles of
$\boldsymbol{L}_{\boldsymbol{k} } \equiv \left(
L_{\boldsymbol{k}, +1, \sigma},
L_{-\boldsymbol{k}, +1, \overline{\sigma} }^\dagger,
L_{\boldsymbol{k}, -1, \sigma},
L_{-\boldsymbol{k}, -1, \overline{\sigma} }^\dagger
\right)^\mathrm{T}$ 
as
\begin{align}
\label{HSC diagonal}
&H_\mathrm{SC}
= \sum_{\boldsymbol{k} } \sum_{\sigma \neq \overline{\sigma} } 
 \left(
\begin{array}{c}
L_{\boldsymbol{k}, +1, \sigma}^\dagger \\
L_{-\boldsymbol{k}, +1, \overline{\sigma} } \\
L_{\boldsymbol{k}, -1, \sigma}^\dagger \\
L_{-\boldsymbol{k}, -1, \overline{\sigma} }
\end{array}
\right)^\mathrm{T}
\nonumber \\
&\times \left(
\begin{array}{cccc}
E_{+1} \left( \boldsymbol{k} \right) & 0 & 0 & 0 \\
0 & - E_{+1} \left( \boldsymbol{k} \right) & 0 & 0 \\
0 & 0 & E_{-1} \left( \boldsymbol{k} \right) & 0 \\
0 & 0 & 0 & -E_{-1} \left( \boldsymbol{k} \right)  \\
\end{array}
\right)
\nonumber \\
&\qquad \qquad \qquad \times
\left(
\begin{array}{c}
L_{\boldsymbol{k}, +1, \sigma}\\
L_{-\boldsymbol{k}, +1, \overline{\sigma} }^\dagger \\
L_{\boldsymbol{k}, -1, \sigma} \\
L_{-\boldsymbol{k}, -1, \overline{\sigma} }^\dagger
\end{array}
\right).
\end{align}
Here, we obtain the excitation energies $\pm E_{+1} \left( \boldsymbol{k} \right)$ and $\pm E_{-1} \left( \boldsymbol{k} \right)$ of the Bogoliubov quasiparticles as
\begin{align}
\label{E_Bogo}
E_{\pm1} \left( \boldsymbol{k} \right)
= \sqrt{
\varepsilon \left( \boldsymbol{k} \right)^2
+ \left| \mathit{\Delta}_{\mp1, \mp1} \left( \boldsymbol{k} \right) \right|^2
}
.
\end{align}
Here, we adopt double-sign correspondence.
The Bogoliubov transformation from the electron state vector of
$\boldsymbol{l}_{\boldsymbol{k} } = \bigl(
l_{\boldsymbol{k}, +1, \sigma},
l_{-\boldsymbol{k}, +1, \overline{\sigma} }^\dagger,
l_{\boldsymbol{k}, -1, \sigma},
l_{-\boldsymbol{k}, -1, \overline{\sigma} }^\dagger
\bigr)^T$
to the Bogoliubov quasiparticles of
$\boldsymbol{L}_{\boldsymbol{k} } = \bigl(
L_{\boldsymbol{k}, +1, \sigma},
L_{-\boldsymbol{k}, +1, \overline{\sigma} }^\dagger,
L_{\boldsymbol{k}, -1, \sigma},
L_{-\boldsymbol{k}, -1, \overline{\sigma} }^\dagger
\bigr)^T$
is expressed in terms of an unitary matrix as
\begin{align}
\label{Vector L}
&\left(
\begin{array}{c}
L_{\boldsymbol{k}, +1, \sigma}\\
L_{-\boldsymbol{k}, +1, \overline{\sigma} }^\dagger \\
L_{\boldsymbol{k}, -1, \sigma} \\
L_{-\boldsymbol{k}, -1, \overline{\sigma} }^\dagger
\end{array}
\right) 
\nonumber \\
&= \left(
\begin{array}{cccc}
u_{+1} \left( \boldsymbol{k} \right) & -v_{+1}^\ast \left( \boldsymbol{k} \right) & 0 & 0 \\
v_{+1} \left( \boldsymbol{k} \right) & u_{+1} \left( \boldsymbol{k} \right) & 0 & 0 \\
0 & 0 & u_{-1} \left( \boldsymbol{k} \right) &  -v_{-1}^\ast \left( \boldsymbol{k} \right) \\
0 & 0 &  v_{-1} \left( \boldsymbol{k} \right) & u_{-1} \left( \boldsymbol{k} \right)  \\
\end{array}
\right)
\left(
\begin{array}{c}
l_{\boldsymbol{k}, +1, \sigma} \\
l_{-\boldsymbol{k}, +1, \overline{\sigma} }^\dagger \\
l_{\boldsymbol{k}, -1, \sigma} \\
l_{-\boldsymbol{k}, -1, \overline{\sigma} }^\dagger 
\end{array}
\right)
\nonumber \\ 
&= \left(
\begin{array}{c}
u_{+1} \left( \boldsymbol{k} \right) l_{\boldsymbol{k}, +, \sigma}
- v_{+1}^\ast \left( \boldsymbol{k} \right) l_{-\boldsymbol{k}, +, \overline{\sigma} }^\dagger \\
v_{-1} \left( \boldsymbol{k} \right) l_{\boldsymbol{k}, +, \sigma}
+ u_{+1} \left( \boldsymbol{k} \right) l_{-\boldsymbol{k}, +, \overline{\sigma} }^\dagger \\
u_{-1} \left( \boldsymbol{k} \right) l_{\boldsymbol{k}, -, \sigma}
- v_{-1}^\ast \left( \boldsymbol{k} \right) l_{-\boldsymbol{k}, -, \overline{\sigma} }^\dagger \\ 
v_{-1} \left( \boldsymbol{k} \right) l_{\boldsymbol{k}, -, \sigma}
+ u_{-1} \left( \boldsymbol{k} \right) l_{-\boldsymbol{k}, -, \overline{\sigma} }^\dagger \\
\end{array}
\right)
.
\end{align}
Taking the constraint of 
$\left| u_{\pm1} \left( \boldsymbol{k} \right) \right|^2 + \left| v_{\pm1} \left( \boldsymbol{k} \right) \right|^2 = 1$
for fermion quasiparticles into account, we set the elements of the Bogoliubov transformation in Eq. (\ref{Vector L}) as
\begin{align}
\label{u}
u_{\pm1} \left( \boldsymbol{k} \right)
&= \frac{ E_{\pm1} \left( \boldsymbol{k} \right) + \varepsilon \left( \boldsymbol{k} \right) } 
{
\sqrt{
\left[ E_{\pm1} \left( \boldsymbol{k} \right) + \varepsilon \left( \boldsymbol{k} \right) \right] + \left| \mathit{\Delta}_{\mp1, \mp1} \left( \boldsymbol{k} \right) \right|^2
} 
}
,\\
\label{v}
v_{\pm1}^\ast \left( \boldsymbol{k} \right)
&= \frac{ \mathit{\Delta}_{\mp1, \mp1}^\ast \left( \boldsymbol{k} \right) } 
{
\sqrt{
\left[ E_{\pm1} \left( \boldsymbol{k} \right) + \varepsilon \left( \boldsymbol{k} \right) \right] + \left| \mathit{\Delta}_{\mp1, \mp1} \left( \boldsymbol{k} \right) \right|^2
}
}
.
\end{align}
Here, $u_{\pm1} \left( \boldsymbol{k} \right)$ is a real number and $v_{\pm1} \left( \boldsymbol{k} \right)$ is a complex number.
The Bogoliubov quasiparticles of $\boldsymbol{L}_{\boldsymbol{k} }$ in Eq. (\ref{HSC diagonal}) obey the fermion commutation relations
\begin{align}
\label{Commutation relation of L 1}
\left\{ L_{\boldsymbol{k}, m, \sigma }, L_{\boldsymbol{k}', m', \sigma' }^\dagger \right\}
&= & \delta_{\boldsymbol{k}, \boldsymbol{k}' } \delta_{m, m'} \delta_{\sigma, \sigma' }
, \\
\label{Commutation relation of L 2}
\left\{ L_{\boldsymbol{k}, m, \sigma }, L_{\boldsymbol{k}', m', \sigma' } \right\}
&= &\left\{ L_{\boldsymbol{k}, m, \sigma }^\dagger, L_{\boldsymbol{k}', m', \sigma' }^\dagger \right\} = 0
.
\end{align}

In the superconducting phase, the energy gaps of Eqs. (\ref{Energy gap 1}) and (\ref{Energy gap 2}) show finite values of
$\mathit{\Delta}_{\pm1, \pm1} \left( \boldsymbol{k} \right) \neq 0$
and
$\mathit{\Delta}_{\pm1, \pm1}^\ast \left( \boldsymbol{k} \right) \neq 0$
for both the ferro-type quadrupole interaction $J_\mathrm{Q} \left( \boldsymbol{k}, \boldsymbol{q} \right) > 0$ and the antiferro-type quadrupole interaction $J_\mathrm{Q} \left( \boldsymbol{k}, \boldsymbol{q} \right) < 0$.
The slight deviation of the anisotropic feature of $\gamma \stackrel{<}{_\sim} 1$ from the ideal $xz$ model is necessary for the manifestation of the superconductivity in the vicinity of the QCP.
The superconducting ground state $\mathit{\Phi}_0$ given by the restricted quadrupole interaction Hamiltonian of Eq. (\ref{Restricted HQQ}) is described in terms of the annihilation operators $L_{- \boldsymbol{k}, \pm1, \overline{\sigma} }L_{\boldsymbol{k}, \pm1, \sigma}$ of the Bogoliubov quasiparticles acting on the vacuum state $\mathit{\Phi}_\mathrm{vac}$.
An alternative expression for the superconducting ground state is obtained in terms of the electron creation operators $l_{\boldsymbol{k}, \pm1, \sigma}^\dagger l_{- \boldsymbol{k}, \pm1, \overline{\sigma} }^\dagger$ of the Cooper pair acting on $\mathit{\Phi}_\mathrm{vac}$.
Consequently, the grand state $\mathit{\Phi}_0$ is expressed as
\begin{align}
\label{Superconducting ground state}
\left| \mathit{\Phi}_0 \right\rangle
&= C \prod_{\boldsymbol{k} } \prod_{m = \pm1} \prod_{\sigma \neq \overline{\sigma} }
L_{-\boldsymbol{k}, m, \overline{\sigma} } L_{\boldsymbol{k}, m, \sigma}
\left| \mathit{\Phi}_\mathrm{vac} \right\rangle
\nonumber \\
&=  C \prod_{\boldsymbol{k} } \prod_{m = \pm1} \prod_{\sigma \neq \overline{\sigma} }
\left[ - v_m^\ast \left( \boldsymbol{k} \right) \right]
\nonumber \\
&\qquad \qquad \times
 \left[ 
u_m \left( \boldsymbol{k} \right)
+ v_m^\ast \left( \boldsymbol{k} \right) 
l_{\boldsymbol{k}, m, \sigma}^\dagger l_{-\boldsymbol{k}, m, \overline{\sigma} }^\dagger 
\right]
\left| \mathit{\Phi}_\mathrm{vac} \right\rangle
.
\end{align}
Here, we take the available operations over the azimuthal quantum numbers of $m = +1$ and $-1$, spin orientations of $\sigma$ and $\overline{\sigma}$, and wavevector $\boldsymbol{k}$.
The coefficient of
$C^{-2} = \prod_{\boldsymbol{k} } \prod_{m = \pm1} \prod_{\sigma \neq \overline{\sigma} }\left| v_m \left( \boldsymbol{k} \right)\right|^2$
stands for the normalized factor of $\mathit{\Phi}_0$.
The superconducting ground state $\mathit{\Phi}_0$ of Eq. (\ref{Superconducting ground state}) consists of the creation operators
$l_{\boldsymbol{k}, \pm1, \sigma}^\dagger l_{-\boldsymbol{k}, \pm1, \overline{\sigma} }^\dagger$
of the Cooper pairs with probability weight density $\left| v_m \left( \boldsymbol{k} \right)\right|^2$ of Eq. (\ref{v}).
This is the coherent state treated in the standard BCS theory \cite{BCS, QTS, Tinkham, Schrieffer}.

The restricted quadrupole interaction Hamiltonian of Eq. (\ref{Restricted HQQ}) brings about the energy gaps $\mathit{\Delta}_{\pm1, \pm1} \neq 0$ of Eq. (\ref{Energy gap 1}) and their complex conjugate energy gaps $\mathit{\Delta}_{\pm1, \pm1}^\ast \neq 0$ of Eq. (\ref{Energy gap 2}).
The corresponding mean fields of the off-diagonal long-range order parameters  of
$\bigl\langle l_{-\boldsymbol{k}, \pm1, \overline{\sigma} } l_{\boldsymbol{k}, \pm1, \sigma } \bigr\rangle$
and 
$\bigl\langle l_{\boldsymbol{k}, \pm1, \sigma }^\dagger l_{-\boldsymbol{k}, \pm1, \overline{\sigma} }^\dagger \bigr\rangle$
are written in terms of the Bogoliubov quasiparticles of Eq. (\ref{Vector L}) as
\begin{align}
\label{For gap eq 1}
\bigl\langle l_{-\boldsymbol{k}, \pm1, \overline{\sigma}} l_{\boldsymbol{k}, \pm1, \sigma} \bigr\rangle
=u_{\pm1}^\ast \left( \boldsymbol{k} \right) v_{\pm1}^\ast \left( \boldsymbol{k} \right)
\left(
1
- \bigl\langle L_{\boldsymbol{k}, \pm1, \sigma}^\dagger L_{\boldsymbol{k}, \pm1, \sigma} \bigr\rangle \right.
\nonumber \\
\quad \left. 
- \bigl\langle L_{-\boldsymbol{k}, \pm1, \sigma}^\dagger L_{-\boldsymbol{k}, \pm1, \sigma} \bigr\rangle 
\right),
\\
\label{For gap eq 2}
\bigl\langle l_{\boldsymbol{k}, \pm1, \sigma}^\dagger l_{-\boldsymbol{k}, \pm1, \overline{\sigma}}^\dagger \bigr\rangle
=u_{\pm1} \left( \boldsymbol{k} \right) v_{\pm1} \left( \boldsymbol{k} \right)
\left(
1
- \bigl\langle L_{\boldsymbol{k}, \pm1, \sigma}^\dagger L_{\boldsymbol{k}, \pm1, \sigma} \bigr\rangle \right.
\nonumber \\
\quad \left.
- \bigl\langle L_{-\boldsymbol{k}, \pm1, \sigma}^\dagger L_{-\boldsymbol{k}, \pm1, \sigma} \bigr\rangle 
\right)
.
\end{align}
Here, we used the fact that the mean-field values for the off-diagonal excitation of
$\bigl\langle L_{\boldsymbol{k}, \pm1, \sigma}^\dagger L_{-\boldsymbol{k}, \pm1, \overline{\sigma} }^\dagger \bigr\rangle$
and
$\bigl\langle L_{-\boldsymbol{k}, \pm1, \overline{\sigma} } L_{\boldsymbol{k}, \pm1, \sigma} \bigr\rangle$
for the Bogoliubov quasiparticles vanish.
The Bogoliubov quasiparticle numbers of
$ \bigl\langle L_{\boldsymbol{k}, \pm1, \sigma}^\dagger L_{\boldsymbol{k}, \pm1, \sigma} \bigr\rangle$
and
$\bigl\langle L_{-\boldsymbol{k}, \pm1, \overline{\sigma} }^\dagger L_{-\boldsymbol{k}, \pm1, \overline{\sigma} } \bigr\rangle$
 in Eqs. (\ref{For gap eq 1}) and (\ref{For gap eq 2}) are calculated in terms of the Fermi distribution function
$f \left( E_{\pm1} \left( \boldsymbol{k} \right) \right) = \left\{ \exp \left [ E_{\pm1} \left( \boldsymbol{k} \right) / k_\mathrm{B}T \right] +1 \right\}^{-1}$ 
for the excitation energy $E_{\pm 1} \left( \boldsymbol{k} \right)$ in Eq. (\ref{E_Bogo}).
By using the alternative expressions for the Bogoliubov-transform elements
$\left| u_{\pm1} \left( \boldsymbol{k} \right) \right|^2 - \left| v_{\pm1} \left( \boldsymbol{k} \right) \right|^2 
= \varepsilon \left( \boldsymbol{k} \right) / E_{\pm1} \left( \boldsymbol{k} \right)$
and
$2u_{\pm1} \left( \boldsymbol{k} \right) v_{\pm1} \left( \boldsymbol{k} \right)
= \mathit{\Delta}_{\mp1, \mp1}^\ast \left( \boldsymbol{k} \right) / E_{\pm1} \left( \boldsymbol{k} \right)$,
we obtain the self-consistent equation for the energy gap of Eq. (\ref{Energy gap 2}) as
\begin{align}
\label{Gap eq}
&\mathit{\Delta}_{\pm 1, \pm 1}^\ast \left( \boldsymbol{k} \right)
\nonumber \\
&= \frac{1}{2} \left( 1-\gamma \right) \sum_{\boldsymbol{q} }
J_\mathrm{Q} \left( \boldsymbol{k}, \boldsymbol{q} \right)
\left\langle
 l_{\boldsymbol{k} + \boldsymbol{q}, \pm1, \sigma } l_{-\boldsymbol{k} - \boldsymbol{q}, \pm1, \overline{\sigma} }
\right\rangle
\nonumber \\
&= \frac{1}{4} \left( 1-\gamma \right) \sum_{\boldsymbol{q} } J_\mathrm{Q} \left( \boldsymbol{k}, \boldsymbol{q} \right)
\frac{ \mathit{\Delta}_{\mp 1, \mp 1}^\ast \left( \boldsymbol{k} + \boldsymbol{q} \right) }
{ \sqrt{ \varepsilon \left( \boldsymbol{k} + \boldsymbol{q} \right)^2 + \left| \mathit{\Delta}_{\mp1, \mp1} \left( \boldsymbol{k} + \boldsymbol{q} \right) \right|^2 } }
\nonumber \\
&\qquad \quad \times
\tanh \left[
\frac{ \sqrt{ \varepsilon \left( \boldsymbol{k} + \boldsymbol{q} \right)^2 + \left| \mathit{\Delta}_{\mp1, \mp1} \left( \boldsymbol{k} + \boldsymbol{q}  \right) \right|^2 } }
{ 2k_\mathrm{B}T }
\right]
.
\end{align}
The quadrupole interaction $J_\mathrm{Q} \left( \boldsymbol{k},\boldsymbol{q} \right)$ and the anisotropic parameter $\gamma$ used in the restricted quadrupole interaction Hamiltonian $H_\mathrm{QQ}'(\gamma)$ of Eq. (\ref{Restricted HQQ}) dominate the self-consistent equation for the superconducting energy gaps of $\mathit{\Delta}_{\pm1, \pm1}^\ast \left( \boldsymbol{k} \right)$ in Eq. (\ref{Gap eq}).
The microscopic mechanism for the Cooper pair formation in the present treatment is in good agreement with the previous theoretical studies based on the quadrupole interaction \cite{Yanagi_1, Kontani, Onari} but are rather different from the theory based on the spin fluctuation \cite{Mazin1, Kuroki, Fernandes}.
The superconducting transition temperature $T_\mathrm{SC}$ determined by Eq. (\ref{Gap eq}) in the present model and the hexadecapole interaction $\widetilde{D}^\mathrm{HH}$ of Eq. (\ref{DHH written by DeltaH}) both contribute to the critical temperature $\mathit{\Theta}_\mathrm{C}$ of the renormalized hexadecapole susceptibility of Eq. (\ref{chiH tilde}) as $\mathit{\Theta}_\mathrm{C} = \widetilde{D}^\mathrm{HH} + T_\mathrm{SC}$ as shown in Eq. (\ref{ThetaC}).

In order to characterize the superconducting state of the present iron pnictide, we calculate the eigenenergy of the restricted quadrupole interaction Hamiltonian
$H_\mathrm{QQ}' \left( \gamma \right)$ of Eq. (\ref{Restricted HQQ}) for the superconducting ground state $\mathit{\Phi}_0$ of Eq. (\ref{Superconducting ground state}) as
\begin{align}
\label{Expectation value of HQQ}
&\left\langle \mathit{\Phi}_0 \left| H_\mathrm{QQ}' \left( \gamma \right) \right| \mathit{\Phi}_0 \right\rangle
\nonumber \\
& \quad
= -\frac{1}{8} \sum_{\boldsymbol{k}, \boldsymbol{q} } \sum_{\sigma \neq \overline{\sigma} } 
J_\mathrm{Q} \left( \boldsymbol{k}, \boldsymbol{q} \right)
\left[
\frac{ \mathit{\Delta}_{-1, -1}^\ast \left( \boldsymbol{k} + \boldsymbol{q} \right) }{ E_{+1} \left( \boldsymbol{k} + \boldsymbol{q} \right) }
\frac{ \mathit{\Delta}_{+1, +1} \left( \boldsymbol{k} \right) }{ E_{-1} \left( \boldsymbol{k} \right) } \right.
\nonumber \\
&\qquad \qquad \qquad \qquad \left.
+
\frac{ \mathit{\Delta}_{+1, +1}^\ast \left( \boldsymbol{k} + \boldsymbol{q} \right) }{ E_{-1} \left( \boldsymbol{k} + \boldsymbol{q} \right) }
\frac{ \mathit{\Delta}_{-1, -1} \left( \boldsymbol{k} \right) }{ E_{+1} \left( \boldsymbol{k} \right) }
\right]
.
\end{align}
The antiferro- and ferro-quadrupole interactions of $J_\mathrm{Q} \left( \boldsymbol{k}, \boldsymbol{q} \right)$ for various wavevectors $\boldsymbol{q}$ of the transverse acoustic phonons over the Brillouin zone participate in the Cooper pair formation.
However, a small-wavenumber limit of $|\boldsymbol{q}| \rightarrow 0$ corresponding to the measured ultrasonic wave particularly plays a significant role to cause the the critical slowing down due to the ferro-hexadecapole ordering.
Therefore, we calculate the quadrupole interaction energy of
$\left\langle \mathit{\Phi}_0 \left| H_\mathrm{QQ}' \left( \gamma \right) \right| \mathit{\Phi}_0 \right\rangle$
of Eq. (\ref{Expectation value of HQQ}) for the Cooper pairs bound by electrons with wavevectors $\boldsymbol{k}$ and $\boldsymbol{k} + \boldsymbol{q}$ for the small-wavenumber limit of $|\boldsymbol{q}| \rightarrow 0$.

In out attempt to examine the interference between the phase $\varphi_{+1,+1} \left( \boldsymbol{k} \right)$ of the energy gap $\mathit{\Delta}_{+1,+1} \left( \boldsymbol{k} \right)$ due to the off-diagonal long-range order parameter
$\left\langle l_{-\boldsymbol{k}, +1, \overline{\sigma} } l_{\boldsymbol{k}, +1, \sigma } \right\rangle$
and the phase $\varphi_{-1,-1} \left( \boldsymbol{k} \right)$ of $\mathit{\Delta}_{-1,-1} \left( \boldsymbol{k} \right)$ due to 
$\left\langle l_{-\boldsymbol{k}, -1, \overline{\sigma} } l_{\boldsymbol{k}, -1, \sigma } \right\rangle$,
we take polar form of the energy gaps.
\begin{align}
\label{Polar form of gap}
\mathit{\Delta}_{\pm1, \pm1} \left( \boldsymbol{k} \right)
= \left| \mathit{\Delta}_{\pm1, \pm1} \left( \boldsymbol{k} \right) \right|
\exp \left[ i \varphi_{\pm1, \pm1} \left( \boldsymbol{k} \right) \right]
.
\end{align}
Thus, we deduce the energy of the restricted quadrupole interaction of Eq. (\ref{Restricted HQQ}) for the superconducting ground state $\mathit{\Phi}_0$ in the small-wavenumber limit of $\left| \boldsymbol{q} \right| \rightarrow 0$ as 
\begin{align}
\label{HQQ by cos}
&
\left\langle \mathit{\Phi}_0 \left| H_\mathrm{QQ}' \left( \gamma \right) \right| \mathit{\Phi}_0 \right\rangle
\nonumber \\
& \qquad
= -\frac{1}{2} \sum_{\boldsymbol{k} } 
J_\mathrm{Q} \left( \boldsymbol{k}, 0 \right)
\frac{ \left| \mathit{\Delta}_{-1, -1} \left( \boldsymbol{k} \right) \right| }{ E_{+1} \left( \boldsymbol{k} \right) }
\frac{ \left| \mathit{\Delta}_{+1, +1} \left( \boldsymbol{k} \right) \right| }{ E_{-1} \left( \boldsymbol{k} \right) }
\nonumber \\ 
&\qquad \qquad \qquad \times
\cos \left[ \varphi_{+1, +1} \left( \boldsymbol{k} \right) - \varphi_{-1, -1} \left( \boldsymbol{k} \right)  \right]
.
\end{align}
The restricted quadrupole interaction energy of Eq. (\ref{HQQ by cos}) depends on 
$\cos \left[ \varphi_{+1,+1} \left( \boldsymbol{k} \right) - \varphi_{-1,-1} \left( \boldsymbol{k} \right) \right]$ 
for the phase difference $\varphi_{+1,+1} \left( \boldsymbol{k} \right) - \varphi_{-1,-1} \left( \boldsymbol{k} \right)$ between the two energy gaps in Eq. (\ref{Polar form of gap}). \cite{Leggett}.
The Cooper pairs dominated by the ferro-type quadrupole interaction with $J_\mathrm{Q} \left( \boldsymbol{k}, 0 \right) > 0$ for the small-wavenumber regime of $|\boldsymbol{k}| < k_\mathrm{b}^\mathrm{Q} $ bring about the state for
$\cos \left[ \varphi_{+1,+1} \left( \boldsymbol{k} \right) - \varphi_{-1,-1} \left( \boldsymbol{k} \right) \right] = 1$
with the phase difference corresponding to the stationary point of 
$\varphi_{+1,+1} \left( \boldsymbol{k} \right) - \varphi_{-1,-1} \left( \boldsymbol{k} \right) = s \pi$
for an even integer $s = 2n$.
In the opposite case, however, the antiferro-type quadrupole interaction of $J_\mathrm{Q} \left( \boldsymbol{k}, 0 \right) < 0$, which is relevant for relatively high wavenumbers of $|\boldsymbol{k}| > k_\mathrm{b}^\mathrm{Q}$, has
$\cos \left[ \varphi_{+1,+1} \left( \boldsymbol{k} \right) - \varphi_{-1,-1} \left( \boldsymbol{k} \right) \right] = -1$
with the stationary point of 
$\varphi_{+1,+1} \left( \boldsymbol{k} \right) - \varphi_{-1,-1} \left( \boldsymbol{k} \right) = s\pi$ for an odd integer $s = 2n + 1$.

The indirect quadrupole interaction coefficient $D^\mathrm{QQ} \left( \boldsymbol{k}, \boldsymbol{q} \right)$ of Eq. (\ref{DQQ2}) in the Hamiltonian of Eq. (\ref{HindQQ_2}) may possess positive or negative sign depending on the sign of its denominator.
Since the coefficient $D^\mathrm{QQ} \left( \boldsymbol{k}, \boldsymbol{q} \right)$ dominates the quadrupole interaction coefficient $J_\mathrm{Q} \left( \boldsymbol{k}, \boldsymbol{q} \right)$, there are expected two cases of the ferro-type quadrupole interaction of $J_\mathrm{Q} \left( \boldsymbol{k}, \boldsymbol{q} \right) > 0$ and antiferro-type of $J_\mathrm{Q} \left( \boldsymbol{k}, \boldsymbol{q} \right) < 0$ in the restricted quadrupole interaction energy of Eq. (\ref{HQQ by cos}).
These both cases of $J_\mathrm{Q} \left( \boldsymbol{k}, \boldsymbol{q} \right) > 0$ and $J_\mathrm{Q} \left( \boldsymbol{k}, \boldsymbol{q} \right) < 0$ are commonly available for the superconducting Cooper pair formation in the system.
The restricted quadrupole interaction $H_\mathrm{QQ}' \left( \gamma \right)$ of Eq. (\ref{Restricted HQQ}) is ruled by the quadrupole interaction coefficient $J_\mathrm{Q} \left( \boldsymbol{k}, \boldsymbol{q} \right)$ and the anisotropic coefficient $\gamma \stackrel{<}{_\sim} 1$ indicating the slight deviation from the ideal $xz$ model.
Consequently, the superconductivity energy gap caused by the restricted quadrupole interaction $H_\mathrm{QQ}' \left( \gamma \right)$ is favorable for the s-like shape superconducting energy gap reflecting the almost isotropic feature in the $x'y'$ plane.

\subsection{Superconducting ground state and hexadecapole ordering}

In order to examine the hexadecapole ordering in the superconducting phase, we investigate whether or not the superconducting ground state $\mathit{\varPhi}_0$ of Eq. (\ref{Superconducting ground state}) due to the restricted quadrupole interaction Hamiltonian $H_\mathrm{QQ}' \left( \gamma \right)$ of Eq. (\ref{Restricted HQQ}) bears the hexadecapole.
The annihilation operators $B_{\boldsymbol{k}, \pm, \sigma, \overline{\sigma} }$ of Eq. (\ref{Fourier transformation of  B 1}) and the creation operators $B_{\boldsymbol{k}, \pm, \sigma, \overline{\sigma} }^\dagger$ of Eq. (\ref{Fourier transformation of  B 2}) describing the hexadecapole $H_{z,  \boldsymbol{k}, \boldsymbol{q} }^\alpha$ of Eq. (\ref{Hexadecapole by Bk}) possess the following expectation values for the superconducting ground state $\mathit{\Phi}_0$ of Eq. (\ref{Superconducting ground state}):
\begin{align}
\label{Expectation value of B 1}
&
\left\langle \mathit{\Phi}_0 \left| B_{\boldsymbol{k}, \pm, \sigma, \overline{\sigma} } \right| \mathit{\Phi}_0 \right\rangle
\nonumber \\
& \quad \qquad
= \frac{1}{2\sqrt{2} }
\left[
e^{\mp i \frac{3\pi}{4} }
\frac{ \mathit{\Delta}_{-1, -1} \left( \boldsymbol{k} \right) } { E_{+1} \left( \boldsymbol{k} \right) }
+ e^{\pm i \frac{3\pi}{4} }
\frac{ \mathit{\Delta}_{+1, +1} \left( \boldsymbol{k} \right) } { E_{-1} \left( \boldsymbol{k} \right) }
\right]
, \\
\label{Expectation value of B 2}
&
\left\langle \mathit{\Phi}_0 \left| B_{\boldsymbol{k}, \pm, \sigma, \overline{\sigma} }^\dagger \right| \mathit{\Phi}_0 \right\rangle
\nonumber \\
& \quad \qquad
= \frac{1}{2\sqrt{2} }
\left[
e^{\pm i \frac{3\pi}{4} }
\frac{ \mathit{\Delta}_{-1, -1}^\ast \left( \boldsymbol{k} \right) } { E_{+1} \left( \boldsymbol{k} \right) }
+ e^{\mp i \frac{3\pi}{4} }
\frac{ \mathit{\Delta}_{+1, +1}^\ast \left( \boldsymbol{k} \right) } { E_{-1} \left( \boldsymbol{k} \right) }
\right]
.
\end{align}
The finite values of
$\bigl\langle \mathit{\Phi}_0 | B_{\boldsymbol{k}, \pm, \sigma, \overline{\sigma} } | \mathit{\Phi}_0 \bigr\rangle \neq 0$
and
$\bigl\langle \mathit{\Phi}_0 | B_{\boldsymbol{k}, \pm, \sigma, \overline{\sigma} }^\dagger | \mathit{\Phi}_0 \bigr\rangle  \neq 0$
are expected with the appearance of the energy gaps $\mathit{\Delta}_{\pm1, \pm1} \left( \boldsymbol{k} \right) \neq 0$ of Eq. (\ref{Energy gap 1}) and $\mathit{\Delta}_{\pm1, \pm1}^\ast \left( \boldsymbol{k} \right) \neq 0$ of Eq. (\ref{Energy gap 2}) below the superconducting transition temperature.
This result shows that the superconducting ground state $\mathit{\Phi}_0$ of Eq. (\ref{Superconducting ground state}) actually bears the off-diagonal long-range order parameters $B_{\boldsymbol{k}, \pm, \sigma, \overline{\sigma} }$ and $B_{\boldsymbol{k}, \pm, \sigma, \overline{\sigma} }^\dagger$ for the Cooper pairs.  

The hexadecapole $H_{z, \boldsymbol{k}, \boldsymbol{q}}^\alpha$ of Eq. (\ref{Hexadecapole by Bk}) is expressed in terms of the difference in the numbers of Cooper pairs as
\begin{align}
\label{Hza by l}
H_{z, \boldsymbol{k}, \boldsymbol{q} }^\alpha
&=\frac{1}{2} \sum_{\sigma \neq \overline{\sigma} }
\left(
B_{\boldsymbol{k} + \boldsymbol{q}, +, \sigma, \overline{\sigma} }^\dagger B_{\boldsymbol{k}, +, \sigma, \overline{\sigma} }
- B_{\boldsymbol{k} + \boldsymbol{q}, -, \sigma, \overline{\sigma} }^\dagger B_{\boldsymbol{k}, -, \sigma, \overline{\sigma} }
\right)
\nonumber \\ 
&= - \frac{i}{2} \sum_{\sigma \neq \overline{\sigma} } 
\left(
l_{\boldsymbol{k} + \boldsymbol{q}, +1, \sigma}^\dagger l_{-\boldsymbol{k} -\boldsymbol{q}, +1, \overline{\sigma} }^\dagger
l_{-\boldsymbol{k}, -1, \overline{\sigma} } l_{\boldsymbol{k}, -1, \sigma} \right.
\nonumber \\
&\left. \qquad \qquad
- l_{\boldsymbol{k} + \boldsymbol{q}, -1, \sigma}^\dagger  l_{-\boldsymbol{k} -\boldsymbol{q}, -1, \overline{\sigma} }^\dagger
l_{-\boldsymbol{k}, +1, \overline{\sigma} } l_{\boldsymbol{k}, +1, \sigma}
\right)
.
\end{align}
Thus, we obtain the expectation value for the hexadecapole for the superconducting ground state $\mathit{\Phi}_0$ as
\begin{align}
\label{Expectation value of Hza}
\left\langle \mathit{\Phi}_0 \left| H_{z, \boldsymbol{k}, \boldsymbol{q} }^\alpha  \right| \mathit{\Phi}_0 \right\rangle
&= -\frac{i}{4}
\left[
\frac{ \mathit{\Delta}_{-1, -1}^\ast \left( \boldsymbol{k} + \boldsymbol{q} \right) } { E_{+1} \left( \boldsymbol{k} + \boldsymbol{q} \right) }
\frac{ \mathit{\Delta}_{+1, +1} \left( \boldsymbol{k} \right) } { E_{-1} \left( \boldsymbol{k} \right) } \right.
\nonumber \\
&\qquad \quad \left.
- \frac{ \mathit{\Delta}_{+1, +1}^\ast \left( \boldsymbol{k} + \boldsymbol{q} \right) } { E_{-1} \left( \boldsymbol{k} + \boldsymbol{q} \right) }
\frac{ \mathit{\Delta}_{-1, -1} \left( \boldsymbol{k} \right) } { E_{+1} \left( \boldsymbol{k} \right) }
\right]
.
\end{align}
The appearance of the energy gaps $\mathit{\Delta}_{\pm1, \pm1} \left( \boldsymbol{k} \right) \neq 0$ in the superconducting phase simultaneously brings about a finite eigenvalue of the hexadecapole
$\bigl\langle \mathit{\Phi}_0 | H_{z, \boldsymbol{k}, \boldsymbol{q} }^\alpha  | \mathit{\Phi}_0 \bigr\rangle \neq 0$.

The critical slowing down of the relaxation time $\tau_\mathrm{H}$ around the superconducting transition point in Fig. \ref{Fig1}(b) is observed via the transverse  ultrasonic waves with wavelengths as long as $\lambda \sim 10$ $\mu$m.
Therefore, we treat the rotation $\omega_{xy} \left( \boldsymbol{q} \right)$ of the transverse acoustic phonons in Eqs. (\ref{Hza by l}) and (\ref{Expectation value of Hza}) for the small-wavenumber limit of $\left| \boldsymbol{q} \right| = 2 \pi / \lambda \rightarrow 0 $.
Furthermore, we assume equivalence in the amplitudes of 
$\left| \mathit{\Delta}_{+1, +1} \left( \boldsymbol{k} \right) \right|
= \left| \mathit{\Delta}_{-1, -1} \left( \boldsymbol{k} \right) \right|
= \left| \mathit{\Delta} \left( \boldsymbol{k} \right) \right|$
for the two energy gaps of Eq. (\ref{Energy gap 1}) and in the excitation energies of
$E_{+1} \left( \boldsymbol{k} \right) = E_{-1} \left( \boldsymbol{k} \right)  = E \left( \boldsymbol{k} \right)$
for the Bogoliubov quasiparticles of Eq. (\ref{Vector L}).
The hexadecapole $H_{z, \boldsymbol{k}, \boldsymbol{q} = 0 }^\alpha $ of Eq. (\ref{Hza by l}) corresponding to the difference in the numbers of Cooper pairs $N_{\boldsymbol{k}, +, \sigma, \overline{\sigma} }$ and $N_{\boldsymbol{k}, -, \sigma, \overline{\sigma} }$ has an expectation value for the superconducting ground state $\mathit{\Phi}_0$ of
\begin{align}
\label{Hza by sin}
\left\langle \mathit{\Phi}_0 \left| H_{z, \boldsymbol{k}, \boldsymbol{q} = 0 }^\alpha  \right| \mathit{\Phi}_0 \right\rangle
&= \left\langle \mathit{\Phi}_0 \left|
\frac{1}{2} \sum_{\sigma \neq \overline{\sigma} }
\left( N_{\boldsymbol{k}, +, \sigma, \overline{\sigma} } - N_{\boldsymbol{k}, -, \sigma, \overline{\sigma} } \right)
\right| \mathit{\Phi}_0 \right\rangle
\nonumber \\
&= \frac{1}{2}
\frac{ \left| \mathit{\Delta} \left( \boldsymbol{k} \right) \right|^2 } { E \left( \boldsymbol{k} \right)^2 }
\sin \left[ \varphi_{+1, +1} \left( \boldsymbol{k} \right) - \varphi_{-1, -1} \left( \boldsymbol{k} \right)  \right]
.
\end{align}
Thus, the hexadecapole ordering corresponding to the finite difference in the numbers of Cooper pairs  
$N_{\boldsymbol{k}, +, \sigma, \overline{\sigma} } - N_{\boldsymbol{k}, -, \sigma, \overline{\sigma} }$
leads to
$\sin \left[ \varphi_{+1, +1} \left( \boldsymbol{k} \right) - \varphi_{-1, -1} \left( \boldsymbol{k} \right)  \right] \neq 0$
for the finite deviation of the phase difference of
$\varphi_{+1, +1} \left( \boldsymbol{k} \right) - \varphi_{-1, -1} \left( \boldsymbol{k} \right)$
from the stationary point of $s\pi$ with an even integer $s = 2n$ for $J_\mathrm{Q} \bigl( \boldsymbol{k}, 0 \bigr) > 0$ or with an odd integer $s = 2n+1$ for $J_\mathrm{Q} \bigl( \boldsymbol{k}, 0 \bigr) < 0$.
On the other hand, when the number of Cooper pairs $N_{\boldsymbol{k}, +, \sigma, \overline{\sigma} }$ is equivalent to the counter number of $N_{\boldsymbol{k}, -, \sigma, \overline{\sigma} }$, we expect the equilibrium state of 
$\sin \left[ \varphi_{+1, +1} \left( \boldsymbol{k} \right) - \varphi_{-1, -1} \left( \boldsymbol{k} \right)  \right] = 0$
in accordance with the phase difference located at the stationary point of $\varphi_{+1, +1} \left( \boldsymbol{k} \right) - \varphi_{-1, -1} \left( \boldsymbol{k} \right) = 2n\pi$ or $(2n+1)\pi$.

The interaction of the hexadecapole $H_{z, \boldsymbol{k}, \boldsymbol{q} }^\alpha$ with the rotation $\omega_{xy} \left( \boldsymbol{q} \right)$ of the transverse acoustic phonons of Eq. (\ref{Hrot by Bk}) generates a perturbation in the quadrupole interaction energy of Eq. (\ref{HQQ by cos}) in the superconducting phase.
The expectation value of the hexadecapole-rotation interaction Hamiltonian $H_\mathrm{rot} \bigl( \omega_{xy} \bigr)$ of Eq. (\ref{Hrot by Bk}) for the superconducting ground state $\mathit{\Phi}_0$ is calculated as
\begin{align}
\label{Expectation value of Hrot}
&
\left\langle \mathit{\Phi}_0 \left| H_\mathrm{rot} \bigl( \omega_{xy} \bigr) \right| \mathit{\Phi}_0 \right\rangle
\nonumber \\
&\quad
= i \left( 1 - \gamma \right) \sum_{\boldsymbol{k}, \boldsymbol{q} } J_\mathrm{Q} \left( \boldsymbol{k}, \boldsymbol{q} \right)
\left[
\frac{ \mathit{\Delta}_{-1, -1}^\ast \left( \boldsymbol{k} + \boldsymbol{q} \right) }{ E_{+1} \left( \boldsymbol{k} + \boldsymbol{q} \right) }
\frac{ \mathit{\Delta}_{+1, +1} \left( \boldsymbol{k} \right) }{ E_{-1} \left( \boldsymbol{k} \right) } \right.
\nonumber \\
&\qquad \qquad \qquad \left. -
\frac{ \mathit{\Delta}_{+1, +1}^\ast \left( \boldsymbol{k} + \boldsymbol{q} \right) }{ E_{-1} \left( \boldsymbol{k} + \boldsymbol{q} \right) }
\frac{ \mathit{\Delta}_{-1, -1} \left( \boldsymbol{k} \right) }{ E_{+1} \left( \boldsymbol{k} \right) }
\right] \omega_{xy} \left( \boldsymbol{q} \right)
.
\end{align}
Since the rotation $\omega_{xy} \left( 0 \right)$ of the transverse acoustic phonons in the small-wavenumber regime of $|\boldsymbol{q}| \rightarrow 0$ participates in the ferro-type hexadecapole interaction, the hexadecapole-rotation interaction energy
$\langle \mathit{\Phi}_0 | H_\mathrm{rot} \bigl( \omega_{xy} \bigr) | \mathit{\Phi}_0 \rangle$
in the superconducting state of Eq. (\ref{Expectation value of Hrot}) is reduced as
\begin{align}
\label{Hrot by sin}
&
\left\langle \mathit{\Phi}_0 \left| H_\mathrm{rot} \bigl( \omega_{xy} \bigr) \right| \mathit{\Phi}_0 \right\rangle
\nonumber \\
&\quad
= - 2\left( 1 - \gamma \right) \sum_{\boldsymbol{k} } 
J_\mathrm{Q} \left( \boldsymbol{k}, \boldsymbol{q} = 0 \right)
\frac{ \left| \mathit{\Delta} \left( \boldsymbol{k} \right) \right|^2 }{ E \left( \boldsymbol{k} \right)^2 }
\nonumber \\
&\qquad \qquad \qquad \qquad \times
\sin \left[ \varphi_{+1, +1} \left( \boldsymbol{k} \right) - \varphi_{-1, -1} \left( \boldsymbol{k} \right) \right]
\omega_{xy} \left( 0 \right)
\nonumber \\
&\quad
= -4 \left( 1 - \gamma \right) \sum_{\boldsymbol{k} }
J_\mathrm{Q} \left( \boldsymbol{k}, \boldsymbol{q} = 0 \right)
\left\langle \mathit{\Phi}_0 \left| H_{z, \boldsymbol{k}, \boldsymbol{q} = 0}^\alpha  \right| \mathit{\Phi}_0 \right\rangle
\omega_{xy} \left( 0 \right)
.
\end{align}
Here, we used the expectation value of the hexadecapole of Eq. (\ref{Hza by sin}) for the superconducting ground state $\mathit{\Phi}_0$ of Eq. (\ref{Superconducting ground state}).
As discussed for Eqs. (\ref{Torque for two-electron state}) and (\ref{Define torque}), the torque $\tau_{xy} \bigl( \boldsymbol{r}_i, \boldsymbol{r}_j \bigr)$ is proportional to  the hexadecapole.
It is meaningful to show the expectation value of the torque $\tau_{xy}$ for the superconducting ground state $\mathit{\Phi}_0$ as follows:
\begin{align}
\label{Expectation value of torque}
&
\left\langle \mathit{\Phi}_0 \left| \tau_{xy} \right| \mathit{\Phi}_0 \right\rangle
\nonumber \\
& \quad
= -4 \left( 1 - \gamma \right) \sum_{\boldsymbol{k} }
J_\mathrm{Q} \left( \boldsymbol{k}, \boldsymbol{q} = 0 \right)
\left\langle \mathit{\Phi}_0 \left| H_{z, \boldsymbol{k}, \boldsymbol{q} = 0 }^\alpha  \right| \mathit{\Phi}_0 \right\rangle.
\end{align}
Thus, we have an elementary expression for the hexadecapole-rotation interaction energy in terms of the torque $\tau_{xy}$ and rotation $\omega_{xy} \left( \boldsymbol{q} = 0 \right)$ as
\begin{align}
\label{Hrot by torque}
\left\langle \mathit{\Phi}_0 \left| H_\mathrm{rot} \bigl( \omega_{xy} \bigr) \right| \mathit{\Phi}_0 \right\rangle
= \left\langle \mathit{\Phi}_0 \left| \tau_{xy} \right| \mathit{\Phi}_0 \right\rangle 
\omega_{xy} \left( 0 \right)
.
\end{align}
Note that the expectation value of the torque $\bigl\langle \mathit{\Phi}_0 | \tau_{xy} | \mathit{\Phi}_0 \bigr\rangle $ defined in Eq. (\ref{Expectation value of torque}) spontaneously becomes finite with the appearance of the hexadecapole ordering $\bigl\langle \mathit{\Phi}_0 | H_{z, \boldsymbol{k}, \boldsymbol{q} }^\alpha  | \mathit{\Phi}_0 \bigr\rangle \neq 0$ due to the superconducting energy gaps of
$| \mathit{\Delta}_{+1, +1} \left( \boldsymbol{k} \right) |
= | \mathit{\Delta}_{-1, -1} \left( \boldsymbol{k} \right) |
= | \mathit{\Delta} \left( \boldsymbol{k} \right) | \neq 0$
in the superconducting phase.

In order to specify the hexadecapole ordering in the superconducting phase far below the transition temperature, we adopt a phenomenological treatment using the microscopic theory described above.
The internal energy $U$ includes both the restricted quadrupole interaction energy of Eq. (\ref{HQQ by cos}) and the hexadecapole-rotation interaction energy for the rotation $\omega_{xy} \left( 0 \right)$ in the small-wavenumber limit of $\left| \boldsymbol{q} \right| \rightarrow 0$ in Eq. (\ref{Expectation value of Hrot}) as
\begin{align}
\label{HQQ + Hrot in SC}
U &= \sum_{ \boldsymbol{k} }U \left( \boldsymbol{k} \right)
\nonumber \\
&
= - \frac{1}{2} \left( 1 - \gamma \right) \sum_{\boldsymbol{k} } 
J_\mathrm{Q} \left( \boldsymbol{k}, 0 \right)
\frac{ \left| \mathit{\Delta} \left( \boldsymbol{k} \right) \right|^2 }{ E \left( \boldsymbol{k} \right)^2 }
\nonumber \\
&\qquad \times 
\left\{
\cos \left[ \varphi_{+1, +1} \left( \boldsymbol{k} \right) - \varphi_{-1, -1} \left( \boldsymbol{k} \right) \right] \right.
\nonumber \\
&\left. \qquad \qquad 
+ 4 \sin \left[ \varphi_{+1, +1} \left( \boldsymbol{k} \right) - \varphi_{-1, -1} \left( \boldsymbol{k} \right) \right] \omega_{xy} \left( 0 \right)
\right\}
\nonumber \\
& \quad
+ \frac{1}{2} C_{66}^0 \omega_{xy} \left( 0 \right)^2 + \frac{1}{4} \beta \omega_{xy} \left( 0 \right)^4
.
\end{align}
Here, we add the harmonic rotation energy of $C_{66}^0 \omega_{xy} \left( 0 \right)^2/2$ and the higher-order term of  $\beta \omega_{xy} \left( 0 \right)^4/4$ with $\beta > 0$ to avoid instability of the system.
Since the present system has the spin-singlet property of $S = 0$, we take the sum of the spin orientation as $\sum_{\sigma \neq \overline{\sigma} } = 2$ in the internal energy $U$ of Eq. (\ref{HQQ + Hrot in SC}).
We seek the minimum of the internal energy $U \left( \boldsymbol{k} \right)$ of Eq. (\ref{HQQ + Hrot in SC}) in finite variations of the phase difference of 
$\delta \varphi \left( \boldsymbol{k} \right)
= \varphi_{+1, +1} \left( \boldsymbol{k} \right) - \varphi_{-1, -1} \left( \boldsymbol{k} \right) - s\pi \neq 0$
from the stationary point of
$s\pi = 2n\pi$ for the ferro-type quadrupole interaction of $J_\mathrm{Q} \left( \boldsymbol{k}, 0\right) > 0$ or $s\pi = (2n+1)\pi$ for the antiferro-type quadrupole interaction of $J_\mathrm{Q} \left( \boldsymbol{k}, 0\right) < 0$ and the finite rotation $\omega_{xy} \left( 0 \right) \neq 0$. 

The internal energy $U \left( \boldsymbol{k} \right)$ characterized by wavevector $\boldsymbol{k}$ in Eq. (\ref{HQQ + Hrot in SC}) is expanded up to the second order of $\delta \varphi \left( \boldsymbol{k} \right)$ as
\begin{align}
\label{HQQ + Hrot in SC with k}
U \left( \boldsymbol{k} \right)
&= - \frac{1}{2} \left( 1 - \gamma \right) J_\mathrm{Q} \left( \boldsymbol{k}, 0 \right) \cos \left( s\pi \right)
\frac{ \left| \mathit{\Delta} \left( \boldsymbol{k} \right) \right|^2 }{ E \left( \boldsymbol{k} \right)^2 }
\nonumber \\
&\qquad \quad \times
\left[ 1 - \frac{1}{2} \delta \varphi \left( \boldsymbol{k} \right)^2 + 4 \delta \varphi \left( \boldsymbol{k} \right) \omega_{xy} \left( 0 \right) \right]
\nonumber \\
&+ \frac{1}{2} \frac{ C_{66}^0 }{ N_\mathrm{C} } \omega_{xy} \left( 0 \right)^2
+  \frac{1}{4} \frac{ \beta }{ N_\mathrm{C} } \omega_{xy} \left( 0 \right)^4
.
\end{align}
Here, $N_\mathrm{C} = \sum_{\boldsymbol{k} }$ is the number of Cooper pairs participating in the superconductivity.
For simplicity, we use the abbreviated notation
$S_\mathrm{Q} \left( \boldsymbol{k} \right)$ proportional to the square of the energy gap $\left| \mathit{\Delta} \left( \boldsymbol{k} \right) \right|^2$ as
\begin{align}
\label{SQ}
S_\mathrm{Q} \left( \boldsymbol{k} \right)
&
= \frac{1}{2} \left( 1 - \gamma \right) J_\mathrm{Q} \left( \boldsymbol{k}, 0 \right) \cos \left( s\pi \right)
\frac{ \left| \mathit{\Delta} \left( \boldsymbol{k} \right) \right|^2 } {E \left( \boldsymbol{k} \right)^2 }
\nonumber \\
&
= \frac{1}{2} \left( 1 - \gamma \right) \left| J_\mathrm{Q} \left( \boldsymbol{k}, 0 \right) \right|
\frac{ \left| \mathit{\Delta} \left( \boldsymbol{k} \right) \right|^2 } {E \left( \boldsymbol{k} \right)^2 }
.
\end{align}
Since the quadrupole interaction of $J_\mathrm{Q} \left( \boldsymbol{k}, 0\right)$ is dominated by the indirect quadrupole interaction of $D^\mathrm{QQ} \left( \boldsymbol{k}, 0\right)$ of Eq. (\ref{DQQ(k,0)}), the ferro-type quadrupole interaction of $J_\mathrm{Q} \left( \boldsymbol{k}, 0\right) > 0$ for the small-wavenumber regime of $|\boldsymbol{k}| < k_\mathrm{b}^\mathrm{Q}$ causes the energy gaps with the phase difference around the stationary point of  $s\pi = 2n\pi$ for $\cos \left( 2n\pi \right) = 1$, while the antiferro-type interaction of $J_\mathrm{Q} \left( \boldsymbol{k}, 0\right) < 0$ with the relatively large-wavenumber regime of $|\boldsymbol{k}| > k_\mathrm{b}^\mathrm{Q}$ brings about the energy gaps with the phase difference around the stationary point of $s\pi = \left(2n+1 \right) \pi$ for $\cos \left[ \left( 2n + 1 \right)\pi \right] = -1$.
The experimentally determined phase diagram as shown in Fig. \ref{Fig4} indicates that the ferro-type quadrupole interaction for the Co concentration $x$ below the QCP of $x_\mathrm{QCP} = 0.061$ changes to the antiferro-type quadrupole interaction with increasing $x$ across $x_\mathrm{QCP}$.
Therefore, the inherent property, that both the antiferro- and ferro-quadrupole interactions participate in the Cooper pair formation in Eq. (\ref{SQ}), is available for the appearance of the superconductivity in the vicinity of the QCP.

Thus, the internal energy $U \left( \boldsymbol{k} \right)$ of Eq. (\ref{HQQ + Hrot in SC with k}) is deduced as
\begin{align}
\label{HQQ + Hrot in SC with k 2}
U \left( \boldsymbol{k} \right)
&
= \frac{1}{2} S_\mathrm{Q} \left( \boldsymbol{k} \right)
\left[ \delta \varphi \left( \boldsymbol{k} \right) - 4 \omega_{xy} \left( 0 \right)   \right]^2
\nonumber \\
&\quad
+ \frac{1}{4} \frac{ \beta }{N_\mathrm{C} }
\left\{
\omega_{xy} \left( 0 \right)^2 - 16 \frac{N_\mathrm{C} }{ \beta }
\left[ S_\mathrm{Q} \left( \boldsymbol{k} \right) - \frac{1}{16} \frac{ C_{66}^0}{ N_\mathrm{C} }
\right]
\right\}^2
\nonumber \\
&\qquad
 - S_\mathrm{Q} \left( \boldsymbol{k} \right)
- 64 \frac{N_\mathrm{C} }{ \beta } 
\left[ S_\mathrm{Q} \left( \boldsymbol{k} \right) - \frac{1}{16} \frac{ C_{66}^0}{ N_\mathrm{C} }
\right]^2
.
\end{align}
The appearance of the energy gap $\left| \mathit{\Delta} \left( \boldsymbol{k} \right) \right| \neq 0$ in the superconducting phase leads to a finite value of $S_\mathrm{Q} \left( \boldsymbol{k} \right) \neq 0$ in Eq. (\ref{SQ}).
The minimum of the internal energy $U \left( \boldsymbol{k} \right)$ of Eq. (\ref{HQQ + Hrot in SC with k 2}) is satisfied by the following constraint for the phase deviation of $\delta \varphi \left( \boldsymbol{k} \right)$:
\begin{align}
\label{Value of omega1}
\delta \varphi \left( \boldsymbol{k} \right)
= 4 \omega_{xy} \left( 0 \right)
.
\end{align}
This result means that the finite phase deviation of
$\delta \varphi \left( \boldsymbol{k} \right) = \varphi_{+1, +1} \left( \boldsymbol{k} \right) - \varphi_{-1, -1} \left( \boldsymbol{k} \right) - s\pi \neq 0$
from the stationary point of $s\pi = 2n\pi$ or $s\pi = (2n+1) \pi$ associated with the hexadecapole ordering brings about the spontaneous rotation of $\omega_{xy} \left( 0 \right) \neq 0$ in the superconducting state.

The energy $U \left( \boldsymbol{k} \right)$ of Eq. (\ref{HQQ + Hrot in SC with k 2}) is optimized by the spontaneous rotation $\omega_{xy} \left( 0 \right)$ as
\begin{align}
\label{Value of omega 2}
\omega_{xy} \left( 0 \right) 
= \pm 4 \sqrt{
\frac{N_\mathrm{C} }{ \beta } 
\left[
S_\mathrm{Q} \left( \boldsymbol{k} \right) - \frac{1}{16} \frac{ C_{66}^0}{ N_\mathrm{C} } 
\right]
}
.
\end{align}
Here, we need the criterion
\begin{align}
\label{Condition of SQ}
S_\mathrm{Q} \left( \boldsymbol{k} \right) - \frac{1}{16} \frac{ C_{66}^0}{ N_\mathrm{C} } > 0
.
\end{align}
When the appropriate magnitude of $S_\mathrm{Q} \left( \boldsymbol{k} \right)$ due to the superconducting energy gap $\left| \mathit{\Delta}\left( \boldsymbol{k} \right) \right|$ in Eq. (\ref{SQ}) satisfies the criterion of Eq. (\ref{Condition of SQ}), the spontaneous rotation $\omega_{xy} \left( 0 \right) \neq 0$ of Eq. (\ref{Value of omega 2}) of the macroscopic superconducting state occurs with respect to the host lattice.
The occurrence of the spontaneous rotation $\omega_{xy} \left( 0 \right) \neq 0$ gives the optimized internal energy associated with the superconductivity bearing the hexadecapole as
\begin{align}
\label{UH}
U_\mathrm{H} \left( \boldsymbol{k} \right)
&=  - S_\mathrm{Q} \left( \boldsymbol{k} \right)
- 64 \frac{N_\mathrm{C} }{ \beta } 
\left[ S_\mathrm{Q} \left( \boldsymbol{k} \right) - \frac{1}{16} \frac{ C_{66}^0}{ N_\mathrm{C} }
\right]^2
\nonumber \\
&=  - S_\mathrm{Q} \left( \boldsymbol{k} \right) - \frac{\beta}{4 N_\mathrm{C} } \omega_{xy} (0)^4
.
\end{align}
The first term $- S_\mathrm{Q} \left( \boldsymbol{k} \right)$ in Eq. (\ref{UH}) is the restricted quadrupole interaction energy with the null phase difference of
$\delta \varphi \left( \boldsymbol{k} \right) = \varphi_{+1, +1} \left( \boldsymbol{k} \right) - \varphi_{-1, -1} \left( \boldsymbol{k} \right) - s\pi = 0$ corresponding to
$\cos \left[ \varphi_{+1, +1} \left( \boldsymbol{k} \right) - \varphi_{-1, -1} \left( \boldsymbol{k} \right) \right] = 1$ for the stationary point of 
$s\pi = 2n \pi$ or 
$\cos \left[ \varphi_{+1, +1} \left( \boldsymbol{k} \right) - \varphi_{-1, -1} \left( \boldsymbol{k} \right) \right] = -1$ for the stationary point of 
$s\pi = (2n+1) \pi$ in Eq. (\ref{HQQ by cos}).

The hexadecapole ordering of $\langle H_z^\alpha \rangle \neq 0$ due to the spontaneous deviation $\delta \varphi \left( \boldsymbol{k} \right) \neq 0$ from the stationary point of $s\pi = 2n \pi$ or $s\pi = (2n+1) \pi$ leads to the spontaneous rotation $\omega_{xy} \left( 0 \right) \neq 0$ with respect to the container of the host lattice.
According to the second term in Eq. (\ref{UH}), the spontaneous rotation $\omega_{xy} \left( 0 \right) \neq 0$ further lowers the internal energy $U\left( \boldsymbol{k} \right)$ from the stationary energy $- S_\mathrm{Q} \left( \boldsymbol{k} \right)$ of the restricted quadrupole interaction for the superconducting ground state of Eq. (\ref{Superconducting ground state}).
In the framework of the phenomenological theory for the second-order transition \cite{Landau and Lifshitz}, the quenching of the pair-electron state with the $A_2$ symmetry of the higher-symmetry space group $D_{4h}^{17}$ associated with the spontaneous rotation $\omega_{xy} (0) \neq 0$ brings about a twisted phase with the lower-symmetry space group $C_{4h}^5$ \cite{Inui Group, International Table}.
The spontaneous breaking of the $A_2$ symmetry associated with $\langle H_z^\alpha \rangle \neq 0$ and $\omega_{xy} \left( 0 \right) \neq 0$ loses the symmetry operations consisting of $2C_2'$, $2C_2''$, $2\sigma_\mathrm{v}$, and $2\sigma_\mathrm{d}$ in the space group $D_{4h}^{17}$ of the mother phase.
Consequently, the macroscopic superconducting state spontaneously twists by $\omega_{xy}(0) \neq 0$ with respect to the host lattice.
The hexadecapole of Eq. (\ref{Hexadecapole}) is carried by the two-electron states of Eq. (\ref{psi+- by y'z and zx'}), which consist of two electrons accommodated in the degenerate $y'z$ and $zx'$ band orbitals mapped on the special unitary group SU(2).
Therefore, the appearance of the hexadecapole ordering $\langle H_z^\alpha \rangle \neq 0$ of Eq. (\ref{Hza by sin}) is explained by the symmetry breaking in the direct product of the special unitary group SU(2) $\otimes$ SU(2).

The critical slowing down in the relaxation time $\tau_\mathrm{H}$ observed via the ultrasonic attenuation coefficient $\alpha_{66}$ of the transverse ultrasonic wave around the superconducting transition is explained in terms of the hexadecapole susceptibility for the ferro-type ordering of the hexadecapole $H_{z, \boldsymbol{k}, \boldsymbol{q} = 0 }^\alpha$.
The finite expectation value of the hexadecapole $H_{z, \boldsymbol{k}, \boldsymbol{q} = 0 }^\alpha$ of Eq. (\ref{Hza by sin}) means that the superconducting ground state of Eq. (\ref{Superconducting ground state}) actually bears the hexadecapole.
This result convincingly accounts for why the critical slowing down due to the hexadecapole ordering is observed around the superconducting transition.
The direct detection of the spontaneous rotated state $\omega_{xy} \left( 0 \right) \neq 0$ due to the hexadecapole ordering in the superconducting phase is strongly required in future.

As is presented in Fig. \ref{Fig1}(b), the onset of the critical slowing down in the relaxation time $\tau_\mathrm{H}$ appears in the normal phase far above the superconducting transition temperature of $T_\mathrm{SC} = 23$ K, where fermion quasiparticles are relevant but the off-diagonal long-range ordering of the Cooper pairs does not exist.
This means that the hexadecapole $H_{z, \boldsymbol{k}, \boldsymbol{q} }^\alpha$ of Eq. (\ref{Hexadecapole by Bk}) in the normal phase is actually carried by the fermion quasiparticles but not by the Cooper pairs.
The indirect hexadecapole interaction $D^\mathrm{HH} \left( \boldsymbol{k}, \boldsymbol{q} \right)$ of Eq. (\ref{DHH}) between two-electron states mediated by the rotation $\omega_{xy} \left( \boldsymbol{q} \right)$ of the transverse acoustic phonons with the small-wavenumber limit of $|\boldsymbol{q}| \rightarrow 0$ favors the ferro-type hexadecapole ordering.
This ferro-type hexadecapole interaction reduces the energy of the two-electron states bearing the hexadecapole in both normal and superconducting phases.
This plausibly explains why the system specially favors the Cooper pairs bearing the hexadecapole among the various types of Cooper pairs due to the quadrupole interaction of Eq. (\ref{HQQ by l}).

The indirect hexadecapole interaction Hamiltonian of Eq. (\ref{HindHH}) responsible for the hexadecapole ordering of $H_{z, \boldsymbol{k}, \boldsymbol{q} = 0 }^\alpha$ is mapped to the Ising model.
The critical index of $z\nu = 1$ determined by the relaxation time $\tau_\mathrm{H}$ in the normal phase above the superconducting transition temperature for $x = 0.071$ is in good agreement with mean field theory \cite{Suzuki} and reasonable consistent with the three-dimensional Ising model with $z\nu = 1.2$ for $\nu = 0.63$ and $z = 2.04$ \cite{G. S. Pawley, S. Wansleben}.
The critical index of $z\nu = 1/3$ in the superconducting phase below $T_\mathrm{SC}$, however, considerably deviates from the standard scaling theory with the same critical indices in both the normal and ordered phases \cite{Halperin and Hohenberg}.
This inconsistency of $z\nu = 1/3$ below $T_\mathrm{SC}$ and $z\nu = 1$ above $T_\mathrm{SC}$ is accounted for by the inherent properties of the present system where the superconductivity and the hexadecapole ordering simultaneously appear.

The quenching of the $A_2$ symmetry of the hexadecapole ordering $\langle H_z^\alpha \rangle \neq 0$ accompanied the superconducting transition for $x = 0.071$ changes the symmetry of the mother tetragonal phase with the space group $D_{4h}^{17}$ to the ordered tetragonal phase with the space group $C_{4h}^5$, while the quenching of the $B_2$ symmetry of the quadrupole ordering $\langle O_{x'^2-y'^2} \rangle \neq 0$ accompanying the structural transition for $x = 0.036$ gives the orthorhombic phase with the space group $D_{2h}^{23}$.
Note that the space groups $C_{4h}^5$ for the hexadecapole ordered phase and $D_{2h}^{23}$ for the quadrupole ordered phase are subfamilies of the space group $D_{4h}^{17}$ for the mother tetragonal phase.
The orthorhombic phase of the space group $D_{2h}^{23}$ due to the ferro-quadrupole ordering changes to the superconducting phase below $T_\mathrm{SC} =$ 16.4 K for $x = 0.036$.
The orthorhombicity of $\delta = (a - b)/(a + b) \sim 10^{-3}$ for the lattice parameters $a$ and $b$ in Ba(Fe$_{1-x}$Co$_x$)$_2$As$_2$ with $x = 0.047$-$0.063$ reveals the tendency of $\delta \rightarrow 0$ with decreasing temperature in the superconducting phase \cite{Nandi}.
This reentrant property favoring the tetragonal lattice instead of the distorted orthorhombic lattice in the superconducting phase is consistent with the tetragonal structure $C_{4h}^5$ proposed as the hexadecapole ordering phase in the present paper.
In short, we conclude that the unconventional superconductivity accompanying the hexadecapole ordering in the present system is a common feature across the QCP for $x = 0.061$.

\section{Conclusion}

 In the present work, we investigated the order parameter dynamics around the superconducting and structural transitions in the iron pnictide Ba(Fe$_{1-x}$Co$_x$)$_2$As$_2$ by means of ultrasonic measurements.
The critical slowing down of the relaxation time $\tau$ due to freezing of the order parameter fluctuations associated with the superconducting and structural transitions was verified by the divergence of the ultrasonic attenuation coefficients.
In the analysis of the experiments, we employed the significant fact that the transverse ultrasonic waves used to measure the elastic constant $C_{66}$ simultaneously induce the rotation $\omega_{xy}$ of the antisymmetric tensor and the strain $\varepsilon_{xy}$ of the symmetric tensor.
Taking the band calculations on the iron pnictide into account, we suppose that the degenerate $y'z$ and $zx'$ bands carrying the electric quadrupoles $O_{x'^2 - y'^2}$ and $O_{x'y'}$ and the angular momentum $l_z$ play an essential role in the appearance of the superconducting and structural transitions in the system.
We attempted to clarify the order parameters caused by the symmetry breaking of the space group $D_{4h}^{17}$ of the tetragonal mother phase and the special unitary group SU(2) of the degenerate $y'z$ and $zx'$ bands.

The structural transition in the compound $x = 0.036$ is caused by the ferro-quadrupole ordering of $O_{x'^2 - y'^2}$, which was identified as one of the generator elements of SU(2).
The interaction of the quadrupole $O_{x'^2 - y'^2}$ to the strain $\varepsilon_{xy}$ leads to a structural transition due to the ferro-quadrupole ordering.
The critical slowing down of the relaxation time $\tau_\mathrm{Q}$ by the ultrasonic attenuation coefficient $\alpha_{66}$ and the large softening of the elastic constant $C_{66}$ are well described in terms of the divergence of the quadrupole susceptibility.
The quenching of the quadrupole $O_{x'^2 - y'^2}$ belonging to the $B_2$ irreducible representation of the tetragonal mother phase of the high-symmetry space group $D_{4h}^{17}$ brings about a distorted orthorhombic phase with the low-symmetry space group $D_{2h}^{23}$.

The rotation $\omega_{xy}$ of the transverse ultrasonic waves or transverse acoustic phonons twists the electronic states with angular momentum $l_z$.
The rotation operator of $\exp \bigl[ - i l_z \omega_{xy} \bigr]$ acting on the Hamiltonian $H$ leads to the perturbation of $H_\mathrm{rot} \bigl( \omega_{xy} \bigr) = i \left[ l_z, H \right] \omega_{xy} =\tau_{xy}\omega_{xy}$, where the torque $\tau_{xy}$ is described by the Heisenberg equation $i\hbar \partial l_z/ \partial t =  \left[ l_z, H \right] = i \tau_{xy}$.
For the CEF Hamiltonian of $H = H_\mathrm{CEF}$, the torque $\tau_{xy}$ corresponding to the hexadecapole of one-electron states, however, vanishes because of the rotational invariance for states in the central force in the absence of an external magnetic field.

The rotation operation on the anisotropic quadrupole interaction Hamiltonian consisting of $O_{x'^2 - y'^2}$ and $O_{x'y'}$ gives the interaction of the hexadecapole
$H_z^\alpha \bigl( \boldsymbol{r}_i, \boldsymbol{r}_j \bigr)
= O_{x'y'} \bigl( \boldsymbol{r}_i \bigr) O_{x'^2 - y'^2} \bigl( \boldsymbol{r}_j \bigr) 
+ O_{x'^2 - y'^2} \bigl( \boldsymbol{r}_i \bigr) O_{x'y'} \bigl( \boldsymbol{r}_j \bigr)$
to the rotation $\omega_{xy}$ of the transverse ultrasonic waves.
The critical slowing down of the relaxation time $\tau_\mathrm{H}$ around the superconducting transition is well described in terms of the divergence of the hexadecapole susceptibility.
This result indicates that the ferro-type hexadecapole ordering is caused by the quenching of the two-electron state belonging to the $A_2$ irreducible representation of the tetragonal phase with space group $D_{4h}^{17}$.

Supposing that the quadrupole interaction of $O_{x'^2 - y'^2}$ mediated by the strain $\varepsilon_{xy} \left( \boldsymbol{q} \right)$ of the transverse acoustic phonons with a small wavevector favors the attractive force for pairs of electrons, we solve the superconducting Hamiltonian based on the anisotropic quadrupole interaction.
The superconducting ground state consisting of two different energy gaps for the Cooper pairs possesses the energy dependence of $\cos \left[  \varphi_{+1, +1} \left( \boldsymbol{k} \right) - \varphi_{-1, -1} \left( \boldsymbol{k} \right) \right]$ for the phase difference of $ \varphi_{+1, +1} \left( \boldsymbol{k} \right) - \varphi_{-1, -1} \left( \boldsymbol{k} \right)$ between the two energy gaps.
The Cooper pair bearing the hexadecapole proportional to $\sin \left[  \varphi_{+1, +1} \left( \boldsymbol{k} \right) - \varphi_{-1, -1} \left( \boldsymbol{k} \right) \right]$ couples to the rotation of $\omega_{xy} \left( \boldsymbol{q} \right)$ of the transverse acoustic phonon.
The hexadecapole interaction mediated by the rotation $\omega_{xy} \left( \boldsymbol{q} \right)$ with the small-wavenumber limit of $\left| \boldsymbol{q} \right| \rightarrow 0$ causes the ferro-type hexadecapole ordering in the superconducting phase, which brings about the spontaneous rotation $\omega_{xy} \left( 0 \right) \neq 0$ of the macroscopic superconducting state with respect to the host tetragonal lattice.
This is in good agreement with the ground scenario that the critical slowing down around the superconducting transition is caused by the hexadecapole ordering.
The simultaneous symmetry breaking of the U(1) gauge for the off-diagonal long-range ordering of the Cooper pairs and the freezing of the $A_2$ symmetry of space group $D_{4h}^{17}$ for the ferro-type hexadecapole ordering specify the unconventional superconductivity of the present iron pnictide.

The present experiments on the iron pnictide Ba(Fe$_{1-x}$Co$_x$)$_2$As$_2$ showed that ultrasonic measurements by transverse ultrasonic waves carrying rotation as well as strain are a powerful means of detecting the hexadecapole and quadrupole effects of quantum systems with orbital degrees of freedom.
We hope that the measurements of the hexadecapole as well as the quadrupole by the transverse ultrasonic waves will lead to the discovery of various exotic phenomena associated with magnetic and quadrupole orderings and unconventional superconductivity in strongly correlated electron physics.

\section*{Acknowledgments}
The authors specially thank Prof. Hidetoshi Fukuyama, Prof. Yoshiaki \=Ono, and Prof. Haruhiko Suzuki for stimulating and helpful discussions in the present work.
We are grateful for financial support by Grants-in-Aid for Specially Promoted Research entitled ``Strongly correlated quantum phase associated with charge fluctuation'' (JSPS KAKENHI Grant Number JP18002008), 
Young Scientists (B) entitled ``Elucidation of electric quadrupole effects in multiband superconductors by ultrasonic measurement'' (JP26800184), 
and Scientific Research (C) entitled ``Elucidation of superconductivity in multiband superconductors by ultrasonic measurements under hydrostatic pressures'' (JP17K05538)
from the Ministry of Education, Culture, Sports, Science and Technology of Japan. 
This work was also supported in part by Grants-in-Aid for Scientific Research (A) (JP26247059), Scientific Research (B) (JP21340097), Young Scientists (B) (JP15K17704), Challenging Exploratory Research (JP23654116 and JP15K13518), and the Strategic Young Researcher Overseas Visits Program for Accelerating Brain Circulation from the Japan Society for the Promotion of Science.

Most of this work is based on a thesis submitted to partially fulfill the requirement for the Ph.D degree in Science of Ryosuke Kurihara of the Graduate School of Science and Technology, Niigata University, Niigata (2016).


\end{document}